\documentclass[conference]{IEEEtran}

\usepackage[utf8]{inputenc}
\usepackage[bookmarks=false,colorlinks=true]{hyperref}
\usepackage{amsmath,amssymb,amsfonts,cite,algorithmic,textcomp,subfigure,booktabs,multirow,enumitem,xcolor,graphicx,microtype}

\graphicspath{{images/}}
\hypersetup{urlcolor=magenta,linkcolor=red,citecolor=red}
\newtheorem{theorem}{Theorem}

\begin{document}

\title{Modeling Updates of Scholarly Webpages Using Archived Data}

\author{
\IEEEauthorblockN{\textbf{Yasith Jayawardana}}
\IEEEauthorblockA{Computer Science Department \\
Old Dominion University \\
Norfolk, VA, USA \\
\texttt{yasith@cs.odu.edu}}
\and
\IEEEauthorblockN{\textbf{Alexander C. Nwala}}
\IEEEauthorblockA{Center for Complex Networks and Systems Research \\
Luddy School of Informatics, Computing, and Engineering \\
Indiana University, Bloomington, IN, USA \\
\texttt{anwala@iu.edu}}
\and
\IEEEauthorblockN{\textbf{Gavindya Jayawardena}}
\IEEEauthorblockA{Computer Science Department \\
Old Dominion University \\
Norfolk, VA, USA \\
\texttt{gavindya@cs.odu.edu}}
\and
\IEEEauthorblockN{\textbf{Jian Wu}}
\IEEEauthorblockA{Computer Science Department \\
Old Dominion University \\
Norfolk, VA, USA \\
\texttt{jwu@cs.odu.edu}}
\and
\IEEEauthorblockN{\textbf{Sampath Jayarathna, Michael L. Nelson}}
\IEEEauthorblockA{Computer Science Department \\
Old Dominion University \\
Norfolk, VA, USA \\
\texttt{\{sampath,mln\}@cs.odu.edu}}
\and
\IEEEauthorblockN{\textbf{C. Lee Giles}}
\IEEEauthorblockA{Information Sciences \&~Technology \\
Pennsylvania State University \\
University Park, PA, USA \\
\texttt{giles@ist.psu.edu}}
}

\IEEEoverridecommandlockouts
\IEEEpubid{\makebox[\columnwidth]{978-1-7281-6251-5/20/\$31.00~\copyright2020 IEEE \hfill} \hspace{\columnsep}\makebox[\columnwidth]{ }}

\maketitle

\begin{abstract}
The vastness of the web imposes a prohibitive cost on building large-scale search engines with limited resources.
Crawl frontiers thus need to be optimized to improve the coverage and freshness of crawled content.
In this paper, we propose an approach for modeling the dynamics of change in the web using archived copies of webpages.
To evaluate its utility, we conduct a preliminary study on the scholarly web using 19,977 seed URLs of authors' homepages obtained from their Google Scholar profiles.
We first obtain archived copies of these webpages from the Internet Archive (IA), and estimate when their actual updates occurred.
Next, we apply maximum likelihood to estimate their mean update frequency ($\lambda$) values.
Our evaluation shows that $\lambda$ values derived from a short history of archived data provide a good estimate for the true update frequency in the short-term, and that our method provides better estimations of updates at a fraction of resources compared to the baseline models.
Based on this, we demonstrate the utility of archived data to optimize the crawling strategy of web crawlers, and uncover important challenges that inspire future research directions.
\end{abstract} 


\section{Introduction}
The sheer size of the Web makes it impossible for small crawling infrastructures to crawl the entire Web to build a general search engine comparable to Google or Bing.
Instead, it is more feasible to build specialized search engines, which employ \textit{focused web crawlers}~\cite{chakrabarti1998automatic, chakrabarti1999focused} to actively harvest webpages or documents of particular topics or types.
Google Scholar, for instance, is a specialized search engine that is especially useful for scientists, technicians, students, and other researchers to find scholarly papers.

The basic algorithm for a focused web crawler is straightforward. The crawl frontier is first initialized with seed URLs that are relevant to the search engine's focus.
Next, the crawler visits webpages referenced by seed URLs, extracts hyperlinks in them, selects hyperlinks that satisfy preset rules (to ensure that only related webpages are visited), adds them to the crawl frontier, and repeats this process until the crawl frontier exhausts~\cite{OlstonC10}.
Although this works for relatively short seed lists, it does not scale for large seed lists.
For instance, the crawler may not finish visiting all webpages before they change.
Given such circumstances, re-visiting web pages that have not changed since their last crawl is a waste of time and bandwidth.
It is therefore important to select and prioritize a subset of seeds for each crawl, based on their likeliness to change in the future.

Without sufficient crawl history, it is difficult to accurately predict when a webpage will change.
Web archives, such as the well-known Internet Archive's~(IA) Wayback Machine~\cite{tofel2007wayback} and others, preserve webpages as they existed at particular points in time for later replay.
The IA has been collecting and saving public webpages since its inception in 1996, and contains archived copies of over 424 billion webpages~\cite{Gomes:2011:SWA:2042536.2042590,2017-ndsa-web-archiving-survey}.
The resulting record of such archived copies is known as a \emph{TimeMap}~\cite{nelson:memento:tr} and allows us to examine each saved copy to determine if a change occurred (not every saved version will represent a change in the webpage).
TimeMaps provide a critical source of information for studying changes in the web.
For example, if a researcher created his website in 2004, via a TimeMap we could retrieve copies of the website observed by the IA between 2004 and 2020, and examine these copies for changes.

In this paper, we propose an approach to model the dynamics of change in the web using archived copies of webpages.
Though such dynamics have been studied in previous papers, e.g.,~\cite{koehler:web-page,cho2003estimating,radinsky2013predicting}, online activities have evolved since then, and to the best of our knowledge, the use of archived data to model these dynamics has not been explored.
While many web archives exist, we use the IA to obtain archived copies of webpages due to its high archival rate, and efficiency of mass queries.
Given a URL, we first obtain its TimeMap from the IA's Wayback Machine, and identify mementos that represent updates.
Next, we use this information to estimate their mean update frequency ($\lambda$).
We then use $\lambda$ to calculate the probability ($p$) of seeing an update $d$ days after it was last updated.
Before each crawl, we repeat this process for each seed URL and use a threshold ($\theta$) on $p$ to select a subset of seed URLs that are most likely to have changed since their next crawl.

Our preliminary analysis demonstrates how this approach can be integrated into a focused web crawler, and its impact on the efficiency of crawl scheduling.
Here, we select the scholarly web as our domain of study, and analyze our approach at both homepage-level (single webpage) and at website-level (multiple webpages).
The former, investigates changes occurring on an author's homepage, while the latter, investigates changes occurring collectively on the homepage and any webpage behind it, e.g., \emph{publications}, \emph{projects}, and \emph{teaching} webpages.
Our contributions are as follows:
\begin{enumerate}
\item
We studied the dynamics of the scholarly web using archived data from the IA for a sample of 19,977 authors' websites.
\item
We verified that the updates to authors' websites and homepages follow a near-Poisson distribution, with spikes that may represent non-stochastic activities.
\item
We developed \textit{History-Aware Crawl Scheduler} (HACS), which uses archived data to find and schedule a subset of seed URLs that are most likely to have changed before the next crawl.
\item
We compared HACS against baseline models for a simulated web crawling task, and demonstrated that it provides better estimations.
\end{enumerate}

\subsection{Crawling the Web}
Although the basic focused web crawling algorithm~\cite{OlstonC10} is simple, challenges in the web, such as scale, content selection trade-offs (e.g., coverage vs freshness), social obligations, and adversaries, makes it infeasible to crawl the web in that manner.
Crawl frontiers should thus be optimized to improve the robustness of web crawlers.
One approach is to reorder the crawl frontier to maximize some goal (e.g., bandwidth, freshness, importance, relevance)~\cite{coffman1998optimal,castillo2004scheduling}.
\textit{Fish-Search}~\cite{de1994information}, for instance, reorders the crawl frontier based on content relevance, and is one of the earliest of such methods.
Given a seed URL and a driving query, it builds a priority queue that prioritizes webpages (and their respective out-links) that match the driving query.
\textit{Shark-Search}~\cite{hersovici1998shark} is an improved version of Fish-Search which uses cosine similarity (number between 0 and 1) to calculate the relevance of a webpage to the driving query, instead of binary similarity (either 0 or 1) used in Fish-Search.
Such algorithms do not require the crawl history to calculate relevance, and can be applied at both the initial crawl and any subsequent crawls.

In incremental crawling, webpages need to be re-visited once they change, to retain the freshness of their crawled copies.
Several methods have been proposed~\cite{shipman2019crawling,gossen2015icrawl}.
Olston et.~al.~\cite{olston2008longevity}, for instance, studied the webpage revisitation policy that a crawler should employ to achieve good freshness.
They considered information longevity, i.e., the lifetime of content fragments that appear and disappear from webpages over time, to avoid crawling ephemeral content such as advertisements, which have limited contribution to the main topic of a webpage.
Such methods require sufficient crawl history to identify ephemeral content, and until sufficient crawl history is generated, the algorithm may yield sub-optimal results.

Algorithms proposed by Cho et al.~\cite{cho1998efficient}, reorders the crawl frontier based on the importance of webpages.
Here, the \textit{query similarity} metric used in Fish-Search and Shark-Search was extended with additional metrics such as, \textit{back-link count}, \textit{forward-link count}, \textit{PageRank}, and \textit{location} (e.g., URL depth, top-level domain).
Alam et al.~\cite{alam2012novel} proposed a similar approach, where the importance of a webpage was estimated using \textit{PageRank}, \textit{partial link structure}, \textit{inter-host links}, \textit{webpage titles}, and \textit{topic relevance} measures.
Although such methods take advantage of the crawl history, the importance of a webpage may not reflect how often it changes.
Thus, such methods favour the freshness of certain content over the others.

Focused web crawlers should ideally discover all webpages relevant to its focus.
However, the coverage that it could achieve depends on the seed URLs used.
Wu et al.~\cite{wu2012whitelist}, for instance, proposed the use of a whitelist and a blacklist for seed URL selection.
The whitelist contains high-quality seed URLs selected from parent URLs in the crawl history, while the blacklist contains seed URLs that should be avoided.
The idea was to concentrate the workforce to exploit URLs with potentially abundant resources.
In addition, Zheng et al.~\cite{zheng2009cikm} proposed a graph-based framework to select seed URLs that maximize the value (or score) of the portion of the web graph ``covered'' by them.
They model this selection as a \emph{Maximum K-Coverage Problem}.
Since this is a NP-hard~\cite{dorit1998analysis} problem, the authors have proposed several greedy and iterative approaches to approximate the optimal solution.
Although this works well for a general web crawler, studies show that the scholarly web has a disconnected structure~\cite{thelwall2003graph}.
Hence, the process of selecting seed URLs for such use cases may benefit from the crawl records of a general web crawler.

CiteSeerX~\cite{giles1998jcdl} is a digital library search engine that has more than 10 million scholarly documents indexed and is growing~\cite{wu2019citeseerx}.
Its crawler, identified as \textit{citeseerxbot}, is an incremental web crawler that actively crawls the scholarly web and harvests scholarly papers in PDF format~\cite{wu2019citeseerx}.
Compared to general web crawlers, crawlers built for the scholarly web has different goals in terms of optimizing the freshness of their content.
The crawl scheduling model used by \textit{citeseerxbot}, which we refer to as the \textit{Last-Obs} model, prioritizes seed URLs based on the \textit{time elapsed since a webpage was last visited}.
In this work, we use the \textit{Last-Obs} model as a baseline to compare with our method.

\subsection{Modeling Updates to a Webpage}

\begin{figure*}[ht]
\centering
\includegraphics[width=.7\linewidth]{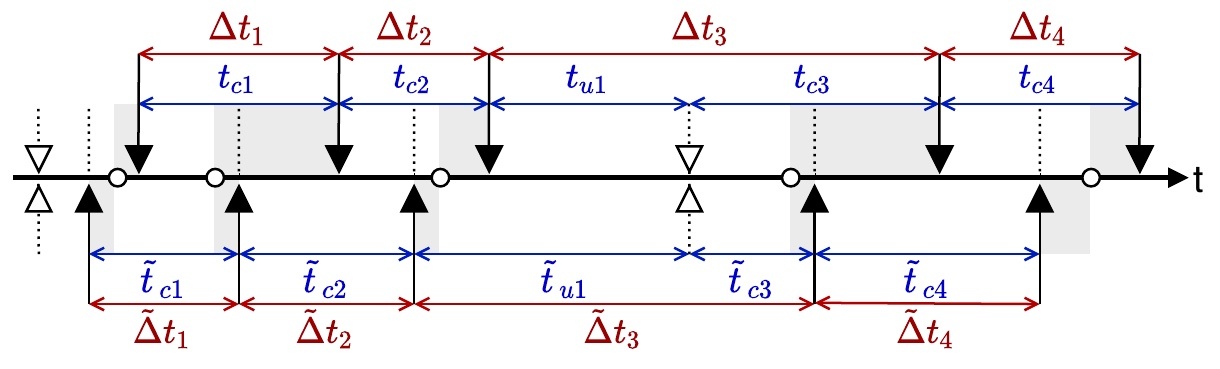}
\caption{
An illustration of accesses ( {$\triangledown$},{$\triangle$} ), accesses with updates ( {$\blacktriangledown$} ), true update occurrences ( {$\circ$} ) and the interpolated update occurrences ( {$\blacktriangle$} ) over time.
Gray shades represent the deviation of the observed and interpolated update occurrences from the true update occurrences. }
\label{fig:changedatetime}
\end{figure*}

Updates to a webpage can be modeled as a Poisson process~\cite{cho2000evolution,cho2003estimating,cho2000synchronizing}.
The model is based on the following theorem.
\begin{theorem}
\label{poissonprocesstheorem}
If $T$ is the time of occurrence of the next event in a Poisson process with rate $\lambda$ (number of events per unit time period), the probability density for $T$ is
\begin{equation}
\label{eq:poisson}
f_T(t)=\lambda e^{-\lambda t},\quad t>0,\quad\lambda>0.
\end{equation}
\end{theorem}
Here, we assume that each update event is independent.
While this assumption is not always true (i.e. certain updates are correlated), as shown later, it is a reasonable estimation.
By integrating $f_T(t)$, we obtain the probability that a certain webpage changes in interval $[t_0, t]$:
\begin{equation}
\label{eq:cumulative_p}
P(\Delta t)=\int^t_{t_0}f_T(t)\,dt=1-e^{-\lambda \Delta t}
\end{equation}
Note that the value of $\lambda$ may vary for different webpages. 
For the same webpage, $\lambda$ may also change over time but for a short period of time, $\lambda$ is approximately constant.
Therefore, by estimating $\lambda$, we calculate how likely a webpage will be updated since its last update at time $t_c$.
Intuitively, $\lambda$ can be estimated using,
\begin{equation}
\label{eq:estimater_biased}
\hat{\lambda}=X/T
\end{equation}
in which $X$ is the number of updates detected during $n$ accesses, and $T$ is the total time elapsed during $n$ accesses.
As proven in~\cite{cho2003estimating}, this estimator is biased and it is more biased when there are more updates than accesses in the interval $T$.
For convenience~\cite{cho2000evolution} defines an intermediate statistical variable $r=\lambda/f$, the ratio of the update frequency to the access frequency.
An improved estimator was proposed below:
\begin{equation}
\label{eq:estimater}
\hat{r}=-\log\left(\frac{\bar{X}+0.5}{n+0.5}\right),\quad\bar{X}=n-X.
\end{equation}
This estimator is much less biased than $X/T$ and i
It is also consistent, meaning that as $n\rightarrow\infty$, the expectation of $\hat{r}$ is $r$.

Unfortunately, since archival rates of the IA depend on its crawl scheduling algorithm and the nature of the webpages themselves, its crawl records have irregular intervals.
As a result, archived copies may not reflect every update that occurred on the live web, and not all consecutive archived copies may reflect an update.
Since both Eq.~(\ref{eq:estimater_biased}) and Eq.~(\ref{eq:estimater}) assume regular access, they cannot be used directly.
To address this limitation, we use a \emph{maximum likelihood estimator} to calculate which $\lambda$ is most likely to produce an observed set of events.
\begin{equation}
\label{eq:mle}
\sum^m_{i=1}\frac{t_{c_i}}{\exp{(\lambda t_{c_i})-1}}=\sum^{n-m}_{j=1}t_{u_j},
\end{equation}
Here, $t_{c_i}$ is the $i$-th time interval where an update was detected, $t_{u_j}$ is the $j$-th time interval where an update was \emph{not} detected, and $m$ is the total number of updates detected from $n$ accesses (see Figure~{\ref{fig:changedatetime}}).
$\lambda$ is calculated by solving Eq.~({\ref{eq:mle}}).
Since this equation is nonlinear, we solve it numerically using Brent's method~\cite{brent1971algorithm}.
There is a special case when $m=n$ (i.e. updates detected at all accesses) where solving Eq.~(\ref{eq:mle}) yields $\lambda=\infty$.
In this case, Eq.({\ref{eq:mle}})'s solution is infinity and Eq.({\ref{eq:estimater}}) is used.

To the best of our knowledge, there has not been an open source crawl scheduler for the scholarly web that takes advantage of the update model above.
With IA providing an excellent, open-accessible resource to model the updates of scholarly webpages, this model can be applied on focused crawl schedulers to save substantial time on crawling and re-visitation.

\section{Methodology}

\subsection{Data Acquisition}
\label{data}
\begin{figure*}[ht]
\centering
\includegraphics[width=.7\linewidth]{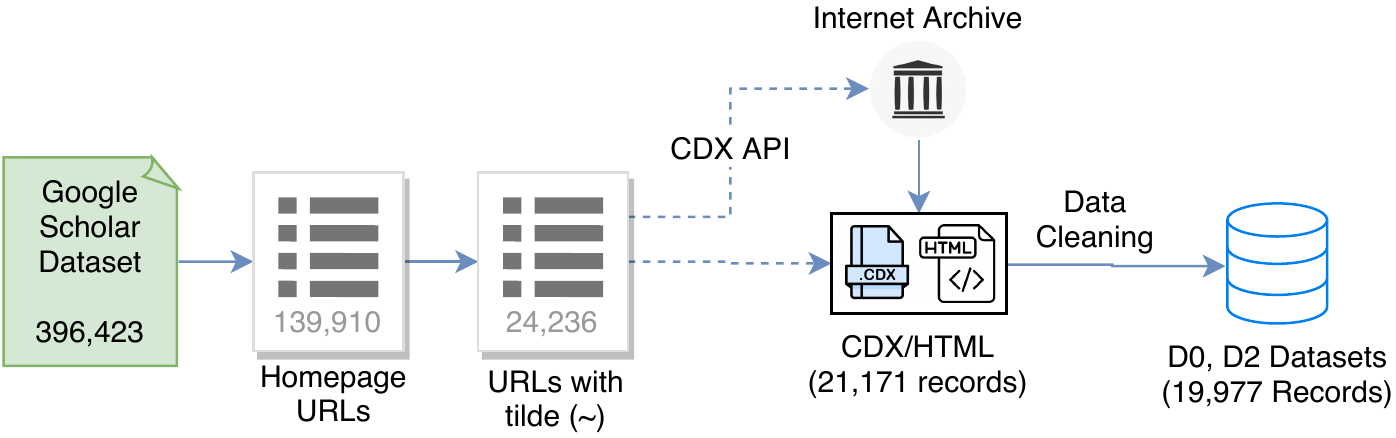}
\caption{Steps followed to acquire and prepare data from IA (depths 0--2).}
\label{fig:flow_of_data}
\end{figure*}

The seed list used in this work was derived from a dataset containing Google Scholar profile records of \textit{396,423} researchers.
This dataset was collected around 2015 by scraping profile webpages in Google Scholar with a long crawl-delay.
The steps for data acquisition and preparation are illustrated in Figure {\ref{fig:flow_of_data}}.

{\bf Step~1:} From the Google Scholar profile records, we discovered \textit{139,910} profiles that provided homepage URLs.
These URLs referenced either individual author homepages, or organizational websites.
Since our study focused on modeling the dynamics of the websites of individual authors, we removed organizational websites.
This was nontrivial using a simple rule-based filter as there were personal homepages that look similar to organizational homepages.
Therefore, we restricted our scope to homepage URLs hosted \textit{within a user directory} of an institution, i.e., URLs with a tilde ($\sim$) in them 
(e.g., \url{foo.edu/~bar/}).
In this manner, we obtained \textit{24,236} homepage URLs.

{\bf Step~2:} Next, we performed a wildcard query on the IA Wayback CDX Server API~\cite{IA_CDX_Server} to obtain TimeMaps for each author website under their homepage URL.
Out of \textit{24,236} websites, we obtained TimeMaps for \textit{21,171} author websites ($87.35\%$ archival rate).
The remaining websites were either not archived, or the CDX Server API returned an error code during access.
The resulting TimeMaps provided information such as the crawl timestamps and URI-Ms of archived copies of each webpage.
From these webpages, we selected webpages at depth~${\leq}2$ (Depth 0 is the homepage).
For instance, for a homepage \url{foo.edu/~bar}, a link to \url{foo.edu/~bar/baz} is of depth 1 and is selected.
However a link to \url{foo.edu/~bar/baz/qux/quux} is of depth 3 and is not selected.

{\bf Step~3:} Next, we generated the \textbf{D0 dataset} and \textbf{D2 dataset}, which we use in our analysis.
First, we de-referenced the URI-Ms of each URL selected in Step 2, and saved their HTML for later use.
When doing so, we dropped inconsistent records such as records with invalid checksum, invalid date, multiple depth 0 URLs, and duplicate captures from our data.
The resulting data, which we refer to as the \textbf{D2 dataset}, contained HTML of \textbf{19,977} websites, totaling \textbf{581,603} individual webpages.
The average number of webpages per website is 227.49.
The minimum and maximum number of webpages per website are 1 and 35,056, respectively.
We selected a subset of the \textbf{D2 dataset} consisting HTML of only the \textbf{19,977} homepages, which we refer to as the \textbf{D0 dataset}.

\begin{figure}[ht]
\centering
\includegraphics[width=\linewidth,trim={0 0 0 20},clip]{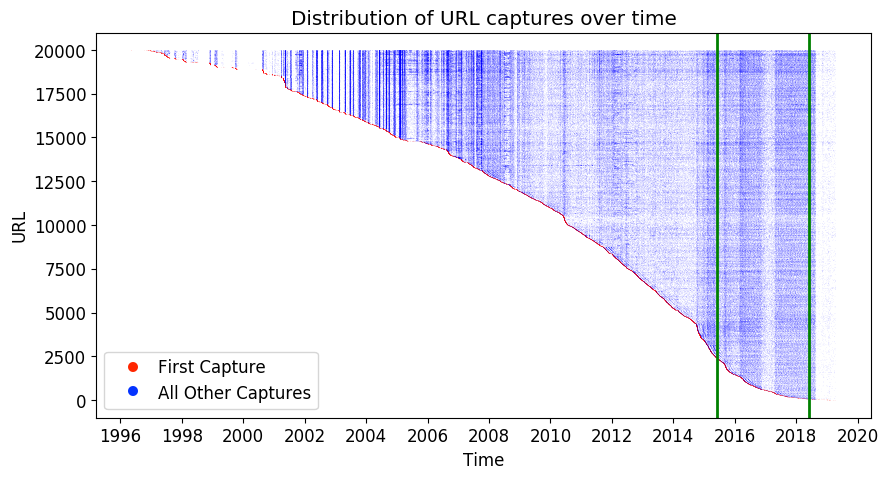}
\caption{
Captures (blue dots) of homepage URLs over time, with URLs sorted by their earliest capture time (red dots).
The captures between 2015-06-01 and 2018-06-01 (green vertical lines) were used for the evaluation.}
\label{fig:scatterplot}
\end{figure}

Figure~{\ref{fig:scatterplot}} shows the distribution of captures in the D0 dataset, sorted by their earliest capture time.
Here, the median crawl interval of $80\%$ of author homepages were between $20-127$ days.
The distribution of capture density over time suggests that the capture densities of IA vary irregularly with time.
For instance, captures during 2015--2018 show a higher density on average than the captures during 2010--2014.
Since high-cadence captures help to obtain a better estimation for the update occurrences, we scoped our analysis to the period between June 1, 2015 and June 1, 2018 (shown by green vertical lines in Figure~{\ref{fig:scatterplot}}).

\subsection{Estimating Mean Update Frequency}
The exact interpretation of update may differ depending on the purpose of study.
We examine a specific type of update -- \textbf{the addition of new links}.
The intuition here is to identify when authors add \textit{new publications} into their webpages, as opposed to identifying when that webpage was updated in general.
We claim that this interpretation of update is more suited to capture such behavior.

For each webpage in datasets \textbf{D0} and \textbf{D2}, we processed each capture $m_i$ to extract links $l(m_i)$ from its HTML, where $l(m_i)$ is the set of links in the $i^{th}$ capture.
Next, we calculated $\left\vert{l^*(m_i)} \right\vert$, i.e., the number of links in a capture $m_i$ that was never seen before $m_i$, for each capture in these datasets.
Formally,
\[l^*(m_i) = l(m_i) - \cup_{k=1}^{i-1}{l(m_k)},\quad i\geq2.\]
and $\cup_{k=1}^{i-1}{l(m_k)}$ is the union of links from captures $m_1$ to $m_{i-1}$.
Finally, we calculated the observed-update intervals $t_{c_i}\,{\in}\,T_c$ and observed non-update intervals $t_{u_j}\,{\in}\,T_u$ based on captures that show link additions, i.e., $l^*(m_i) > 0$ and ones that do not, i.e., $l^*(m_i) = 0$ (see Figure~\ref{fig:changedatetime}).
We estimate $\lambda$ in two ways.

\subsubsection{Estimation Based on Observed Updates}
\label{sec:est_lambda}
For each webpage, we substituted $t_{c_i}$ and $t_{u_j}$ values into Eq.~{(\ref{eq:mle})} or Eq.({\ref{eq:estimater}}) and solved for $\lambda$ using Brent's method to obtain its \textbf{estimated mean observed-update frequency} ($\lambda$).
In this manner, we calculated $\lambda$ for author websites at both homepage-level (using \textbf{D0} dataset) and webpage-level (using \textbf{D2} dataset).

\begin{figure}[ht]
\centering
\includegraphics[width=\linewidth,trim={30 5 40 40},clip]{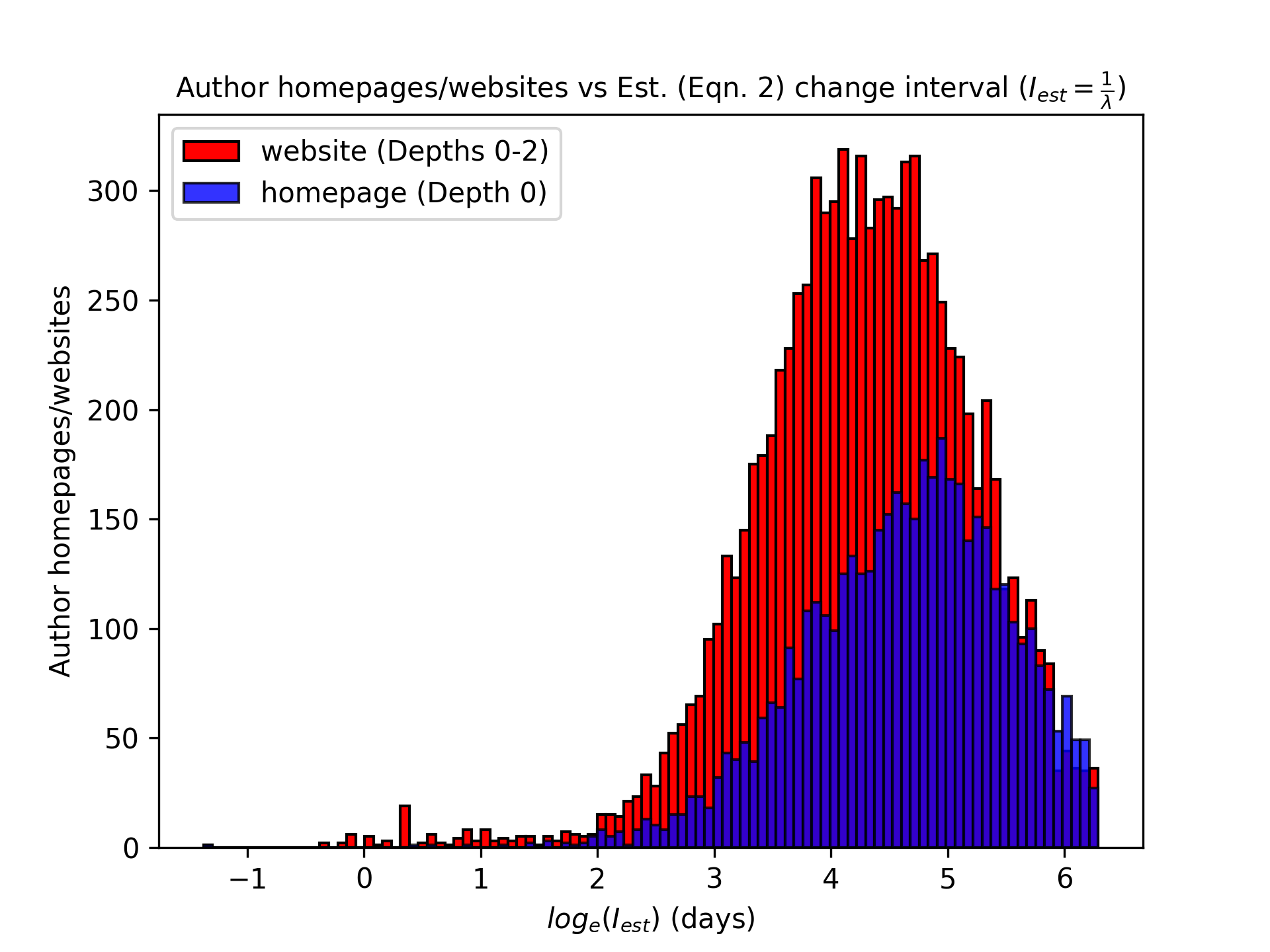}
\caption{
Distribution of $1/\lambda$ of author websites at website-level (red) and homepage-level (blue).
Here, $\lambda$ was calculated using captures from 2015-06-01 to 2018-06-01.
}
\label{fig:est_chg_interval}
\end{figure}

Figure~{\ref{fig:est_chg_interval}} shows the distribution of $I_{\rm est}=1/\lambda$ at both website-level and homepage-level, obtained using captures from 2015-06-01 to 2018-06-01.
Both distributions are approximately log-normal, with a median of $74$ days at website-level, and of $110$ days at homepage-level.
This suggests that most authors add links to their homepage less often than they add links to their website (e.g., \textit{publications}).

\subsubsection{Estimation Based on Interpolated Updates}
\label{sec:validate}
The method described in Section~\ref{sec:est_lambda} calculates the maximum likelihood of observing the updates given by intervals $t_{c_i}$ and $t_{u_j}$.
Intuitively, an update could have occurred at any time between $t(m_{x-1})$~and~$t(m_{x})$, where $t(m_{x})$ is the time of an updated capture, and $t(m_{x-1})$ is the time when the capture before it was taken.
Here, we use an improved method where we first interpolate when a URL was updated.
We define interpolated-update time~($\blacktriangle$) as $(t(m_{x-1})+t(m_{x}))/2$, i.e., the midpoint between $t(m_{x})$ and $t(m_{x-1})$.
Next, we obtain the update intervals $\tilde{t}_{c_i}$ and $\tilde{t}_{u_j}$ from these interpolated updates, and use them to calculate the \textbf{estimated mean interpolated-update frequency}~($\tilde{\lambda}$).

\subsection{Distribution of Updates}

\begin{figure}[ht]
\centering
\subfigure[Homepage-level]{
\label{fig:scatterplot_homepage}
\includegraphics[width=\linewidth,trim={20 5 40 40},clip]{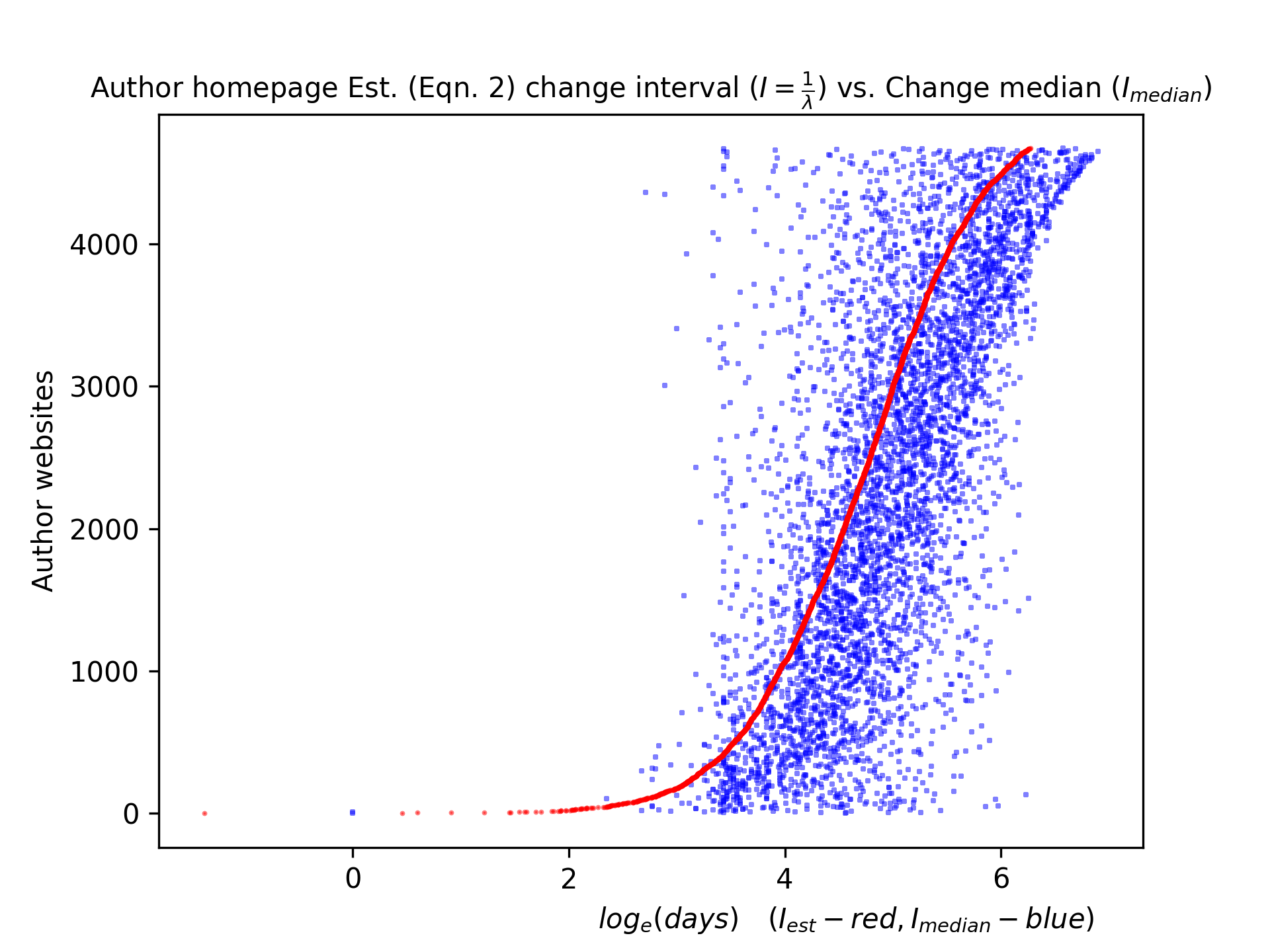}
}
\subfigure[Website-level]{
\label{fig:scatterplot_website}
\includegraphics[width=\linewidth,trim={20 5 40 40},clip]{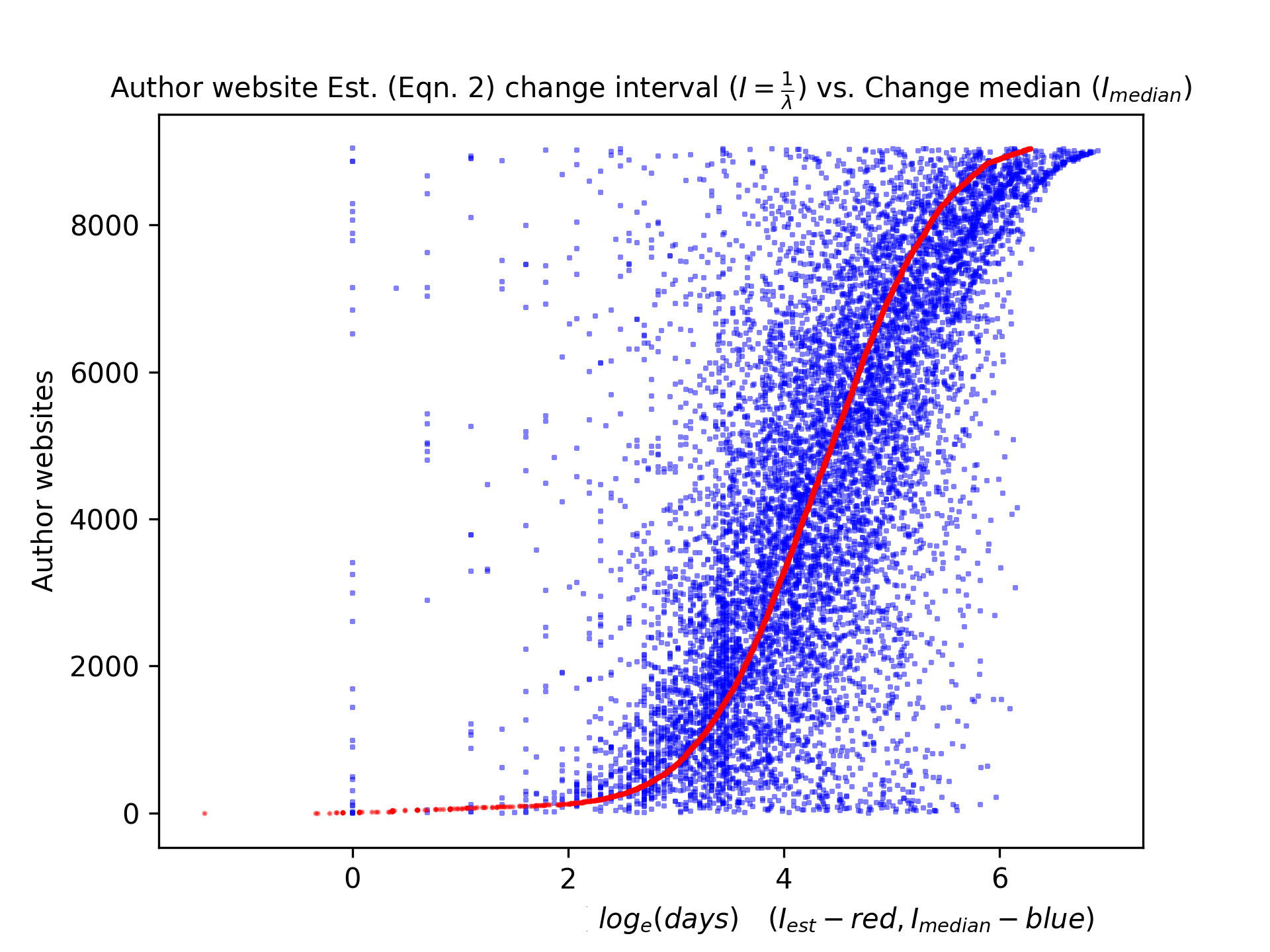}
}
\caption{
The distributions of $1/\tilde{\lambda}$ (red) and the median \textit{interpolated-update interval}~($\tilde{{\Delta}t}$) (blue) of author websites at (a) homepage-level and (b) website-level.
The $y$-axis represents individual author websites, in the increasing order of $1/\tilde{\lambda}$.
}
\label{fig:update-time-est}
\end{figure}

Figure~{\ref{fig:update-time-est}} shows the distribution of $1/\tilde{\lambda}$ (red) and the median \textit{interpolated-update interval}~($\tilde{{\Delta}t}$) (blue) of author websites at both homepage-level and website-level.
It suggests that the distribution of $1/\tilde{\lambda}$ is consistent with the distribution of median $\tilde{{\Delta}t}$ at both homepage-level and website-level.

\subsection{Poisson Distribution}
Next, we observe whether updates to author websites follow a Poisson distribution, at both homepage-level and website-level.
Here, we group author websites by their calculated $1/\tilde{\lambda}$ values into bins having a width of 1 day.
Within each bin, we calculate the probability (y-axis) of finding an author website having an interpolated-update interval~($\tilde{\Delta}{t}$) of $d$ days (x-axis).

\begin{figure*}[ht]
\centering
\subfigure[Homepage-level, $1/\tilde{\lambda}=35$ days]{
\label{fig:poissonplot-page-35d}
\includegraphics[width=.45\linewidth,trim={0 0 0 40},clip]{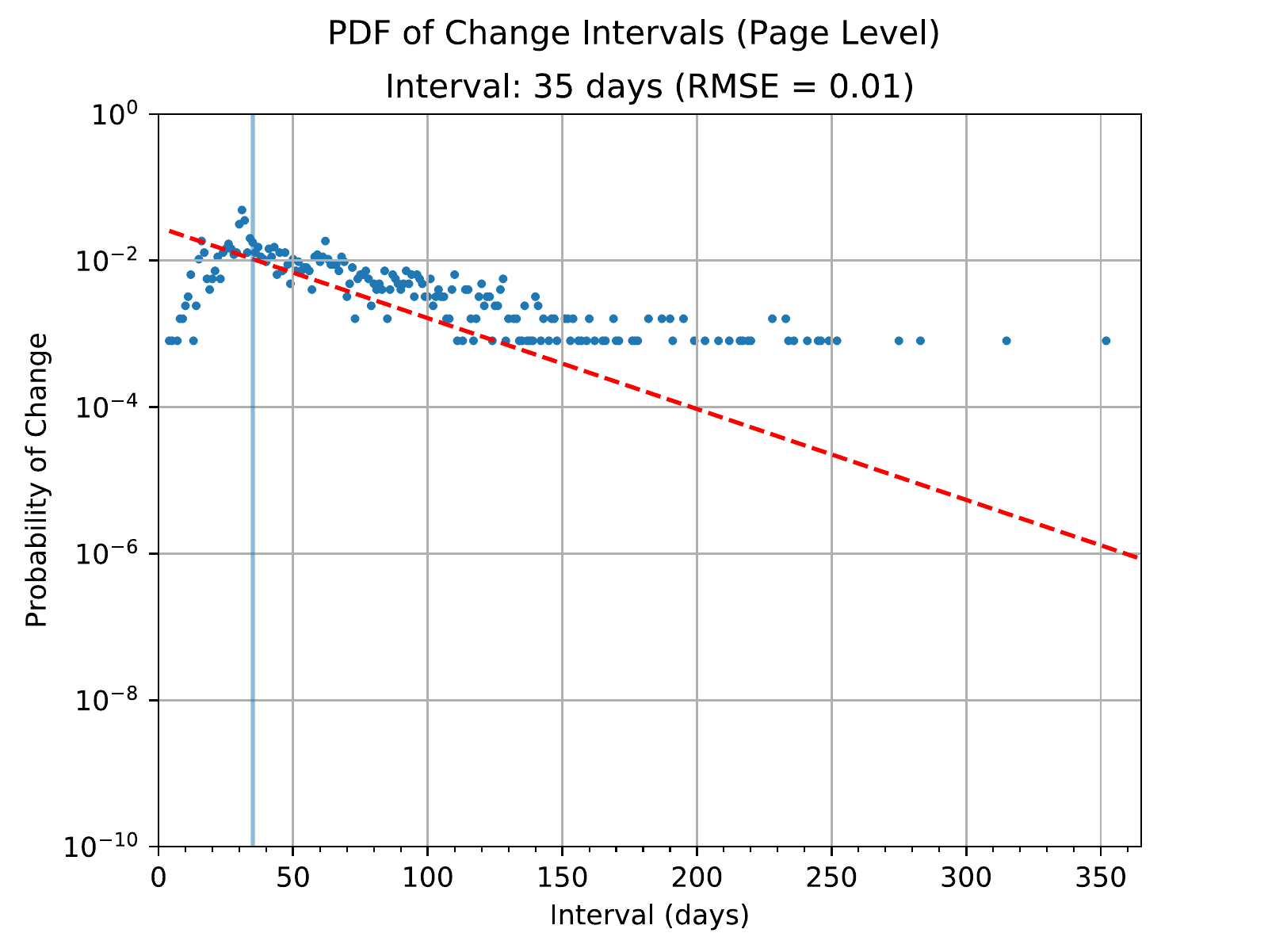}
}
\enskip
\subfigure[Homepage-level, $1/\tilde{\lambda}=70$ days]{
\label{fig:poissonplot-page-70d}
\includegraphics[width=.45\linewidth,trim={0 0 0 40},clip]{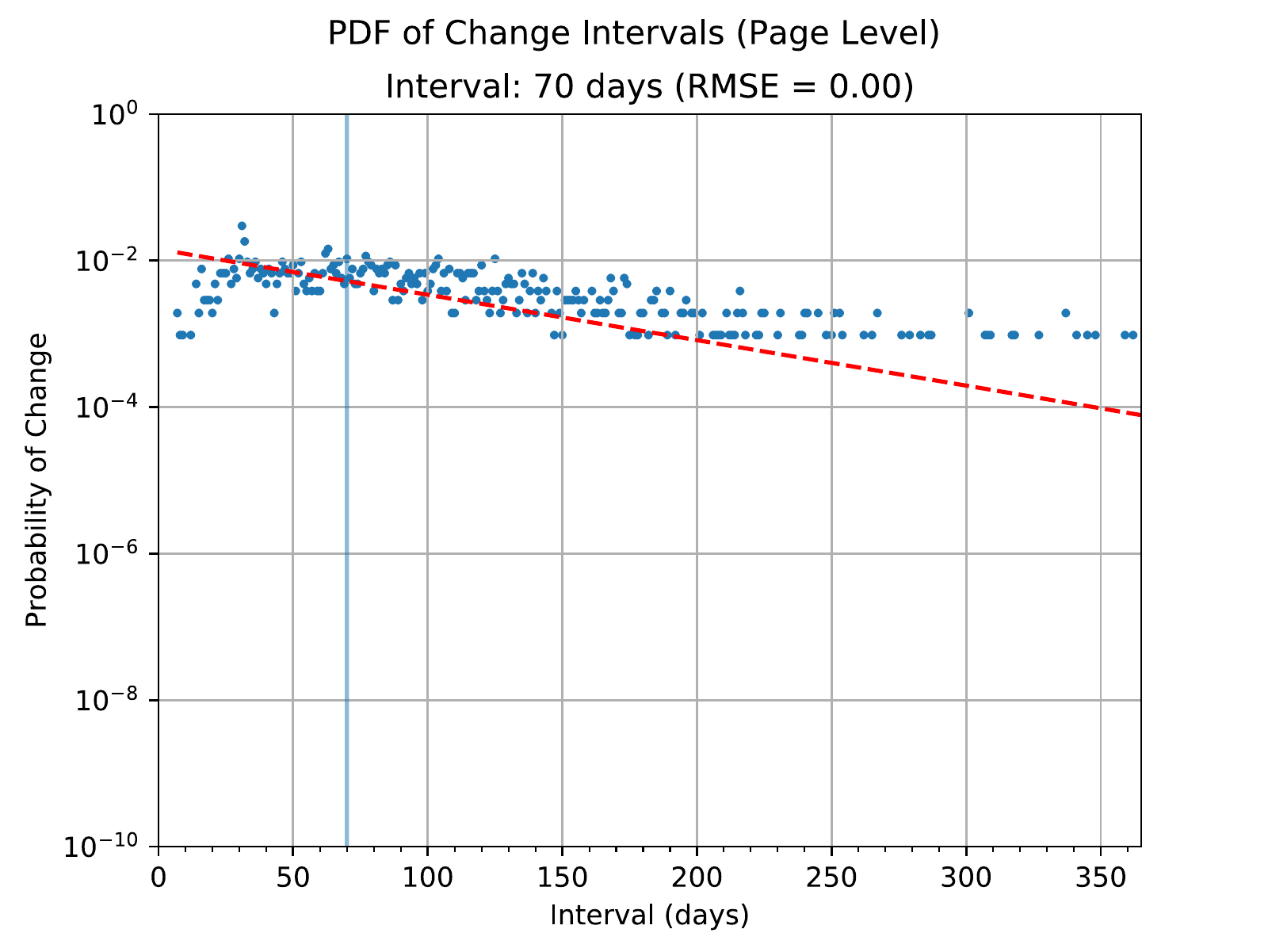}
}
\enskip
\subfigure[Website-level, $1/\tilde{\lambda}=35$ days]{
\label{fig:poissonplot-site-35d}
\includegraphics[width=.45\linewidth,trim={0 0 0 40},clip]{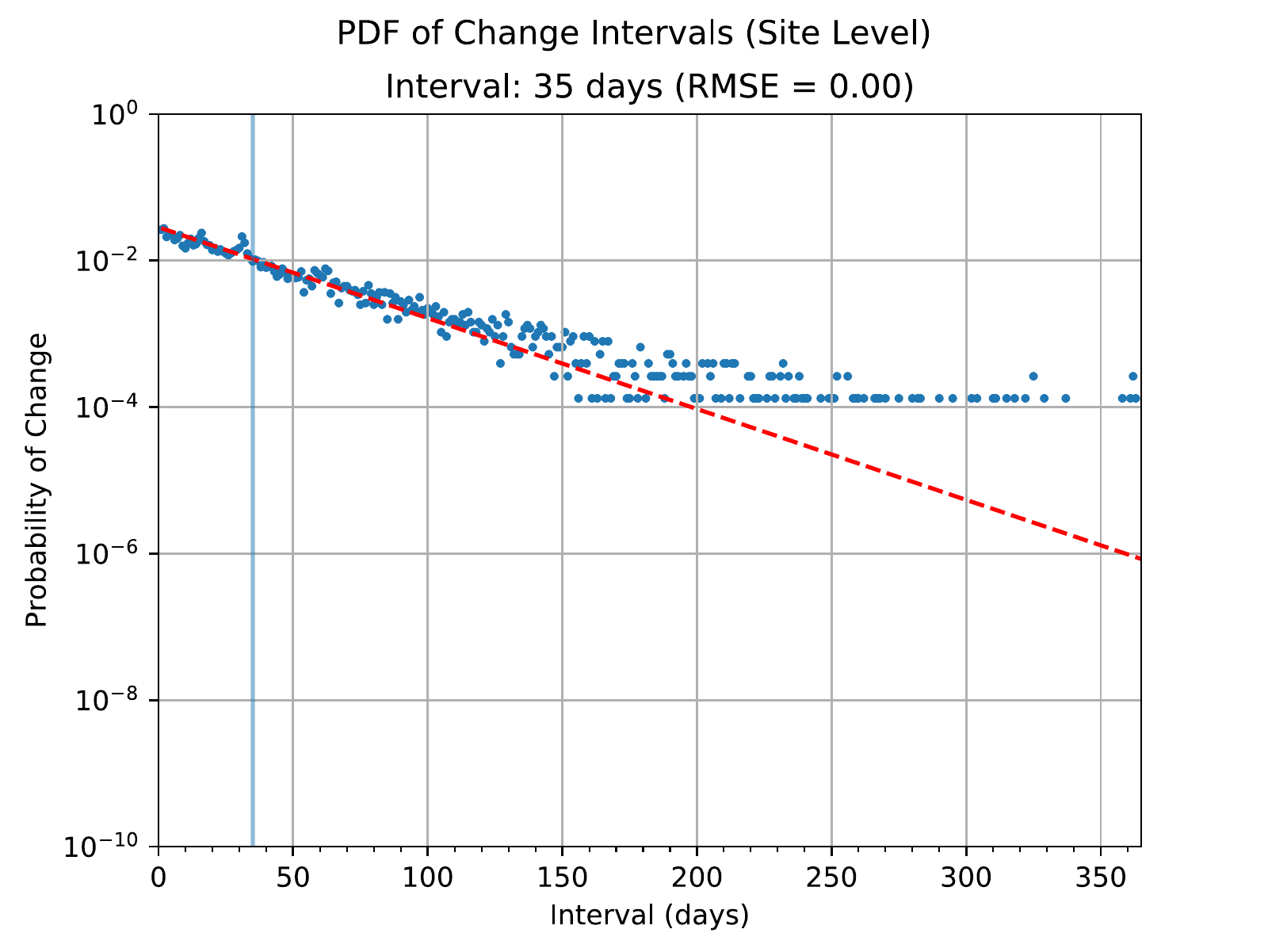}
}
\enskip
\subfigure[Website-level, $1/\tilde{\lambda}=70$ days]{
\label{fig:poissonplot-site-70d}
\includegraphics[width=.45\linewidth,trim={0 0 0 40},clip]{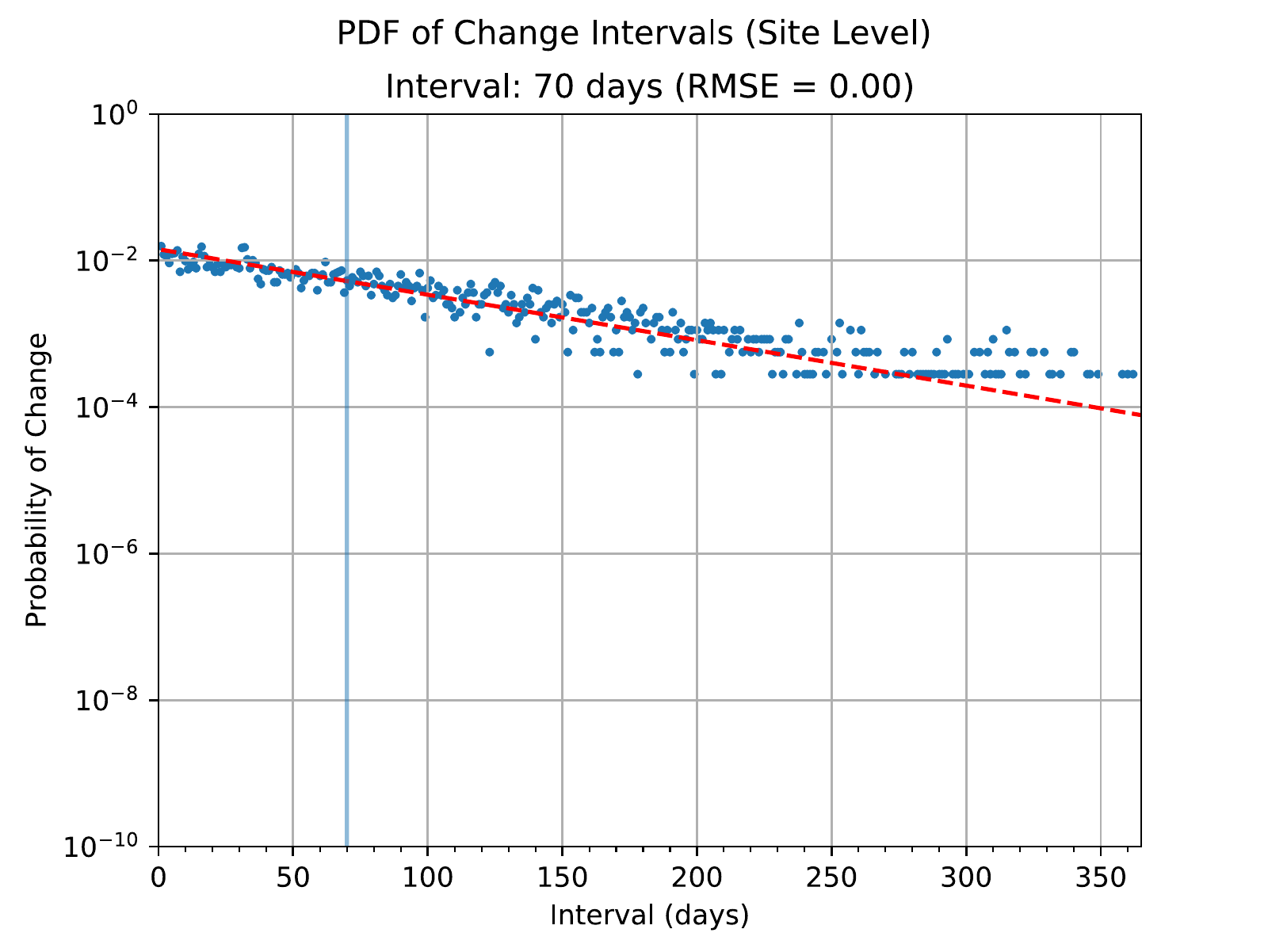}
}
\caption{
Probability (y-axis) of finding author websites with an interpolated-update interval~($\tilde{\Delta}{t}$) of $d$ days (x-axis) at both homepage-level and website-level, among author websites having $1/\tilde{\lambda}$ of 35 days and 70 days, respectively.
The vertical blue line shows where $d=1/\tilde{\lambda}$.}
\label{fig:poissonplot}
\end{figure*}

Figure~{\ref{fig:poissonplot}} shows the probability distributions for homepage-level (using \textbf{D0} dataset) and website-level (using \textbf{D2} dataset), at $1/\tilde{\lambda}=35$ days and $1/\tilde{\lambda}=70$ days, respectively.
The majority of data points follow a power-law distribution in the logarithmic scale, indicating that they fit into a Poisson distribution.
We also observe that at homepage-level, the data points follow a power-law distribution with a positive index when $d$ is (approximately) lower than $1/\tilde{\lambda}$.
We observe sporadic spikes on top of the power law.
This indicates that:
(1) For a given $\tilde{\lambda}$, consecutive changes within short intervals occur less frequently than predicted by a Poisson distribution,
(2) The updates of scholarly webpages are not absolutely random but exhibit a certain level of weak correlation.
Investigating the reasons behind these correlations is beyond the scope of this paper, but presumably, they may reflect collaboration or community-level activities.
Probability distributions for other values of $1/\widetilde{\lambda}$ also exhibit similar patterns (see Figures 15, 16, 17, and 18 in Appendix).

\subsection{Prediction Model}
\label{sec:pred_model}
We formally define our prediction model using two functions, $f$ and $g$.
The function $f:m\rightarrow(\lambda,\tau)$ takes the captures $m$ (i.e. crawl snapshots from the IA) of a website as input, and outputs its estimated mean update frequency $\lambda$ (See Eq.~(\ref{eq:mle})) and last known update time $\tau$.
The function $g:(\lambda,\tau,e)\rightarrow{p}$ takes a website's estimated mean update frequency ($\lambda$), its last known update time ($\tau$), and a time interval ($e$) as input, and outputs the probability ($p$) that the website changes after the time interval $e$ since its last known update time $\tau$.

\section{Evaluation}
\label{sec:evaluation}
Here, we study how archived copies of webpages, and the quasi-Poisson distribution of webpage updates can be leveraged to build a focused crawl scheduler for the scholarly web.

\begin{figure}[ht]
\centering
\includegraphics[width=\linewidth]{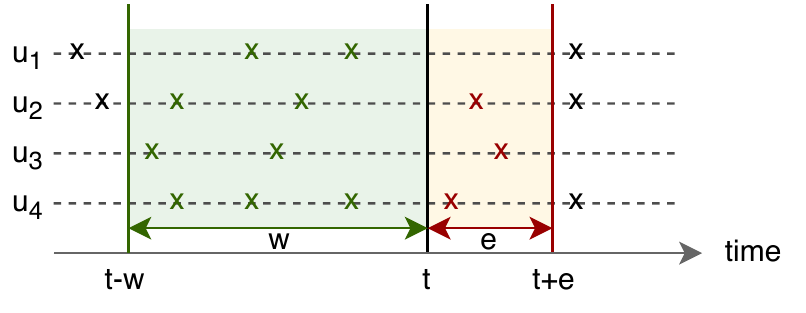}
\caption{An illustration of history size $(w)$, reference point $(t)$, evaluation interval $(e)$, and updates $(\times)$.
For each URL $u_i$, $\lambda$ was estimated using updates between $[t-w,t]$ (green), and the probability of change $(p)$ at $t+e$ was calculated.
In Evaluation 1, the correctness of $p$ (red) was checked using the actual updates between $[t,t+e]$.
In Evaluation 2, URLs were ordered by $p$, and compared against the ordering of those that changed first after $t$.
}
\label{fig:wte}
\end{figure}

Figure~\ref{fig:wte} illustrates our crawl scheduling model, HACS.
For a selected date $t$ between 2015-06-01 and 2018-06-01, we first obtain, from the \textbf{D2} and \textbf{D0}, archived captures of seed URLs within $w$ weeks prior to $t$ (i.e., in the interval $[t-w, t]$~).
Based on these captures, we calculate the \textbf{estimated mean interpolated-update frequency} ($\tilde{\lambda}$) of each seed URL.
Next, we use the $\tilde{\lambda}$ values thus obtained, to calculate the probability ($p$) that each seed URL would exhibit a change $e$ days from $t$ (i.e., by day $t+e$).
Following this, we sort the seed URLs in the decreasing order of $p$, and apply a threshold parameter ($\theta$) to select a subset of seed URLs to be crawled on that date.

\subsection{Simulated Crawl Scheduling Task}
Here, we set $e=1$ week, and advance $t$ across different points in time from 2015-06-01 to 2018-06-01, to simulate a crawl scheduling task.
At each $t$, we use standard IR metrics to evaluate whether the selected subset of seed URLs were the ones that actually changed within the interval $[t, t+e]$.
We also experiment with different values of $w$ (i.e., history size), to determine which $w$ yields an optimal result.

The following metrics are used for evaluating our model in comparison with several baseline models.
First, we look at precision, recall, and $F_1$ to measure how accurately the scheduler selects URLs for a simulated crawl job (see Evaluation~1).
Then, we use $P$@K to evaluate how accurate the scheduler ranks URLs in the order they change (see Evaluation~2).

\subsection{Evaluation 1}
Because most implementations of scholarly web crawlers are not published, we compare with two baseline models,
(1) random URLs (Random), and 
(2) Brute Force (select all URLs).
We introduce a threshold parameter $\theta\in[0,1]$ to select webpages with a probability of change $p\geq\theta$ for crawling.
Formally, we define the scheduling function as,
{\small
\[D_{w,t}{(\theta)} =\{u;~g(\lambda,\tau,1)\geq\theta, (\lambda,\tau)=f(M_{w,t}(u))\mid\forall u \in U \}\]
\[M_{w,t}(u)=\{m_{x}; x \in [t-w, t] \mid \forall m \in M_u\}\]
}
Here, $U$ is the set of all seed URLs, and $M_u$ is the set of captures of a seed URL $u$.
The parameters $w$, $t$, and $\theta$ are the history size, reference point, and threshold, respectively.
The functions $f$ and $g$ are as defined in Section \ref{sec:pred_model}.
For each $(w,t,\theta)$, the following actions are performed:
In the HACS model, we use $D_{w,t}(\theta)$ to select URLs for crawling.
In the Random model, we randomly pick $|D_{w,t}(\theta)|$ URLs from $D_{w,t}(0)$, i.e., all URLs having captures within the time window of $[t-w,t]$.
In the Brute Force model, we mimic the behavior of a hypothetical crawler by picking all URLs from $D_{w,t}(0)$.
The results from each model were compared to the URLs that \textit{actually} changed within the interval $[t,t+e]$.

Following this, we counted the number of true positives (TP), true negatives (TN), false positives (FP), and false negatives (FN) at each $(w,t,\theta)$.
Next, we got rid of the reference point $t$ by macro/micro-averaging over $t$, and calculated Precision $(P)$, Recall $(R)$, and F1 $(F)$ for each $w$ and $\theta$, respectively.
At each $w$, we then calculated the threshold $\theta=\hat{\theta}$ which maximizes $F1$ for both homepage-level and website-level.
Table~{\ref{tab:prf}} shows the results from this evaluation.

\setlength{\tabcolsep}{3pt}
\begin{table*}
\centering\footnotesize
\caption{
Comparison of HACS model to baseline models using Precision~$(P)$, Recall~$(R)$ and $F1$ values at $e=1$ week, and at threshold $\hat{\theta}$ where $F1$ is maximum. Here, $w$ is the history size (in weeks). Maximum values are in bold, and highlighted in blue.}
\begin{tabular}{l|c|cccc|cccc}
\toprule
\multicolumn{10}{c}{Homepage-level} \\
\midrule
\multirow{2}{*}{} &\multirow{2}{*}{$w$} &\multicolumn{4}{c|}{Micro Average} &\multicolumn{4}{c}{Macro Average} \\ \cline{3-10} & &$\hat{\theta}$ &HACS &Random &Brute  &$\hat{\theta}$ &HACS &Random &Brute \\
\midrule
\multirow{12}{*}{$P$} &1   &0.8  &\textcolor{blue}{\textbf{0.759}}  &0.028  &0.014  &0.9  &\textcolor{blue}{\textbf{0.919}}  &0.034  &0.026 \\
                      &2   &0.7  &\textcolor{blue}{\textbf{0.367}}  &0.020  &0.021  &0.8  &\textcolor{blue}{\textbf{0.647}}  &0.018  &0.026 \\
                      &3   &0.7  &\textcolor{blue}{\textbf{0.267}}  &0.031  &0.026  &0.7  &\textcolor{blue}{\textbf{0.305}}  &0.025  &0.029 \\
                      &4   &0.6  &\textcolor{blue}{\textbf{0.175}}  &0.046  &0.037  &0.7  &\textcolor{blue}{\textbf{0.243}}  &0.038  &0.039 \\
                      &5   &0.6  &\textcolor{blue}{\textbf{0.178}}  &0.053  &0.044  &0.7  &\textcolor{blue}{\textbf{0.220}}  &0.047  &0.046 \\
                      &6   &0.6  &\textcolor{blue}{\textbf{0.155}}  &0.047  &0.044  &0.7  &\textcolor{blue}{\textbf{0.186}}  &0.045  &0.045 \\
                      &7   &0.6  &\textcolor{blue}{\textbf{0.134}}  &0.047  &0.043  &0.6  &\textcolor{blue}{\textbf{0.136}}  &0.044  &0.045 \\
                      &8   &0.6  &\textcolor{blue}{\textbf{0.124}}  &0.046  &0.043  &0.6  &\textcolor{blue}{\textbf{0.125}}  &0.044  &0.045 \\
                      &9   &0.7  &\textcolor{blue}{\textbf{0.134}}  &0.050  &0.045  &0.7  &\textcolor{blue}{\textbf{0.139}}  &0.048  &0.047 \\
                      &10  &0.7  &\textcolor{blue}{\textbf{0.127}}  &0.047  &0.045  &0.7  &\textcolor{blue}{\textbf{0.132}}  &0.046  &0.047 \\
                      &11  &0.7  &\textcolor{blue}{\textbf{0.121}}  &0.047  &0.045  &0.7  &\textcolor{blue}{\textbf{0.125}}  &0.046  &0.047 \\
                      &12  &0.7  &\textcolor{blue}{\textbf{0.114}}  &0.050  &0.045  &0.7  &\textcolor{blue}{\textbf{0.118}}  &0.050  &0.046 \\
\midrule
\multirow{12}{*}{$R$} &1   &0.8  &0.500  &0.019  &\textcolor{blue}{\textbf{1.000}}  &0.9  &0.556  &0.026  &\textcolor{blue}{\textbf{1.000}} \\
                      &2   &0.7  &0.332  &0.018  &\textcolor{blue}{\textbf{1.000}}  &0.8  &0.321  &0.007  &\textcolor{blue}{\textbf{1.000}} \\
                      &3   &0.7  &0.291  &0.033  &\textcolor{blue}{\textbf{1.000}}  &0.7  &0.346  &0.025  &\textcolor{blue}{\textbf{1.000}} \\
                      &4   &0.6  &0.426  &0.111  &\textcolor{blue}{\textbf{1.000}}  &0.7  &0.299  &0.043  &\textcolor{blue}{\textbf{1.000}} \\
                      &5   &0.6  &0.445  &0.133  &\textcolor{blue}{\textbf{1.000}}  &0.7  &0.322  &0.070  &\textcolor{blue}{\textbf{1.000}} \\
                      &6   &0.6  &0.445  &0.136  &\textcolor{blue}{\textbf{1.000}}  &0.7  &0.325  &0.083  &\textcolor{blue}{\textbf{1.000}} \\
                      &7   &0.6  &0.448  &0.156  &\textcolor{blue}{\textbf{1.000}}  &0.6  &0.459  &0.147  &\textcolor{blue}{\textbf{1.000}} \\
                      &8   &0.6  &0.454  &0.168  &\textcolor{blue}{\textbf{1.000}}  &0.6  &0.466  &0.164  &\textcolor{blue}{\textbf{1.000}} \\
                      &9   &0.7  &0.335  &0.125  &\textcolor{blue}{\textbf{1.000}}  &0.7  &0.342  &0.122  &\textcolor{blue}{\textbf{1.000}} \\
                      &10  &0.7  &0.342  &0.125  &\textcolor{blue}{\textbf{1.000}}  &0.7  &0.351  &0.124  &\textcolor{blue}{\textbf{1.000}} \\
                      &11  &0.7  &0.348  &0.134  &\textcolor{blue}{\textbf{1.000}}  &0.7  &0.356  &0.134  &\textcolor{blue}{\textbf{1.000}} \\
                      &12  &0.7  &0.349  &0.153  &\textcolor{blue}{\textbf{1.000}}  &0.7  &0.358  &0.152  &\textcolor{blue}{\textbf{1.000}} \\
\midrule
\multirow{12}{*}{$F$1}&1   &0.8  &\textcolor{blue}{\textbf{0.603}}  &0.022  &0.028  &0.9  &\textcolor{blue}{\textbf{0.750}}  &0.678  &0.044 \\
                      &2   &0.7  &\textcolor{blue}{\textbf{0.349}}  &0.019  &0.041  &0.8  &\textcolor{blue}{\textbf{0.420}}  &0.221  &0.048 \\
                      &3   &0.7  &\textcolor{blue}{\textbf{0.279}}  &0.032  &0.051  &0.7  &\textcolor{blue}{\textbf{0.306}}  &0.087  &0.055 \\
                      &4   &0.6  &\textcolor{blue}{\textbf{0.248}}  &0.065  &0.071  &0.7  &\textcolor{blue}{\textbf{0.255}}  &0.070  &0.074 \\
                      &5   &0.6  &\textcolor{blue}{\textbf{0.254}}  &0.076  &0.084  &0.7  &\textcolor{blue}{\textbf{0.253}}  &0.071  &0.087 \\
                      &6   &0.6  &\textcolor{blue}{\textbf{0.230}}  &0.070  &0.084  &0.7  &\textcolor{blue}{\textbf{0.228}}  &0.067  &0.086 \\
                      &7   &0.6  &\textcolor{blue}{\textbf{0.206}}  &0.072  &0.083  &0.6  &\textcolor{blue}{\textbf{0.205}}  &0.073  &0.086 \\
                      &8   &0.6  &\textcolor{blue}{\textbf{0.194}}  &0.072  &0.082  &0.6  &\textcolor{blue}{\textbf{0.194}}  &0.071  &0.086 \\
                      &9   &0.7  &\textcolor{blue}{\textbf{0.191}}  &0.071  &0.086  &0.7  &\textcolor{blue}{\textbf{0.192}}  &0.072  &0.090 \\
                      &10  &0.7  &\textcolor{blue}{\textbf{0.186}}  &0.068  &0.086  &0.7  &\textcolor{blue}{\textbf{0.187}}  &0.069  &0.089 \\
                      &11  &0.7  &\textcolor{blue}{\textbf{0.180}}  &0.069  &0.086  &0.7  &\textcolor{blue}{\textbf{0.181}}  &0.067  &0.089 \\
                      &12  &0.7  &\textcolor{blue}{\textbf{0.172}}  &0.075  &0.085  &0.7  &\textcolor{blue}{\textbf{0.173}}  &0.074  &0.088 \\
\midrule
\bottomrule
\end{tabular}
\qquad
\begin{tabular}{l|c|cccc|cccc}
\toprule
\multicolumn{10}{c}{Website-level} \\
\midrule
\multirow{2}{*}{} &\multirow{2}{*}{$w$} &\multicolumn{4}{c|}{Micro Average} &\multicolumn{4}{c}{Macro Average} \\ \cline{3-10} & &$\hat{\theta}$ &HACS &Random &Brute  &$\hat{\theta}$ &HACS &Random &Brute \\
\midrule
\multirow{12}{*}{$P$} &1   &0.5  &\textcolor{blue}{\textbf{0.195}}  &0.103  &0.099  &0.5  &\textcolor{blue}{\textbf{0.191}}  &0.096  &0.098 \\
                      &2   &0.5  &\textcolor{blue}{\textbf{0.185}}  &0.104  &0.099  &0.5  &\textcolor{blue}{\textbf{0.181}}  &0.099  &0.099 \\
                      &3   &0.5  &\textcolor{blue}{\textbf{0.164}}  &0.099  &0.096  &0.5  &\textcolor{blue}{\textbf{0.162}}  &0.095  &0.096 \\
                      &4   &0.5  &\textcolor{blue}{\textbf{0.158}}  &0.097  &0.096  &0.5  &\textcolor{blue}{\textbf{0.157}}  &0.094  &0.096 \\
                      &5   &0.5  &\textcolor{blue}{\textbf{0.150}}  &0.098  &0.096  &0.5  &\textcolor{blue}{\textbf{0.150}}  &0.097  &0.096 \\
                      &6   &0.5  &\textcolor{blue}{\textbf{0.139}}  &0.094  &0.092  &0.5  &\textcolor{blue}{\textbf{0.139}}  &0.093  &0.093 \\
                      &7   &0.5  &\textcolor{blue}{\textbf{0.130}}  &0.091  &0.089  &0.5  &\textcolor{blue}{\textbf{0.131}}  &0.091  &0.090 \\
                      &8   &0.5  &\textcolor{blue}{\textbf{0.123}}  &0.089  &0.087  &0.5  &\textcolor{blue}{\textbf{0.124}}  &0.089  &0.088 \\
                      &9   &0.5  &\textcolor{blue}{\textbf{0.118}}  &0.087  &0.085  &0.5  &\textcolor{blue}{\textbf{0.120}}  &0.088  &0.087 \\
                      &10  &0.5  &\textcolor{blue}{\textbf{0.113}}  &0.084  &0.084  &0.5  &\textcolor{blue}{\textbf{0.115}}  &0.085  &0.085 \\
                      &11  &0.5  &\textcolor{blue}{\textbf{0.108}}  &0.083  &0.082  &0.5  &\textcolor{blue}{\textbf{0.110}}  &0.084  &0.084 \\
                      &12  &0.5  &\textcolor{blue}{\textbf{0.104}}  &0.081  &0.080  &0.5  &\textcolor{blue}{\textbf{0.106}}  &0.082  &0.082 \\
\midrule                    
\multirow{12}{*}{$R$} &1   &0.5  &0.435  &0.230  &\textcolor{blue}{\textbf{1.000}}  &0.5  &0.444  &0.228  &\textcolor{blue}{\textbf{1.000}} \\
                      &2   &0.5  &0.447  &0.253  &\textcolor{blue}{\textbf{1.000}}  &0.5  &0.457  &0.257  &\textcolor{blue}{\textbf{1.000}} \\
                      &3   &0.5  &0.460  &0.277  &\textcolor{blue}{\textbf{1.000}}  &0.5  &0.471  &0.280  &\textcolor{blue}{\textbf{1.000}} \\
                      &4   &0.5  &0.475  &0.292  &\textcolor{blue}{\textbf{1.000}}  &0.5  &0.486  &0.297  &\textcolor{blue}{\textbf{1.000}} \\
                      &5   &0.5  &0.482  &0.315  &\textcolor{blue}{\textbf{1.000}}  &0.5  &0.493  &0.320  &\textcolor{blue}{\textbf{1.000}} \\
                      &6   &0.5  &0.488  &0.329  &\textcolor{blue}{\textbf{1.000}}  &0.5  &0.498  &0.334  &\textcolor{blue}{\textbf{1.000}} \\
                      &7   &0.5  &0.492  &0.345  &\textcolor{blue}{\textbf{1.000}}  &0.5  &0.502  &0.349  &\textcolor{blue}{\textbf{1.000}} \\
                      &8   &0.5  &0.494  &0.356  &\textcolor{blue}{\textbf{1.000}}  &0.5  &0.504  &0.361  &\textcolor{blue}{\textbf{1.000}} \\
                      &9   &0.5  &0.497  &0.367  &\textcolor{blue}{\textbf{1.000}}  &0.5  &0.507  &0.373  &\textcolor{blue}{\textbf{1.000}} \\
                      &10  &0.5  &0.501  &0.374  &\textcolor{blue}{\textbf{1.000}}  &0.5  &0.510  &0.378  &\textcolor{blue}{\textbf{1.000}} \\
                      &11  &0.5  &0.505  &0.385  &\textcolor{blue}{\textbf{1.000}}  &0.5  &0.512  &0.389  &\textcolor{blue}{\textbf{1.000}} \\
                      &12  &0.5  &0.507  &0.393  &\textcolor{blue}{\textbf{1.000}}  &0.5  &0.514  &0.397  &\textcolor{blue}{\textbf{1.000}} \\
\midrule                    
\multirow{12}{*}{$F$1}&1   &0.5  &\textcolor{blue}{\textbf{0.269}}  &0.142  &0.180  &0.5  &\textcolor{blue}{\textbf{0.262}}  &0.132  &0.177 \\
                      &2   &0.5  &\textcolor{blue}{\textbf{0.261}}  &0.148  &0.181  &0.5  &\textcolor{blue}{\textbf{0.256}}  &0.141  &0.179 \\
                      &3   &0.5  &\textcolor{blue}{\textbf{0.242}}  &0.145  &0.175  &0.5  &\textcolor{blue}{\textbf{0.238}}  &0.140  &0.174 \\
                      &4   &0.5  &\textcolor{blue}{\textbf{0.237}}  &0.146  &0.175  &0.5  &\textcolor{blue}{\textbf{0.234}}  &0.141  &0.175 \\
                      &5   &0.5  &\textcolor{blue}{\textbf{0.229}}  &0.150  &0.175  &0.5  &\textcolor{blue}{\textbf{0.227}}  &0.146  &0.175 \\
                      &6   &0.5  &\textcolor{blue}{\textbf{0.216}}  &0.146  &0.168  &0.5  &\textcolor{blue}{\textbf{0.215}}  &0.143  &0.169 \\
                      &7   &0.5  &\textcolor{blue}{\textbf{0.206}}  &0.144  &0.163  &0.5  &\textcolor{blue}{\textbf{0.206}}  &0.143  &0.164 \\
                      &8   &0.5  &\textcolor{blue}{\textbf{0.197}}  &0.142  &0.159  &0.5  &\textcolor{blue}{\textbf{0.197}}  &0.141  &0.161 \\
                      &9   &0.5  &\textcolor{blue}{\textbf{0.191}}  &0.141  &0.157  &0.5  &\textcolor{blue}{\textbf{0.192}}  &0.140  &0.159 \\
                      &10  &0.5  &\textcolor{blue}{\textbf{0.184}}  &0.137  &0.154  &0.5  &\textcolor{blue}{\textbf{0.186}}  &0.137  &0.156 \\
                      &11  &0.5  &\textcolor{blue}{\textbf{0.178}}  &0.136  &0.151  &0.5  &\textcolor{blue}{\textbf{0.180}}  &0.136  &0.154 \\
                      &12  &0.5  &\textcolor{blue}{\textbf{0.173}}  &0.134  &0.149  &0.5  &\textcolor{blue}{\textbf{0.175}}  &0.134  &0.151 \\
\midrule
\bottomrule
\end{tabular}
\label{tab:prf}
\end{table*}

\begin{figure*}[ht]
\centering
\subfigure[Homepage-level, History = 1 week]{
\label{fig:f1-page-1w}
\includegraphics[width=.45\linewidth,trim={0 0 0 40},clip]{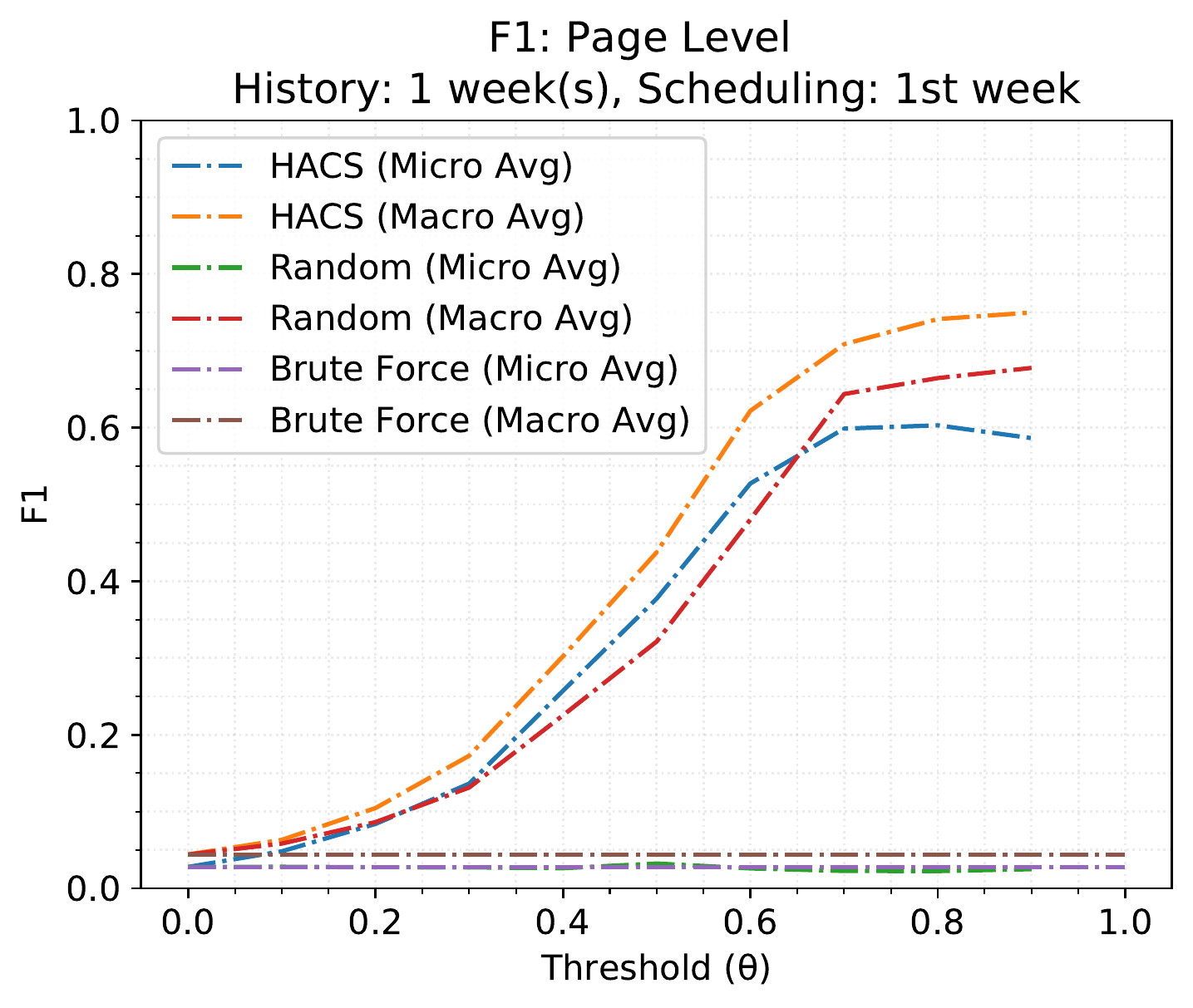}
}
\enskip
\subfigure[Homepage-level, History = 2 weeks]{
\label{fig:f1-page-2w}
\includegraphics[width=.45\linewidth,trim={0 0 0 40},clip]{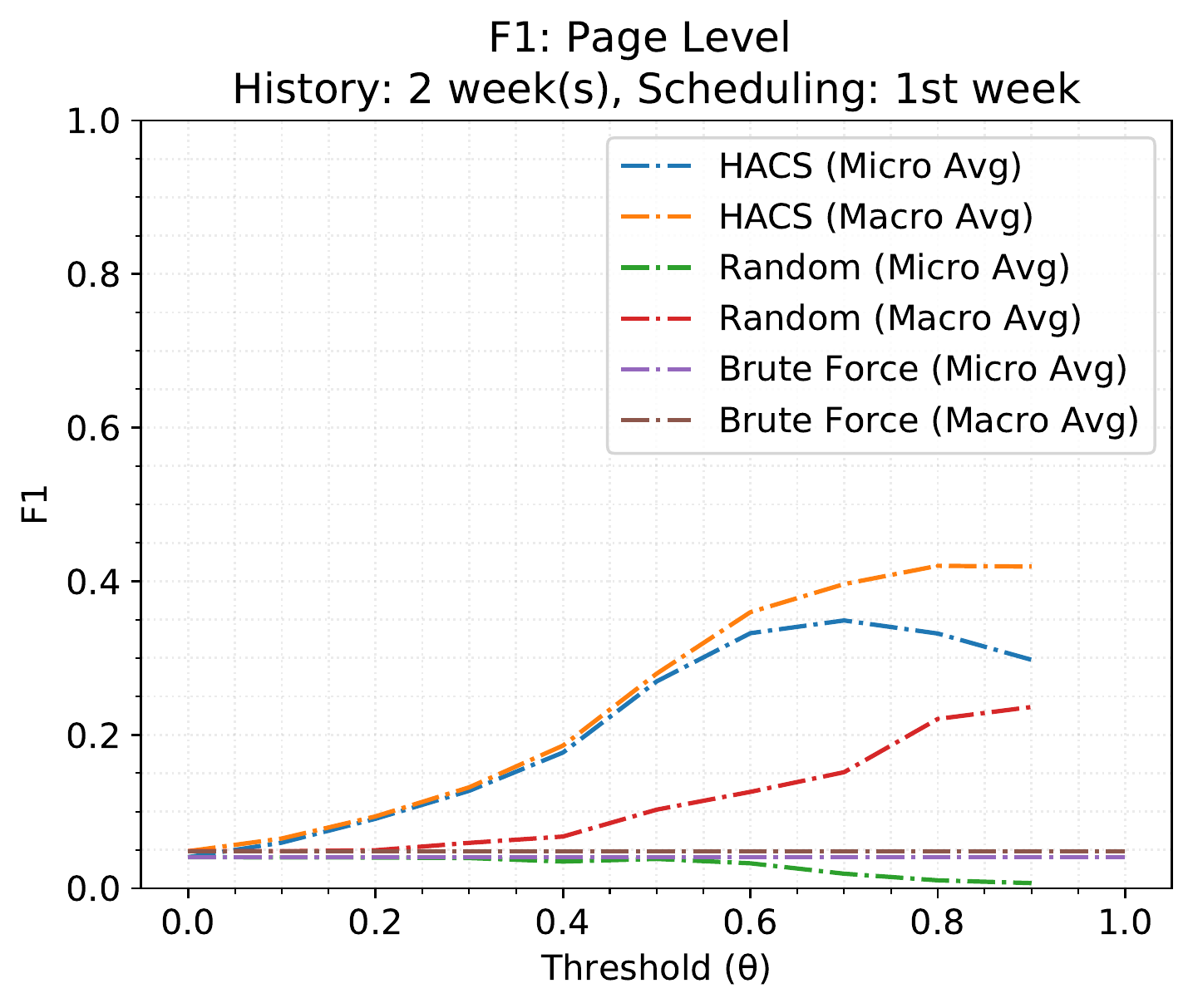}
}
\enskip
\subfigure[Website-level, History = 1 week]{
\label{fig:f1-site-1w}
\includegraphics[width=.45\linewidth,trim={0 0 0 40},clip]{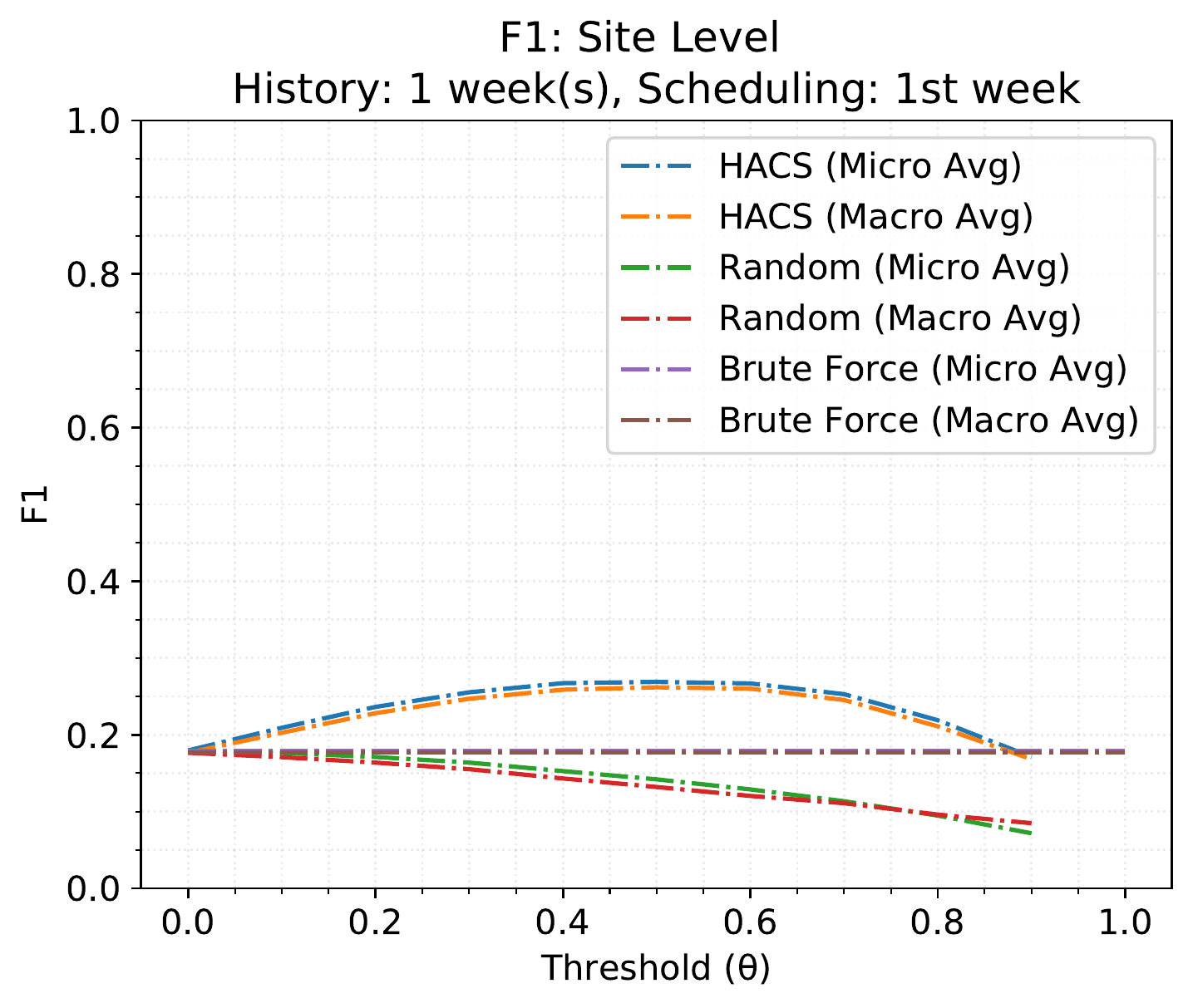}
}
\enskip
\subfigure[Website-level, History = 2 weeks]{
\label{fig:f1-site-2w}
\includegraphics[width=.45\linewidth,trim={0 0 0 40},clip]{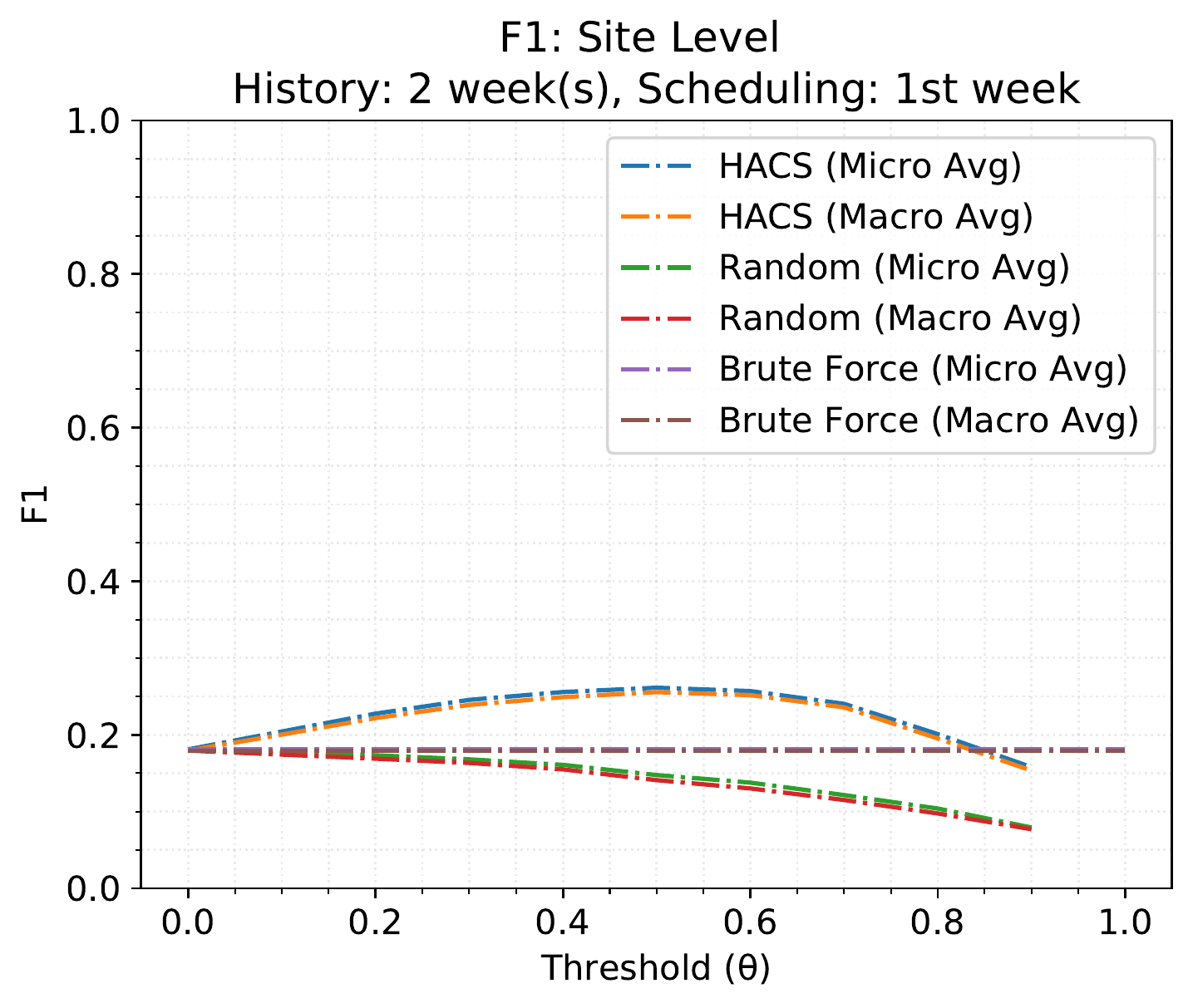}
}
\caption{
${F_1}$ vs Threshold (${\theta}$).
\textnormal{The HACS model produced a higher ${F_1}$ than other baseline models.
This lead is more visible at the homepage-level than the website-level.
As $\theta$ increases, the ${F_1}$ of the HACS model increases up to $\theta=\hat{\theta}$, and then drops as $\theta$ further increases.
This drop is more visible at the website-level than the homepage-level.
The macro-average ${F_1}$ of Random model follows the HACS model with a similar trend at the Homepage-level, History = 1 week.}
}
\label{fig:f1}
\end{figure*}

\begin{figure*}[ht]
\centering
\subfigure[Homepage-level, History = 1 week]{
\label{fig:p-page-1w}
\includegraphics[width=.45\linewidth,trim={0 0 0 40},clip]{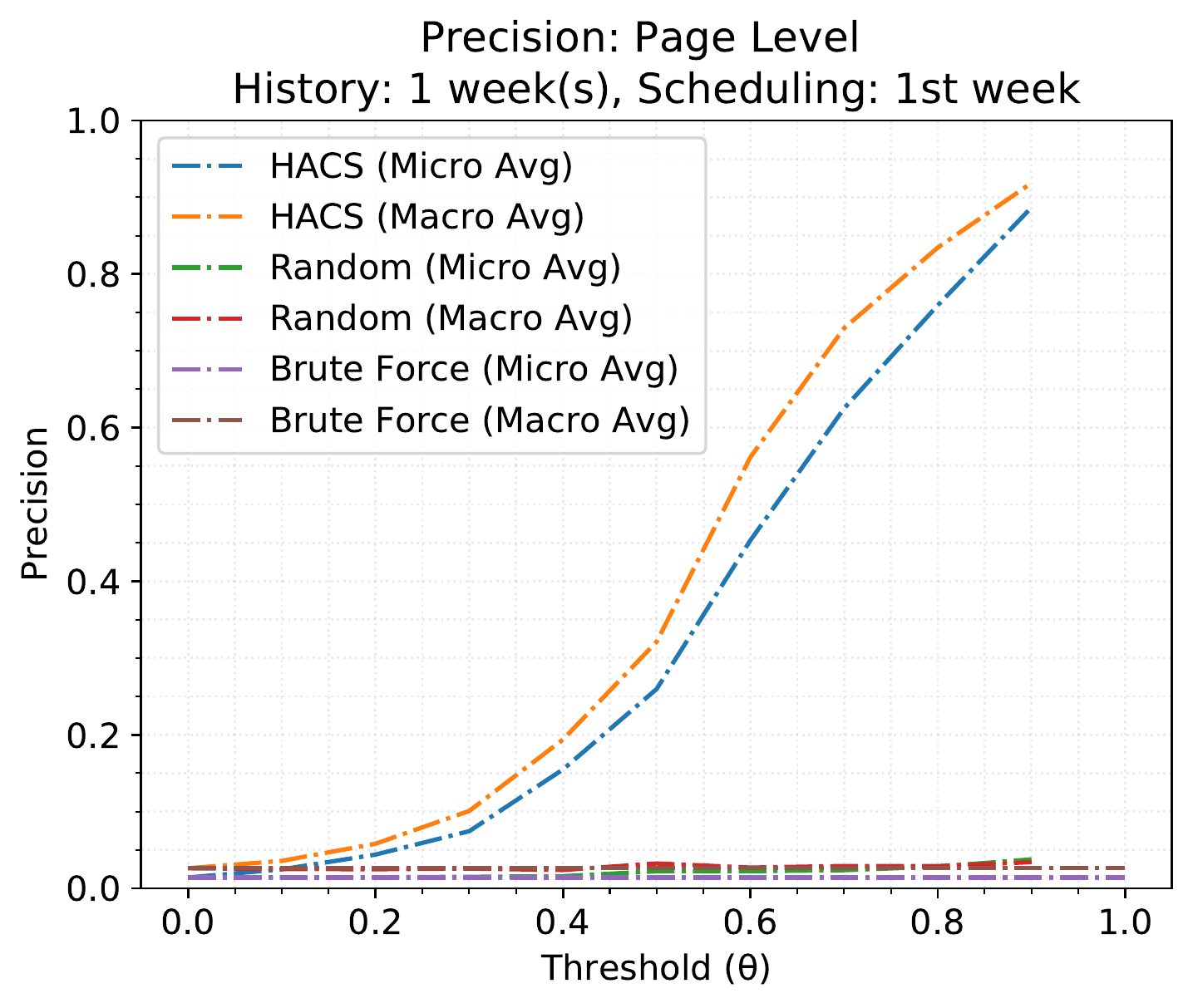}
}
\enskip
\subfigure[Homepage-level, History = 2 weeks]{
\label{fig:p-page-2w}
\includegraphics[width=.45\linewidth,trim={0 0 0 40},clip]{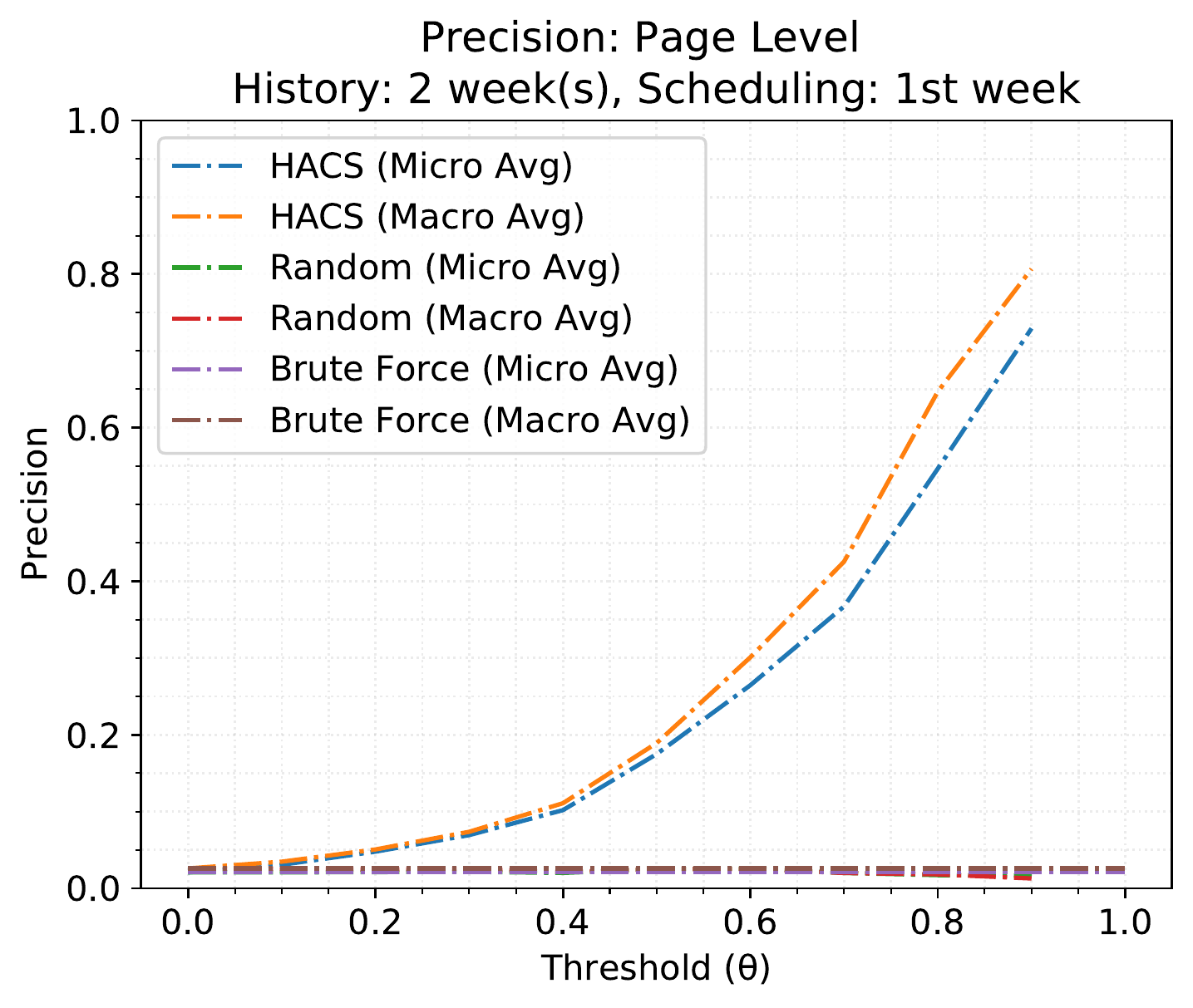}
}
\enskip
\subfigure[Website-level, History = 1 week]{
\label{fig:p-site-1w}
\includegraphics[width=.45\linewidth,trim={0 0 0 40},clip]{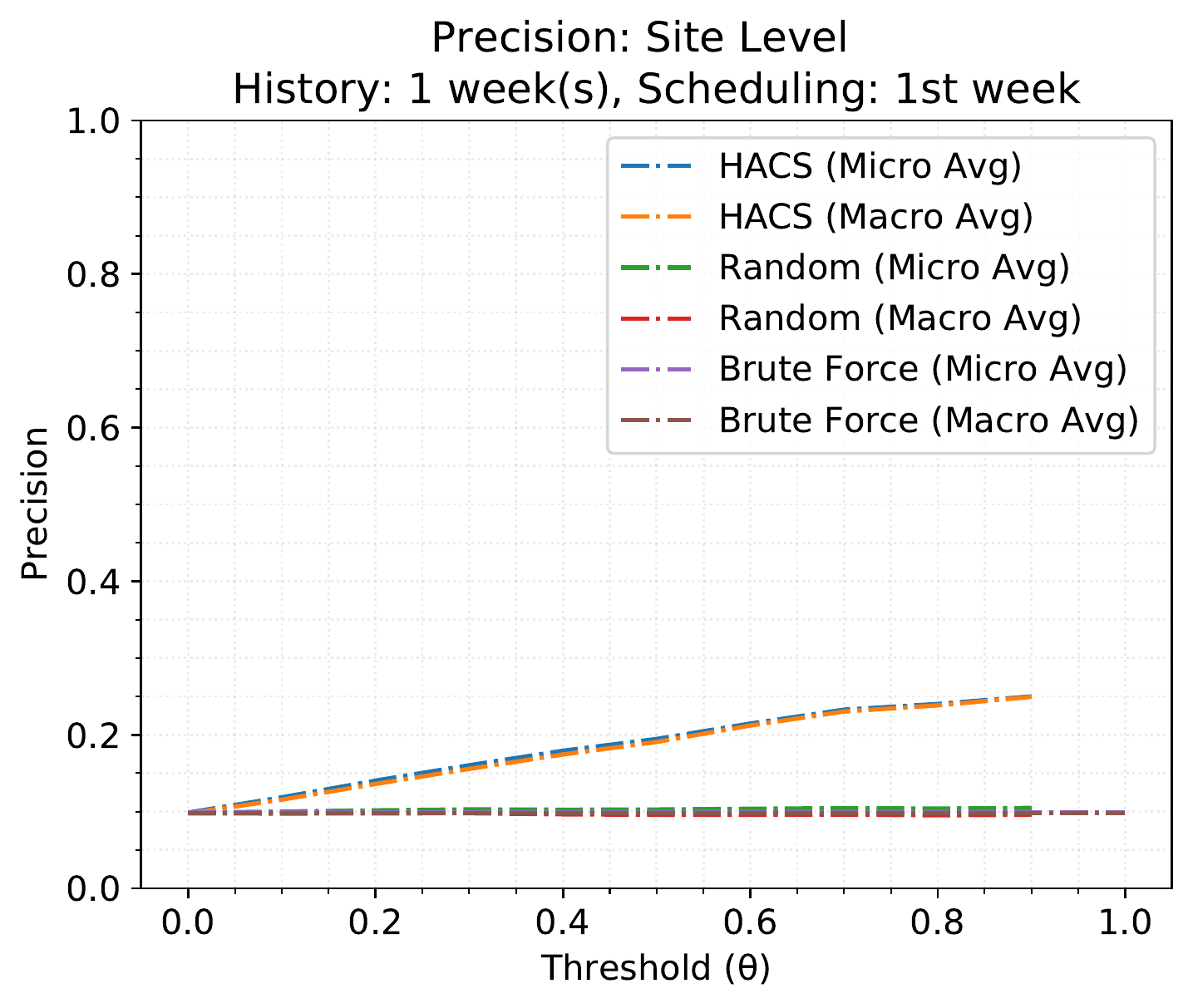}
}
\enskip
\subfigure[Website-level, History = 2 weeks]{
\label{fig:p-site-2w}
\includegraphics[width=.45\linewidth,trim={0 0 0 40},clip]{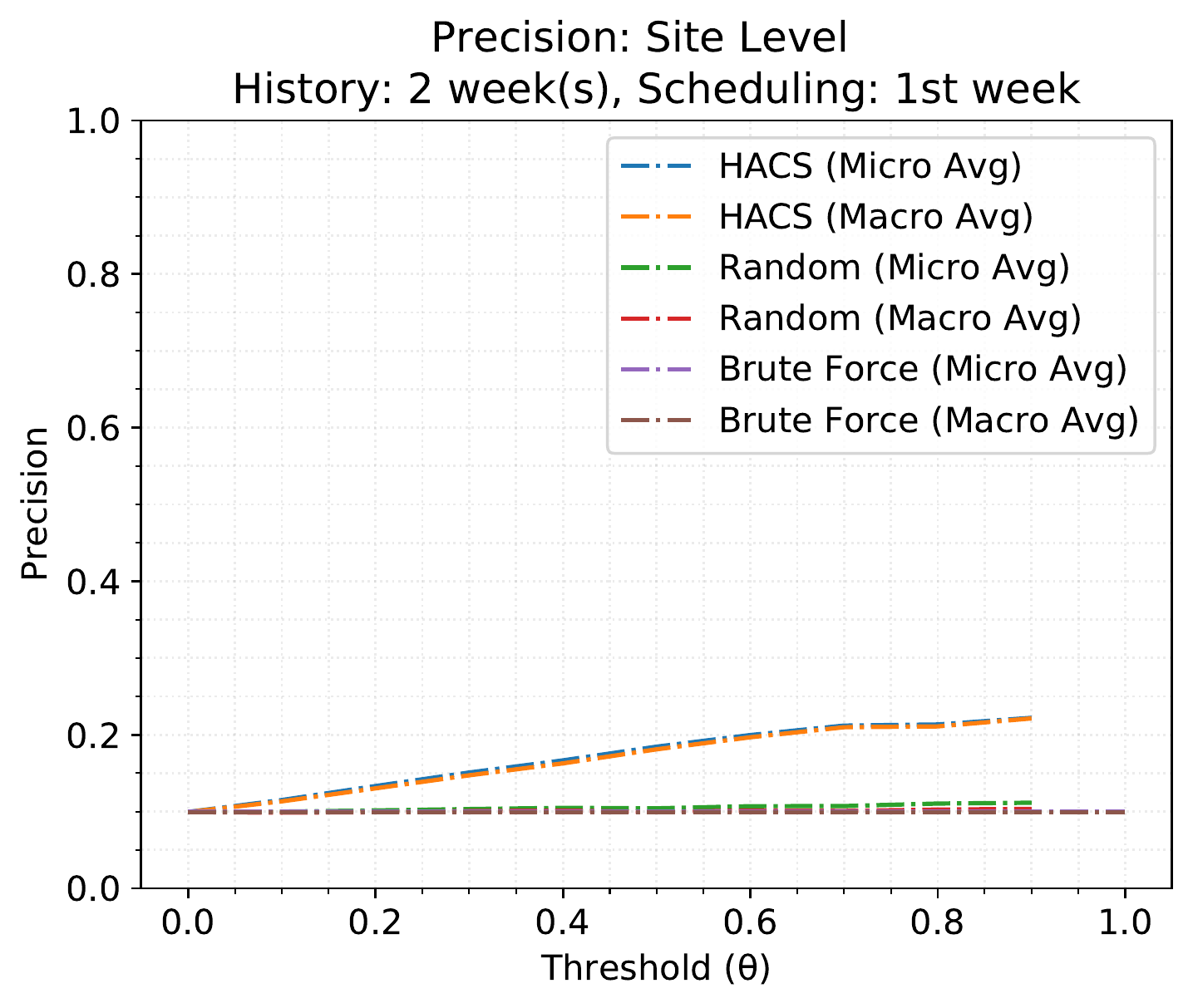}
}
\caption{
Precision (${P}$) vs Threshold (${\theta}$).
\textnormal{The HACS model produced a higher ${P}$ than other baseline models, and increases with $\theta$.
This lead is more visible at homepage-level than website-level.
Both Random and Brute Force models have a low ${P}$, regardless of $\theta$.}
}
\label{fig:p}
\end{figure*}

\begin{figure*}[ht]
\centering
\subfigure[Homepage-level, History = 1 week]{
\label{fig:r-page-1w}
\includegraphics[width=.45\linewidth,trim={0 0 0 40},clip]{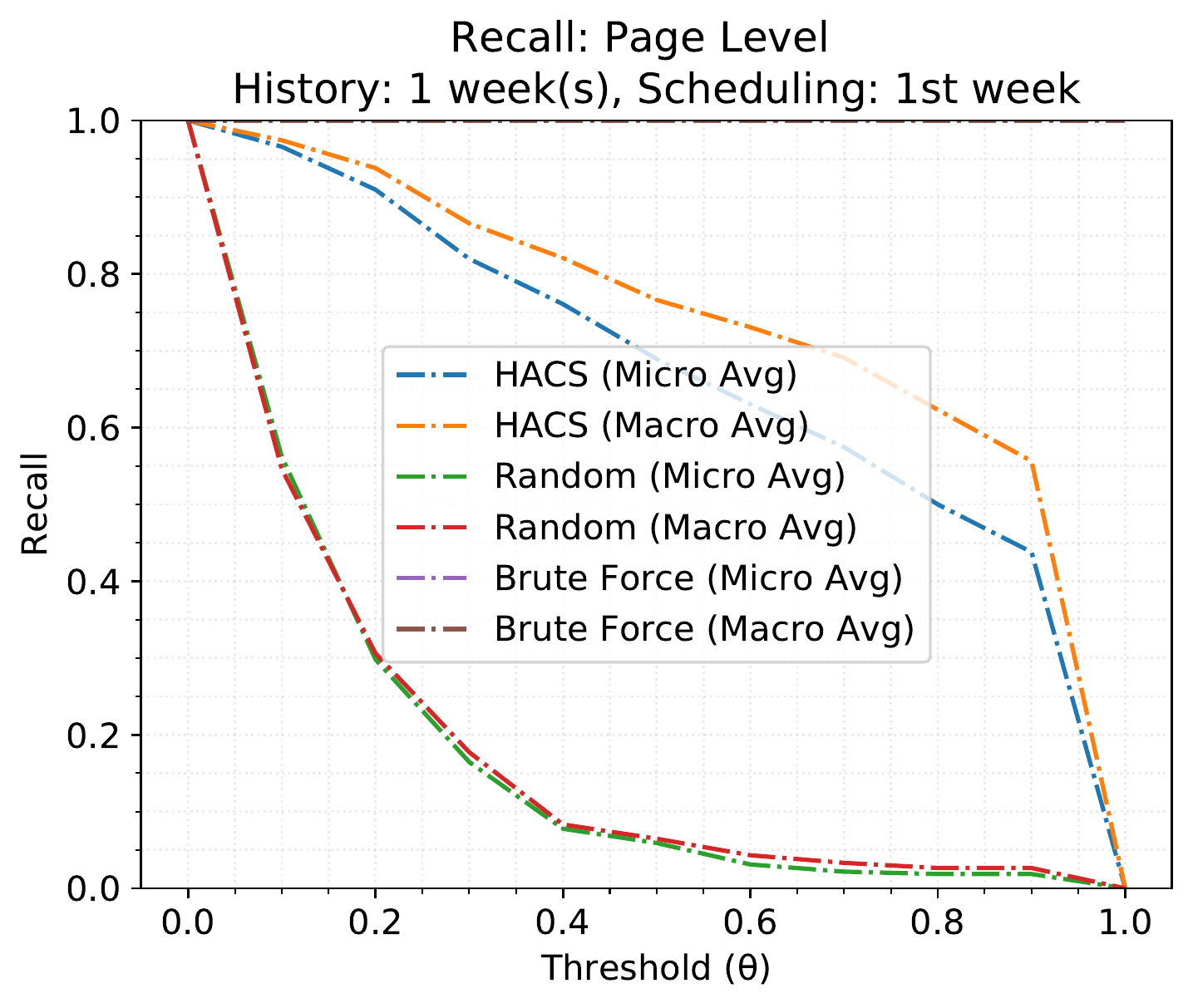}
}
\enskip
\subfigure[Homepage-level, History = 2 weeks]{
\label{fig:r-page-2w}
\includegraphics[width=.45\linewidth,trim={0 0 0 40},clip]{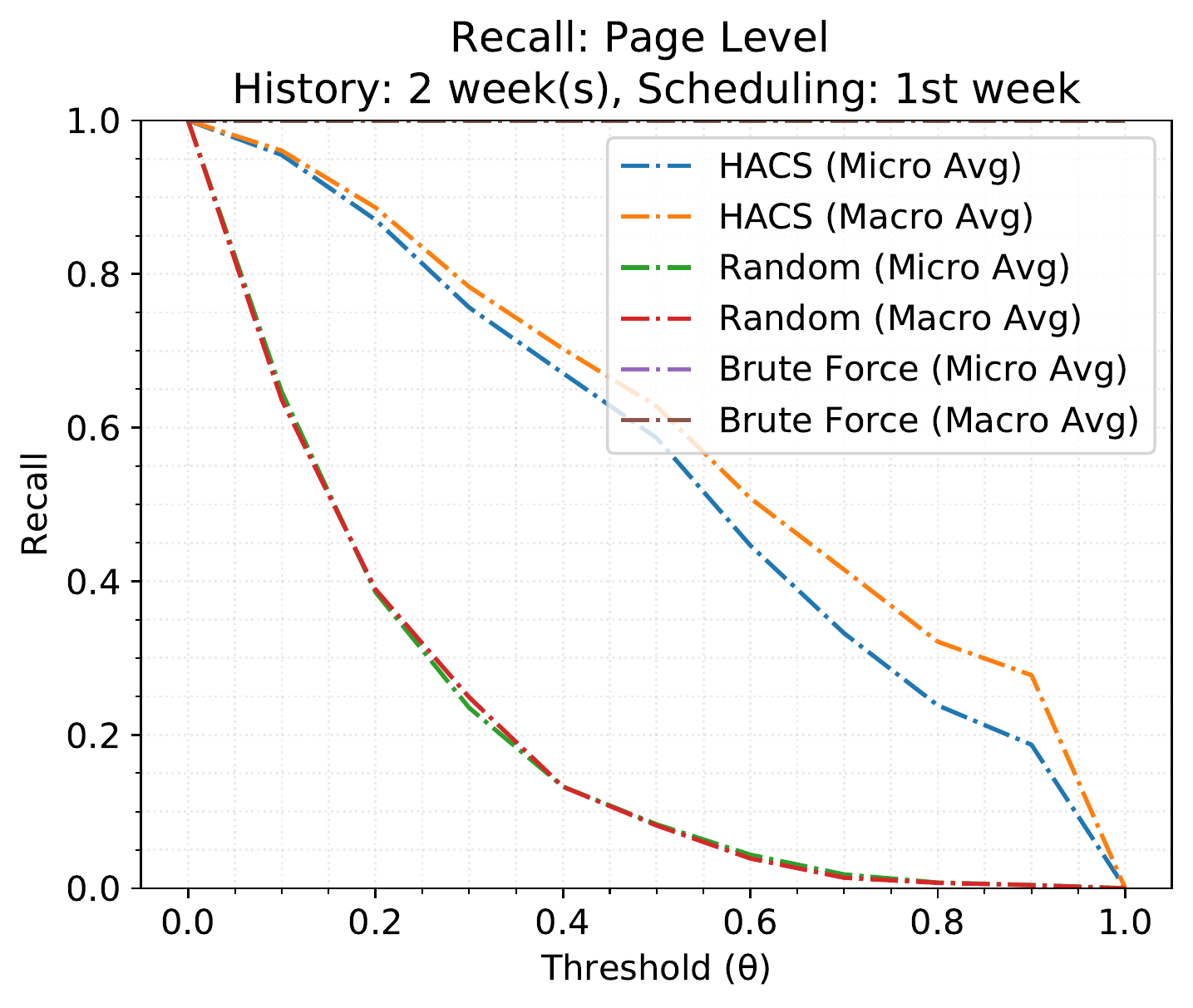}
}
\enskip
\subfigure[Website-level, History = 1 week]{
\label{fig:r-site-1w}
\includegraphics[width=.45\linewidth,trim={0 0 0 40},clip]{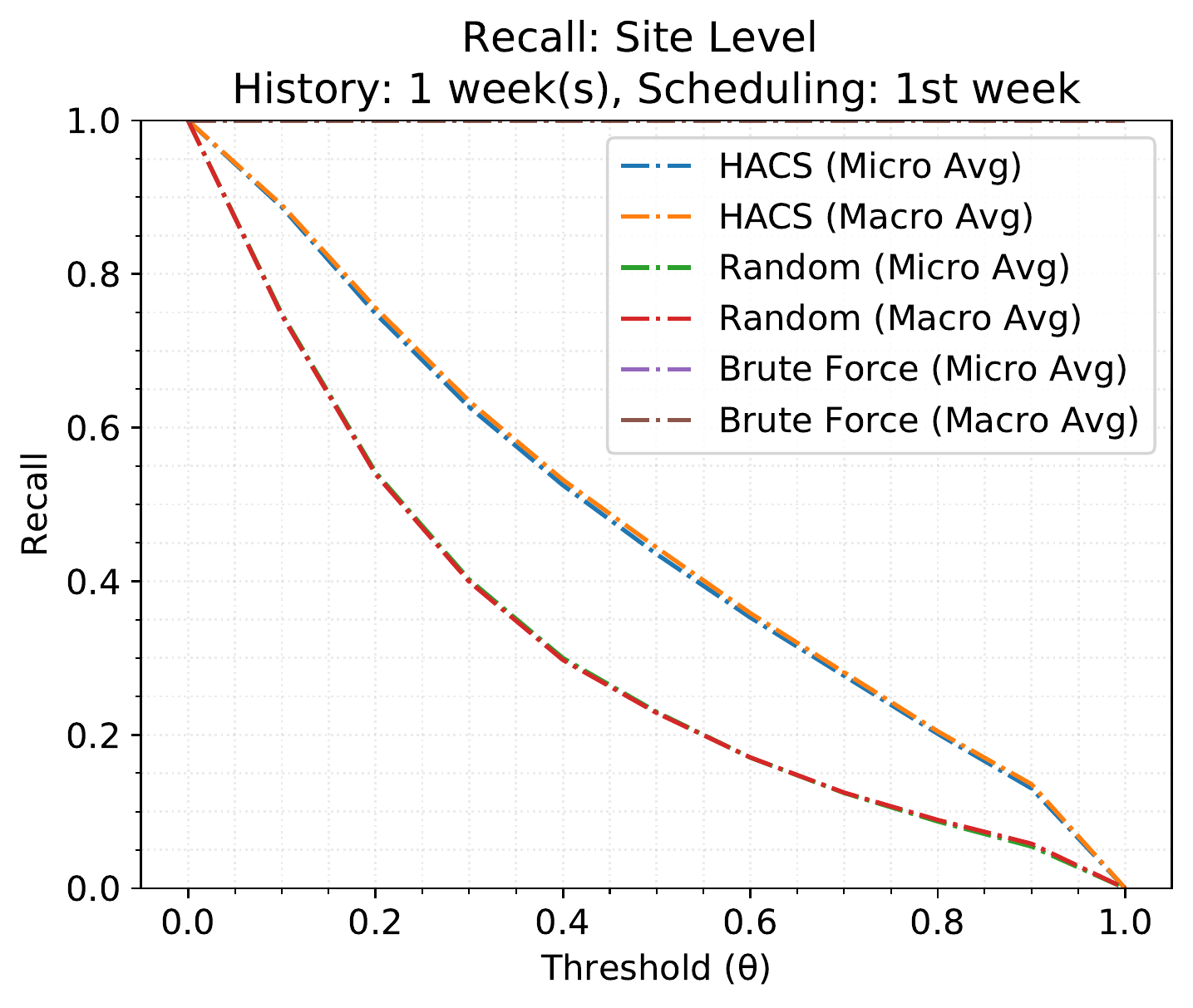}
}
\enskip
\subfigure[Website-level, History = 2 weeks]{
\label{fig:r-site-2w}
\includegraphics[width=.45\linewidth,trim={0 0 0 40},clip]{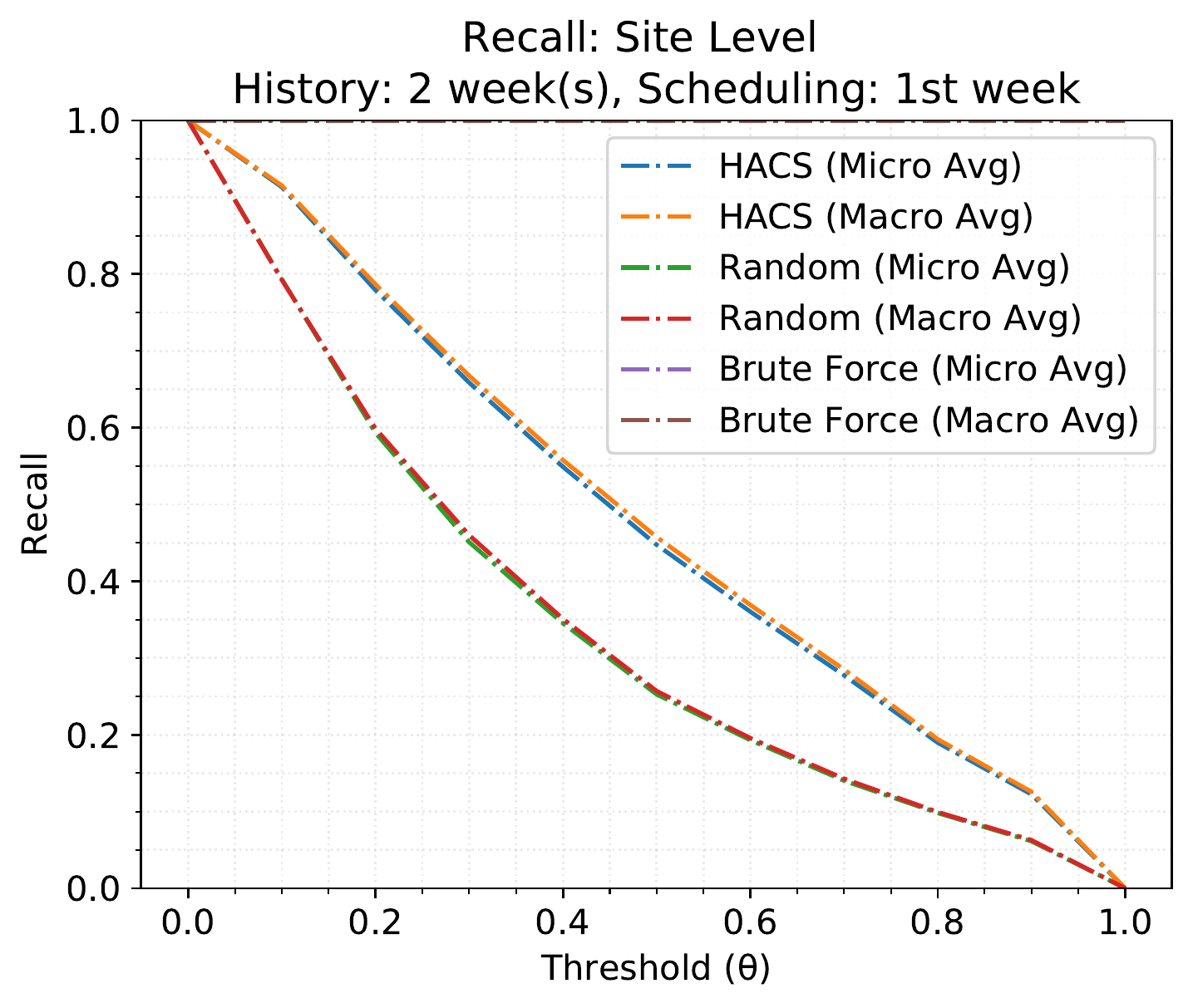}
}
\caption{
Recall (${R}$) vs Threshold (${\theta}$).
\textnormal{The HACS model produced a higher ${R}$ at lower values of $\theta$, and reaches $0$ as $\theta$ increases.
The HACS model has a much higher ${R}$ than Random model at $0<\theta<1$.
This lead is more visible at homepage-level than website-level.
The Brute Force model has a constant ${R}$ of $1.00$.}
}
\label{fig:r}
\end{figure*}

We also show how $P$, $R$ and $F1$ changes with $\theta\in[0,1]$ for both homepage-level and website-level updates.
Figures~{\ref{fig:f1}},{\ref{fig:p}}, and {\ref{fig:r}} illustrate these results at $w=1$ and $w=2$ (also, results at $w=3$ given in Figures 12, 13, and 14 in Appendix).

\subsection{Evaluation 2}
\begin{figure}[ht]
\centering
\label{fig:corr-page}
\includegraphics[width=\linewidth,trim={20 0 0 20},clip]{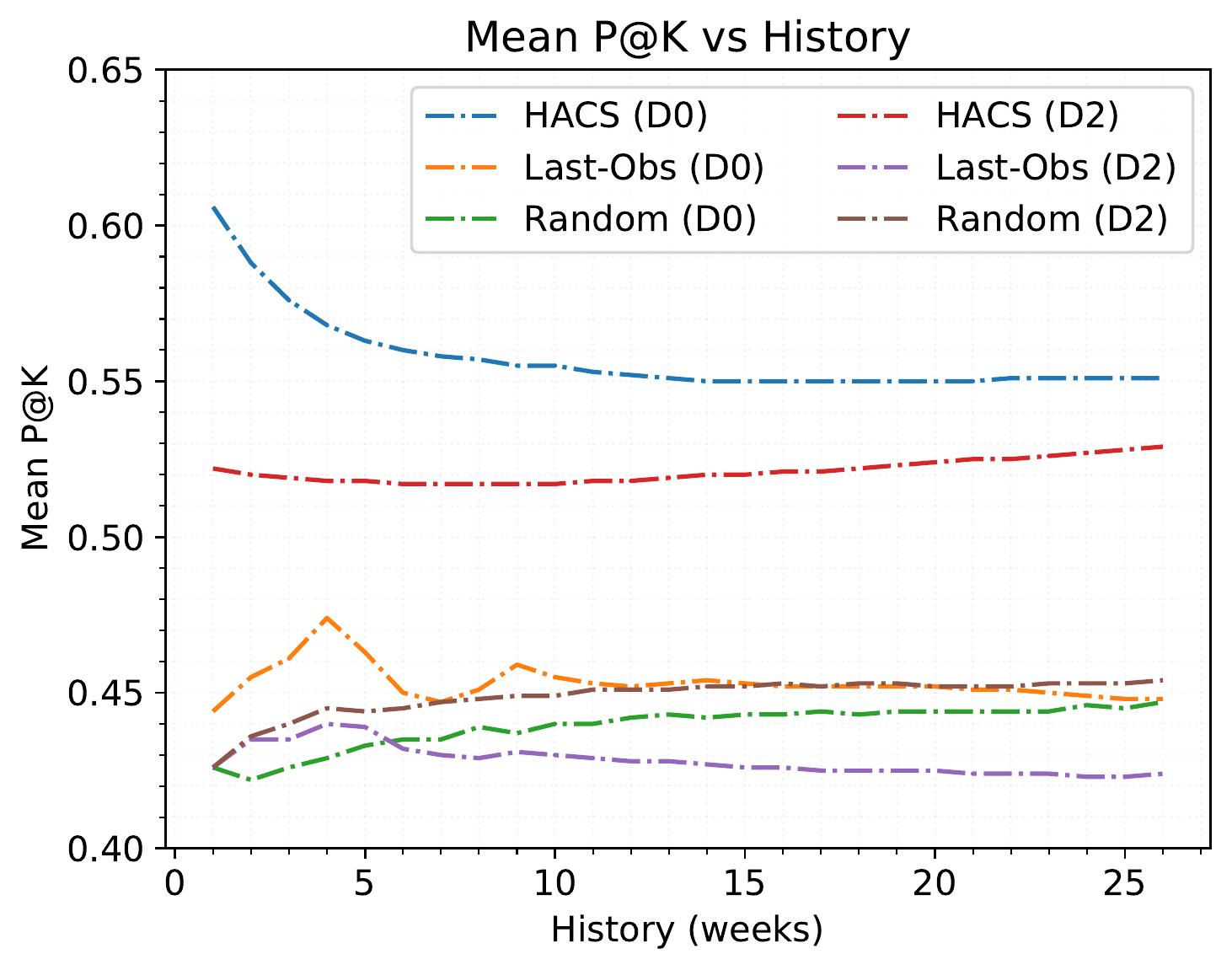}
\caption{
Mean Weighted $P$@K of rankings of HACS, Last-Obs, and Random models to the expected ranking, at different history ($w$) sizes. The HACS model outperforms the Last-Obs and Random models at both homepage-level and website-level.}
\label{fig:corr}
\end{figure}

Here, the HACS model was compared against two baseline models: Last-Obs and Random.
In the HACS model, URLs that have a higher probability of change on the crawl date $(t+e)$ are ranked higher.
In the Last-Obs model, URL ranks are determined by the date they were last accessed.
Here, URLs that have not been updated the longest (i.e. larger $(t-\tau)$) are ranked higher.
In the Random model, URLs are ranked randomly.
By comparing the URL rankings from each model to the \textit{expected} URL ranking (where URLs that were updated closer to $t$ were ranked higher), we calculate a weighted $P$@$K$ over all $K$.
Here, the weights were obtained via a logarithmic decay function to increase the contribution from lower $K$ values.
This weighted $P$@$K$ provides a quantitative measure of whether URLs that were actually updated first were ranked higher.
Next, we get rid of the reference point $t$ by calculating the mean weighted $P$@$K$ over all $t$, at each history size $w$.
In this manner, we obtain the mean weighted $P$@$K$ of each model when different history sizes ($w$) are used.
Figure~{\ref{fig:corr}} shows the results from this evaluation.

\section{Results}
The results in Table~{\ref{tab:prf}} indicate that the $P$ and $F_1$ values of HACS model are higher than the Random and Brute Force models for all values of $w$ (history size in weeks).
This lead is higher when $w$ is lower.
However, this difference becomes less significant as $w$ increases.
The Brute Force method had a consistent $R$ of $1.00$, since it crawls all URLs at all times.
However, this model is impractical due to resource constraints.
The HACS model produced a higher $R$ than the Random model at all $w$.
Also, $\hat{\theta}\in[0.7,0.9]$ for homepage-level and $\hat{\theta}\in[0.5,0.5]$ for website-level indicates the optimal ranges for $\theta$.

From Figure~{\ref{fig:f1}}, as $\theta$ increases, the $F_1$ score of HACS model increases until $\theta=\hat{\theta}$, and then drops as $\theta$ increases further.
At $\hat{\theta}$, the HACS model yields the highest micro-average $F1$ score at both the homepage-level and the website-level.
This trend is more prominent at the homepage-level than the website-level.
In terms of macro-average $F1$, the Random model closely follows the HACS model at homepage-level when $w=1$.
However, the HACS model yields better $F1$ scores in all other cases.
The Brute Force model gives constant $F1$ scores at both homepage-level and website-level, as it selects all seed URLs regardless of $\theta$.

When comparing precision $P$, Figure~{\ref{fig:p}} shows that both micro-average and macro-average $P$'s of HACS model increases as $\theta$ increases.
This is expected as the URL selection becomes stricter as $\theta$ increases, which, in turn, generates less false positives.
Similar to $F1$, the lead in $P$ of the HACS model is more noticeable at homepage-level than website-level.
Nevertheless, the HACS model yields higher $P$ than other models in all cases.
The Brute Force model has a constant $P$, as it selects all URLs regardless of $\theta$.
However, $P$ of Brute Force model is lower than HACS model at both homepage-level and website-level.
Interestingly, the $P$ of both Brute Force and Random models remain close to each other.
At $\theta=0.0$ (i.e. when no threshold is applied), all models give the same results, as they select all seed URLs.

When comparing results of $R$, Figure~{\ref{fig:r}} shows that both micro-average $R$ and macro-average $R$ decreases as $\theta$ increases.
This is expected as the URL selection becomes stricter as $\theta$ increases, which, in turn, generates less false negatives.
The Brute Force model has a constant $R$ of $1.00$, as it selects all URLs regardless of $\theta$.
At $\theta=0.0$ (i.e. when no threshold is applied), all models give $R=1.00$ as they select all seed URLs.
At $\theta=1.0$, both HACS and Random models give $R=0.00$, as they select no URLs here.
For $\theta$ values other than these, the HACS model consistently yields better $R$ than Random model at both homepage-level and website-level.
However, this lead is less significant at website-level than at homepage-level, and diminishes as $w$ increases.

When comparing the average P@K results,
Figure~{\ref{fig:corr}} shows that the HACS model yields a better average P@K than the Last-Obs and Random models at both homepage-level and website-level, for all values of $w$.
However, the HACS model yields a higher average P@K for lower values of $w$ than for higher values of $w$.
As $w$ increases, the average P@K of all models become approximately constant.
At homepage-level, the Last-Obs model yields a better average P@K than the Random model for lower values of $w$.
At website-level, however, it yields a worse average P@K than the Random model for higher values of $w$.

\section{Discussion}
From Table~{\ref{tab:prf}}, the $P$, $R$, and $F1$ values obtained from the HACS model are greater than the baseline models at both the homepage-level and the website-level, when the optimal threshold $\hat{\theta}$ is selected.
Figure~{\ref{fig:f1}} shows that regardless of the $\theta$ selected, the HACS model performs better than the baseline models.
Also, the $P$ of the HACS model increases as $\theta$ increases.
This indicates that the HACS model predicted a higher probability ($p$) for the URLs that got updated first during $[t,t+e]$.
This is also confirmed by the higher mean weighted $P$@$K$ values obtained by the HACS model (see Figure~{\ref{fig:corr}}).
Since $R$ decreases with increasing $\theta$ while $P$ increases with increasing $\theta$, it is imperative that an optimal $\theta$ value should be selected.
Results in Table~\ref{tab:prf} show that selecting $\theta=\hat{\theta}$ (which maximizes $F1$) provides a good compromise between precision and recall, yet perform better than the baseline models.

The $P$ and $R$ of the Brute Force model is constant irrespective of $\theta$.
Though this model yields the highest $R$ (which is $1.00$), it consumes a significant amount of resources to crawl everything.
This approach does not scale well to a large number of seed URLs.
It also yields a lower $P$ and $F1$ than the HACS model across all $w$, at both homepage-level and website-level.
These results suggest that the HACS model, which yields a much higher $P$ and $F1$ at a marginal reduction in $R$, is more suited for a resource-constrained environment.

Recall that the archival of webpages is both irregular and sparse (See Figure \ref{fig:scatterplot}).
In our sample, authors updated their homepages every $141.5$ days on average, and their websites every $75$ days on average.
Note that here, an update to a webpage means adding a new link into it.
Authors may update their homepages or websites by updating content or adding external links.
Content updates can be studied in a similar way by comparing the checksum of webpages.
Since CDX files only contain mementos of webpages within the same domain, taking external links into consideration may require other data sources.
The better performance of the HACS model in estimating the mean update frequency $(\lambda)$ for homepages may be attributed to the fact that homepages undergo fewer changes than websites.

From Table~{\ref{tab:prf}}, the best micro-average $F1$ measure obtained at homepage-level and website-level were $0.603$ and $0.269$, respectively.
Similarly, the best macro-average $F1$ measures obtained at homepage-level and website-level were $0.750$ and $0.262$, respectively.
In both cases, these $F1$ measures originated from the HACS model when $w=1$ and $\theta\in[0.5,0.9]$.

Figure~{\ref{fig:f1}} demonstrates the efficiency of our model.
As the threshold $\theta$ increases, the number of false positives is reduced, thereby increasing the precision.
Here, we note that even a small increase in precision matters, because for a large number of seed URLs, even the slightest increase in precision attributes to a large decrease in false positives.
If crawling is performed on a regular basis, the HACS model could be utilized to pick seed URLs that have most likely been updated.
This, based on the above results, would improve collection freshness while using resources and bandwidth more effectively.

\section{Conclusion}
\label{conclusion}
We studied the problem of improving the efficiency of a focused crawl scheduler for the scholarly web.
By analyzing the crawl history of seed URLs obtained from the IA, we fit their change information into a Poisson model and estimated the probability that a webpage would update (addition of new links) by the next crawl.
Finally, our scheduler automatically generates a list of seed URLs most likely to have changed since the last crawl.
Our analysis found that the estimated mean update frequency (or equivalently, update interval) follow a log-normal distribution.
For the 19,977 authors we studied from Google Scholar, new links were added on an average interval of 141.5 days for a homepage, and 75 days for a website.
We also observed that the median crawl interval of $80\%$ of author homepages was between 20--127 days.
Our evaluation results show that our scheduler achieved better results than the baseline models when $\theta$ is optimized.
To encourage reproducible research, our research dataset consisting of HTML, CDX files, and evaluation results have been made publicly available\footnote{\url{https://github.com/oduwsdl/scholarly-change-rate}}.

In the future, we will investigate different types of updates, such as the addition of a scholarly publication in PDF format.
Additionally, author websites could be crawled regularly to ensure that updates are not missed, and its effect on the estimation of mean update frequency could be evaluated.
We will also generalize this work into more domains by exploring non-scholarly URLs.

\section{Acknowledgement}
This work was supported in part by the National Science Foundation and the Dominion Graduate Scholarship from the College of Science at the Old Dominion University.

\bibliographystyle{IEEEtran}
\bibliography{main}

\appendix
\section{Appendix: Supplementary Figures}
\label{app:supfigs}

This section documents additional results obtained from the evaluation of HACS model against our baselines, and the verification of the stochastic nature of scholarly webpage updates for more interval sizes.

\begin{figure*}[ht]
\centering
\subfigure[Homepage-level, History = 3 weeks]{
\includegraphics[width=.44\linewidth,trim={0 0 0 40},clip]{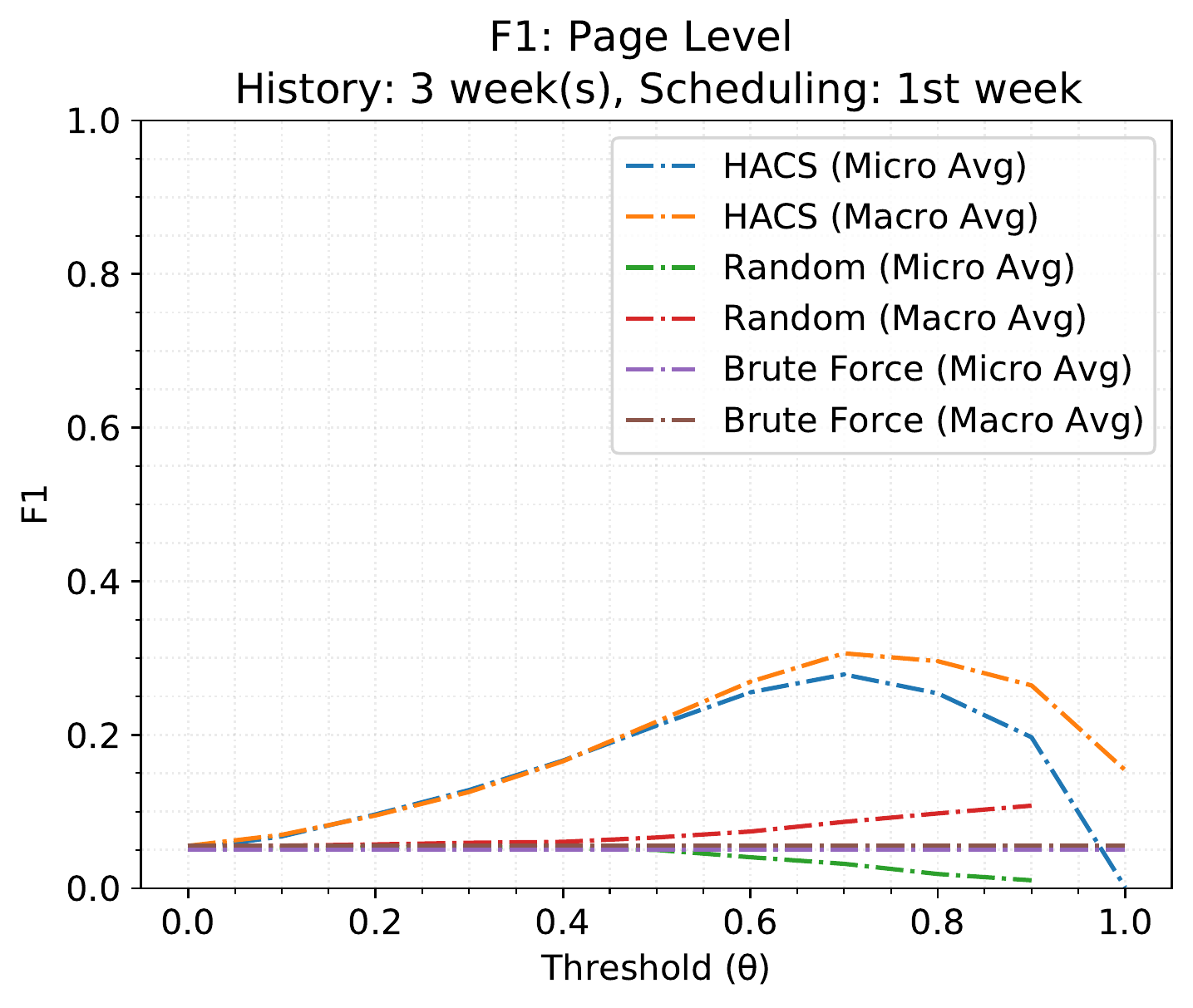}
}
\enskip
\subfigure[Website-level, History = 3 weeks]{
\includegraphics[width=.44\linewidth,trim={0 0 0 40},clip]{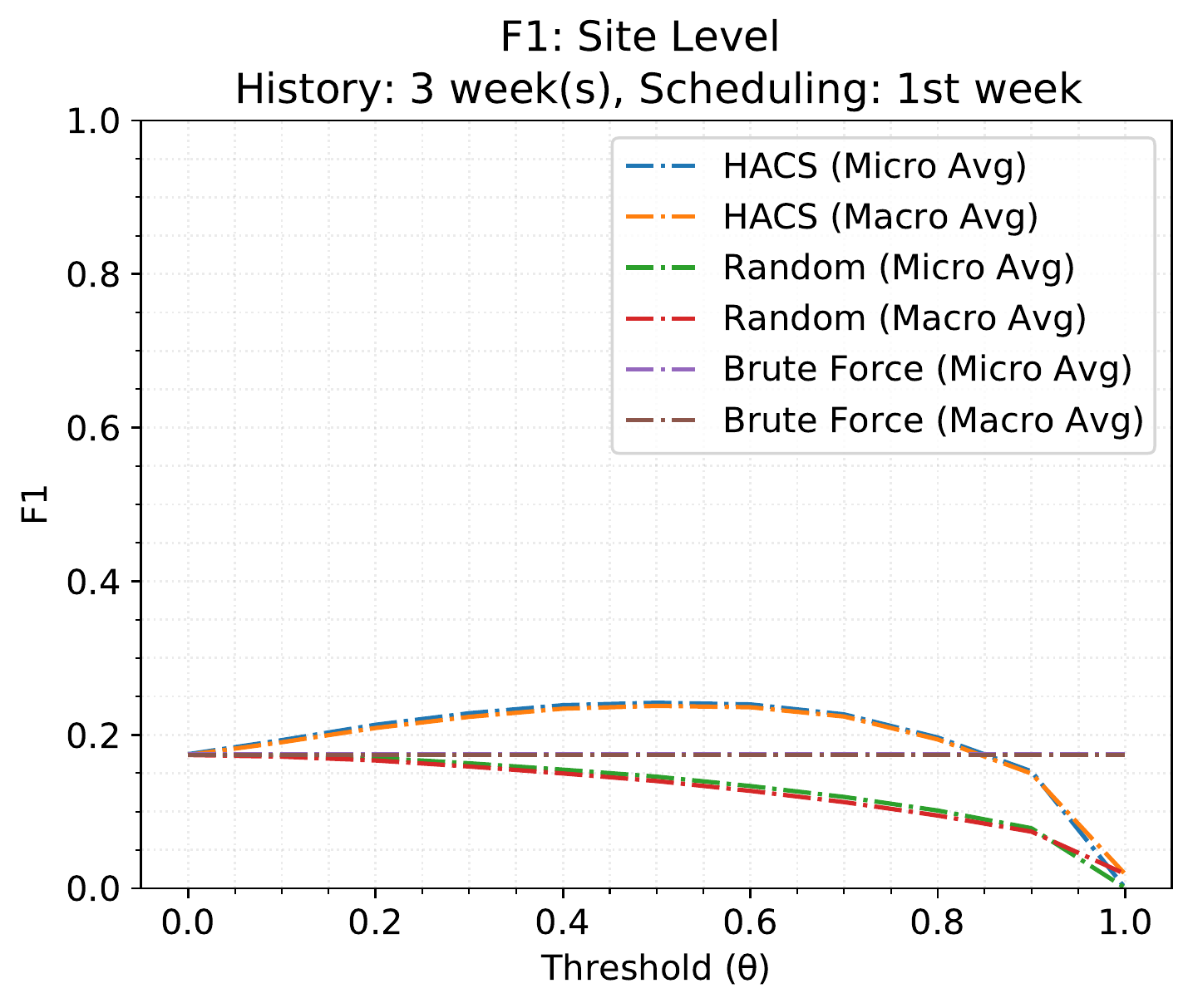}
}
\caption{${F1}$ vs Threshold (${\theta}$) when a history of 3 weeks is used}
\end{figure*}

Figure~12 illustrates the $F_1$ vs Threshold ($\theta$) of each model, when a history size of 3 weeks is used.
Here too, the HACS model produced a higher $F_1$ than other baseline models.
This lead is more visible at the homepage-level than the website-level.
However, compared to a history size of 1 week and 2 weeks, this lead is less prominent at both homepage-level and webpage-level.
As $\theta$ increases, the ${F_1}$ of the HACS model increases up to $\theta=\hat{\theta}$, and then drops as $\theta$ further increases.
This drop is more visible at the website-level than the homepage-level.
The macro-average ${F_1}$ of Random model follows the HACS model with a similar trend at the Homepage-level, History = 1 week.

\begin{figure*}[ht]
\centering
\subfigure[Homepage-level, History = 3 weeks]{
\includegraphics[width=.44\linewidth,trim={0 0 0 40},clip]{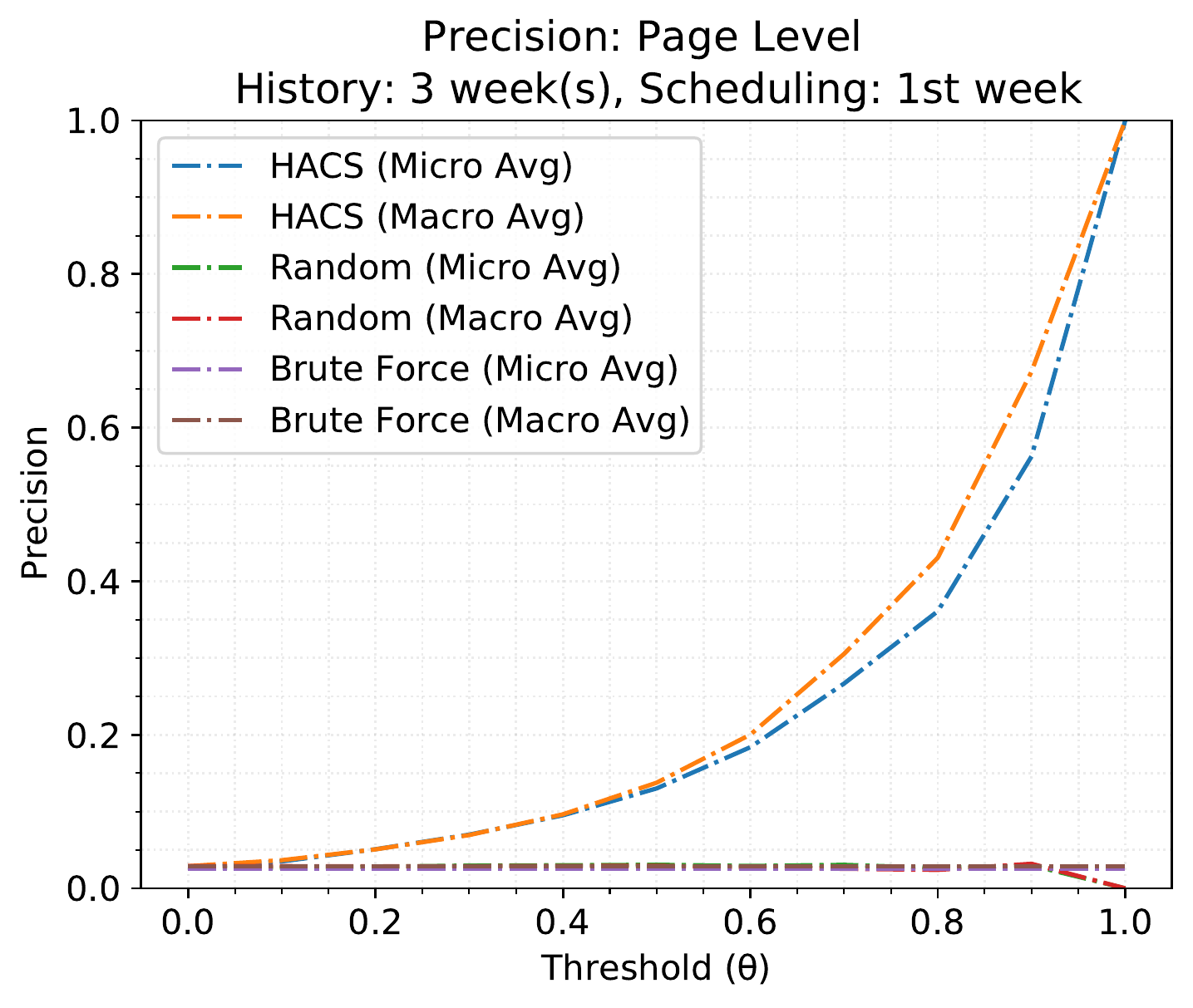}
}
\enskip
\subfigure[Website-level, History = 3 weeks]{
\includegraphics[width=.44\linewidth,trim={0 0 0 40},clip]{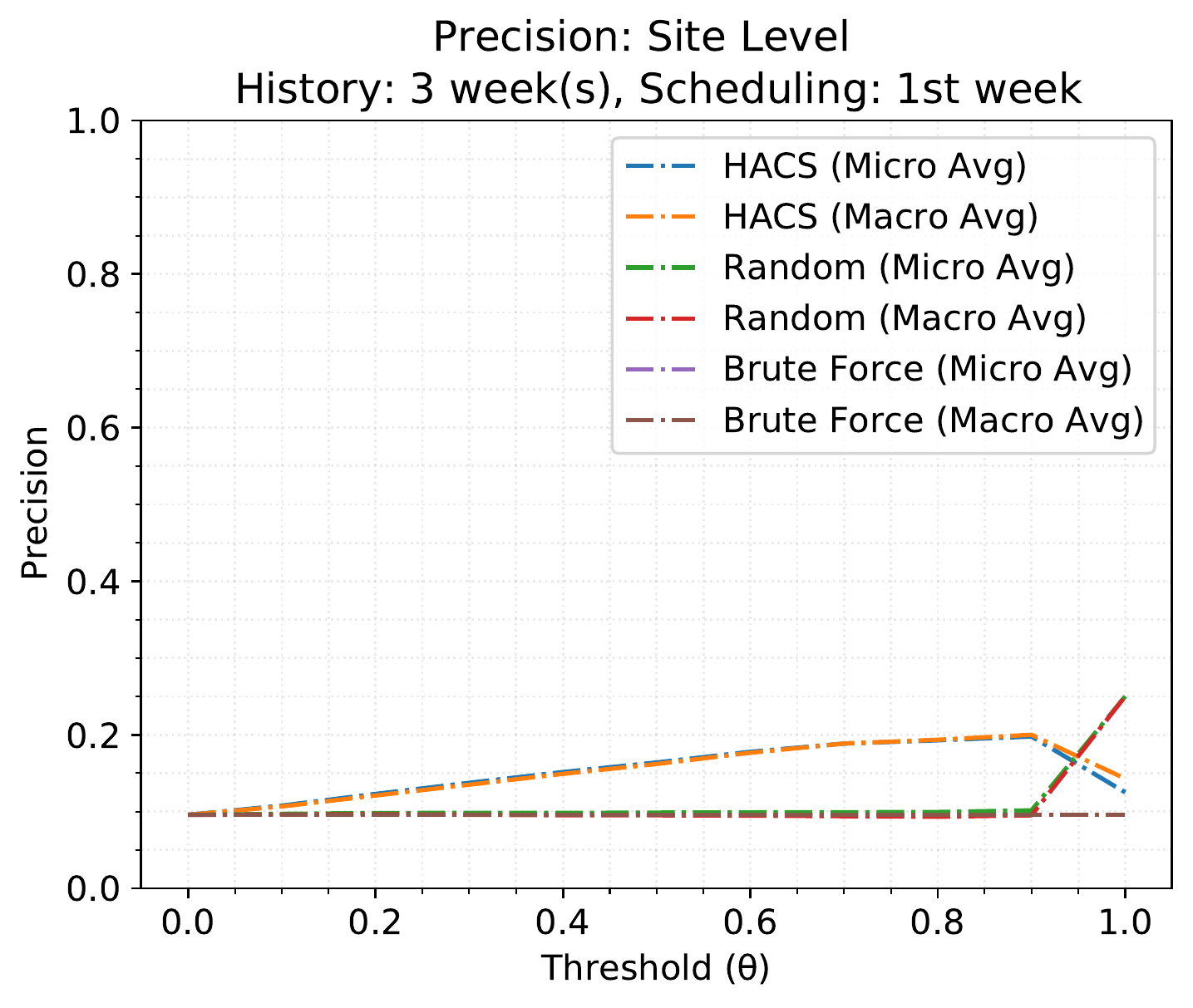}
}
\caption{Precision (${P}$) vs Threshold (${\theta}$) when a history of 3 weeks is used}
\end{figure*}

Figure 13 illustrates the Precision ($P$) vs Threshold ($\theta$) of each model, when a history size of three weeks is used.
Here too, the HACS model produced a higher P than the baselines for all values of $\theta$ at homepage level, and for $\theta \leq 0.95$ at website level.
This lead is more visible at homepage-level than website-level.
However, compared to a history size of 1 week and 2 weeks, this lead is less prominent at both homepage-level and webpage-level.
Both Random and Brute Force models have a low ${P}$, regardless of $\theta$.

\begin{figure*}[ht]
\centering
\subfigure[Homepage-level, History = 3 weeks]{
\includegraphics[width=.44\linewidth,trim={0 0 0 40},clip]{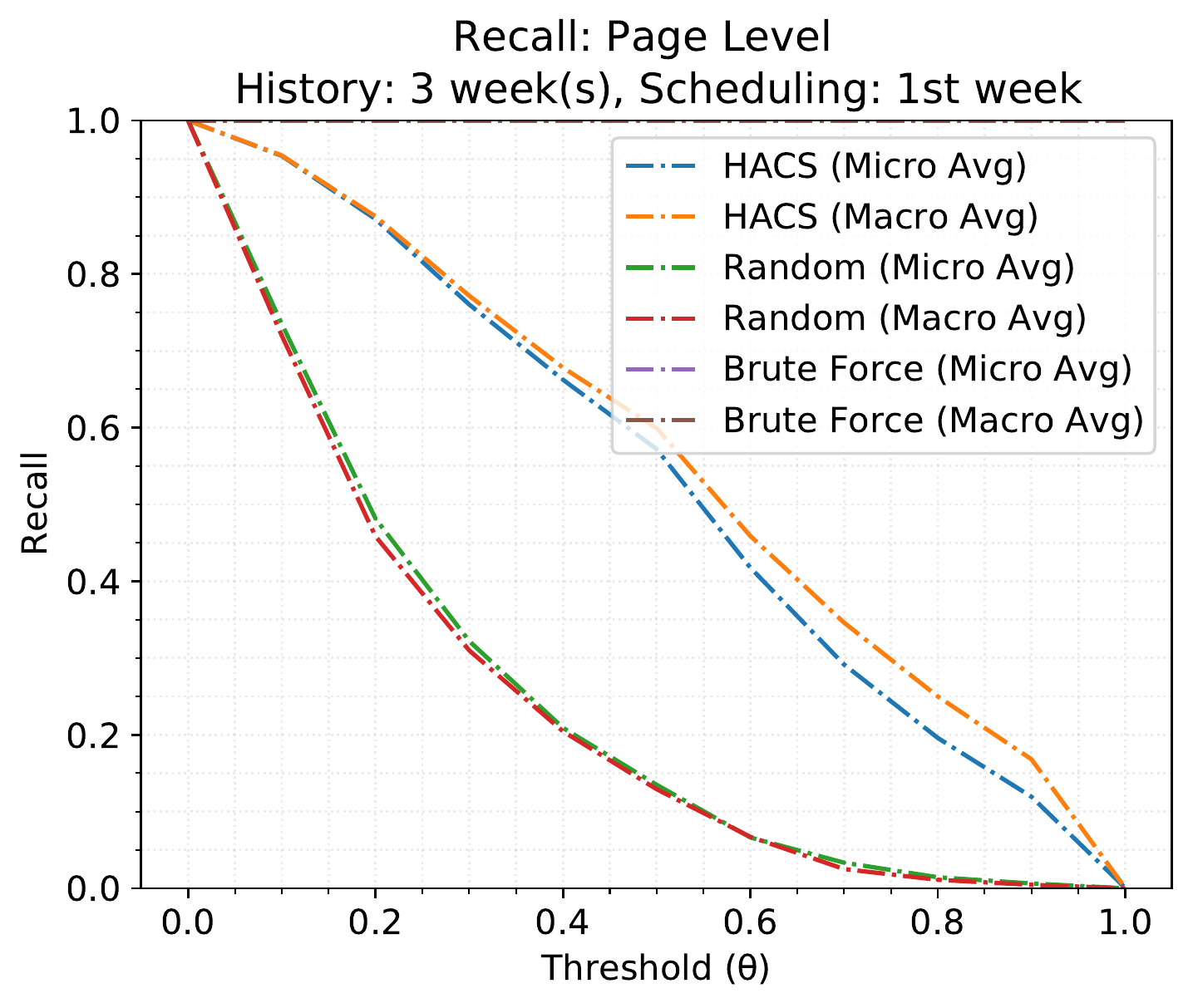}
}
\subfigure[Website-level, History = 3 weeks]{
\includegraphics[width=.44\linewidth,trim={0 0 0 40},clip]{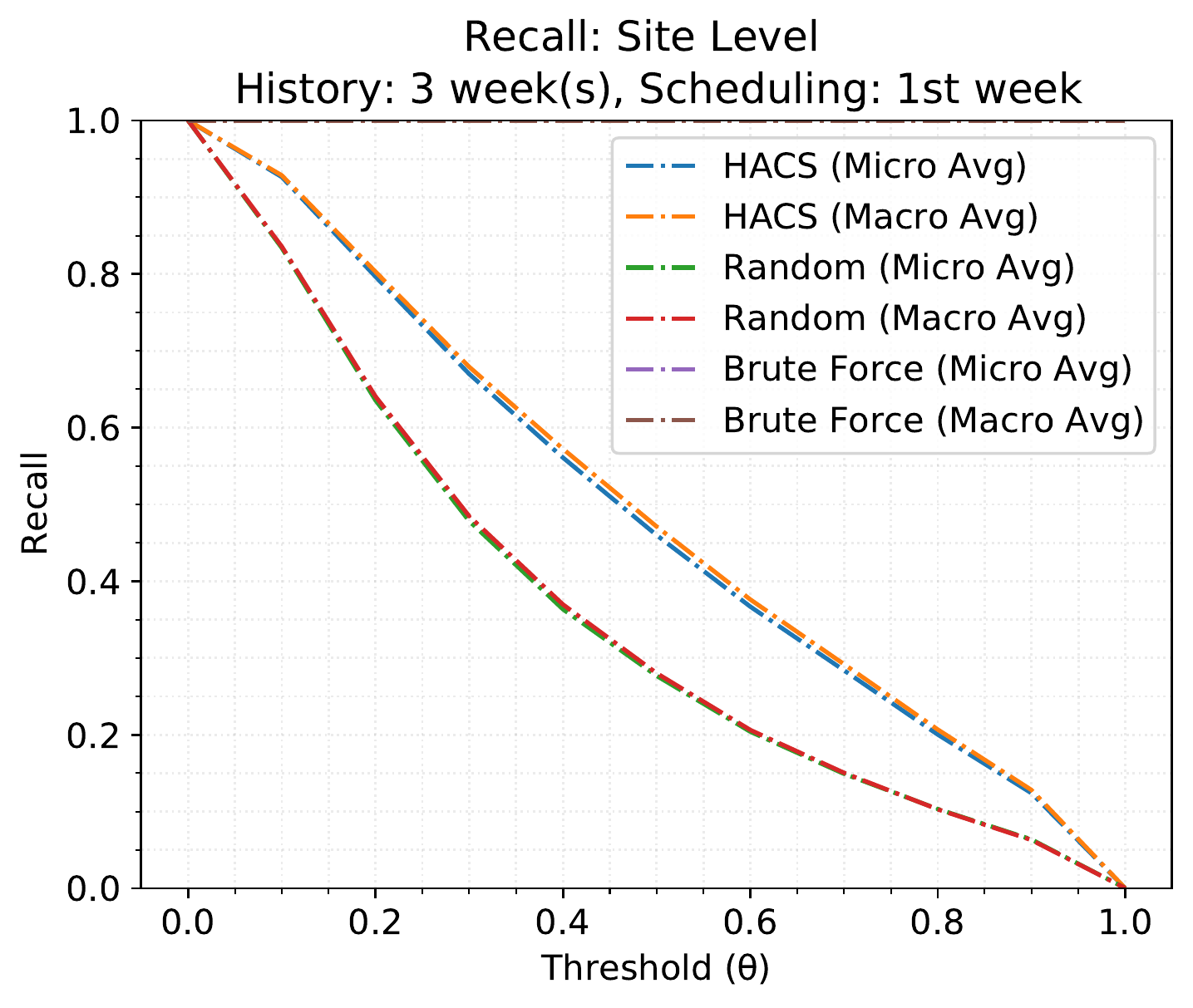}
}
\caption{Recall (${R}$) vs Threshold (${\theta}$) when a history of 3 weeks is used}
\end{figure*}

Figure 14 illustrates the Recall ($R$) vs Threshold ($\theta$) of each model, when a history size of three weeks is used.
Here too, the HACS model produced a higher $R$ than other baseline models for all values of $\theta$, at both homepage level and website level.
This lead is more visible at homepage-level than website-level.
However, compared to a history size of 1 week and 2 weeks, this lead is less prominent at both homepage-level and webpage-level.
The Brute Force model has a consistent ${R}$ of $1.0$, as it selects all seed URLs regardless.
The Random model has a low ${R}$, regardless of $\theta$.

\begin{figure}[ht]
\centering
\subfigure[Homepage-level, $1/\widetilde{\lambda}=7$ days]{
\includegraphics[width=.44\linewidth,trim={0 0 0 40},clip]{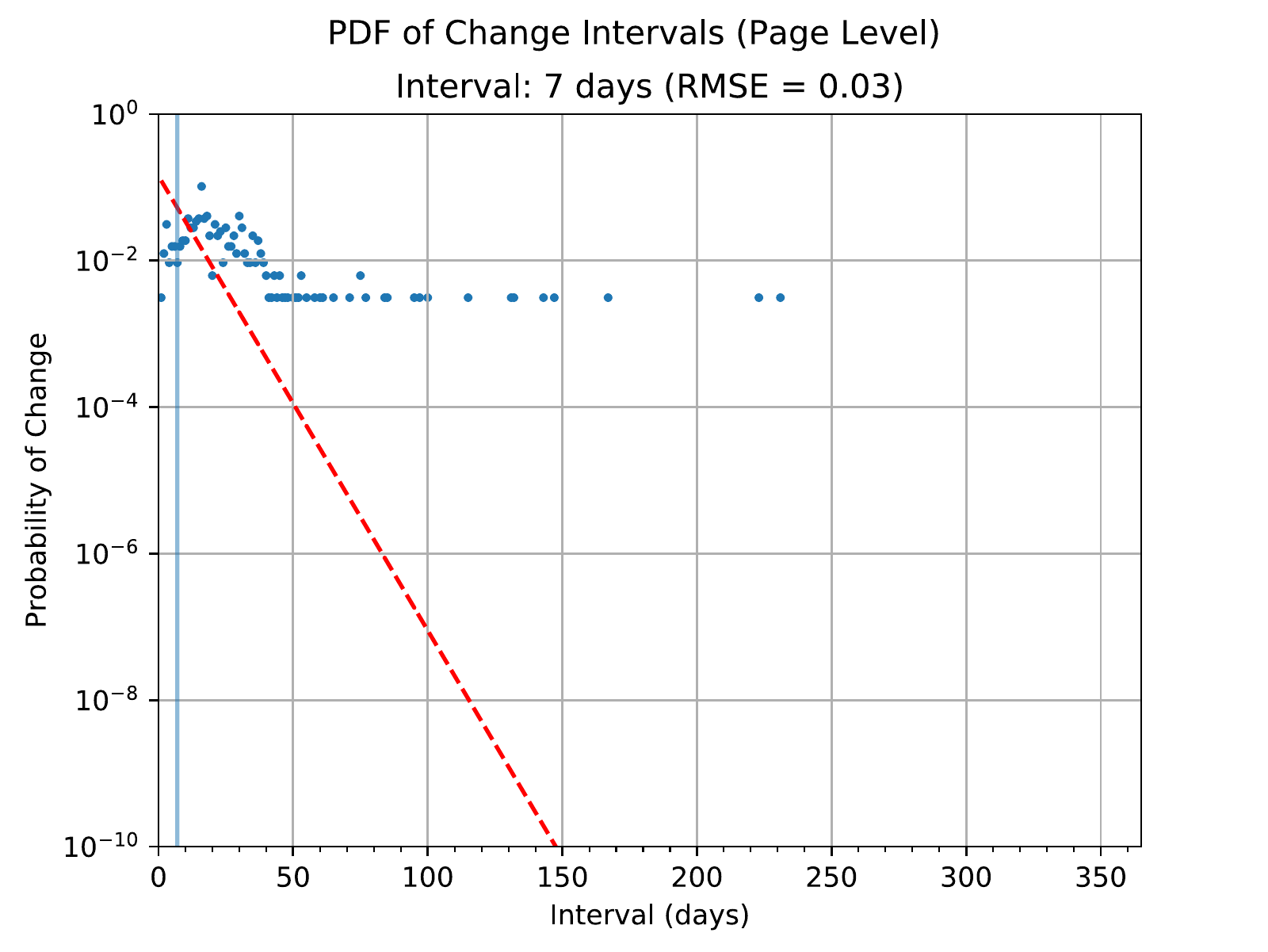}
}
\enskip
\subfigure[Homepage-level, $1/\widetilde{\lambda}=14$ days]{
\includegraphics[width=.44\linewidth,trim={0 0 0 40},clip]{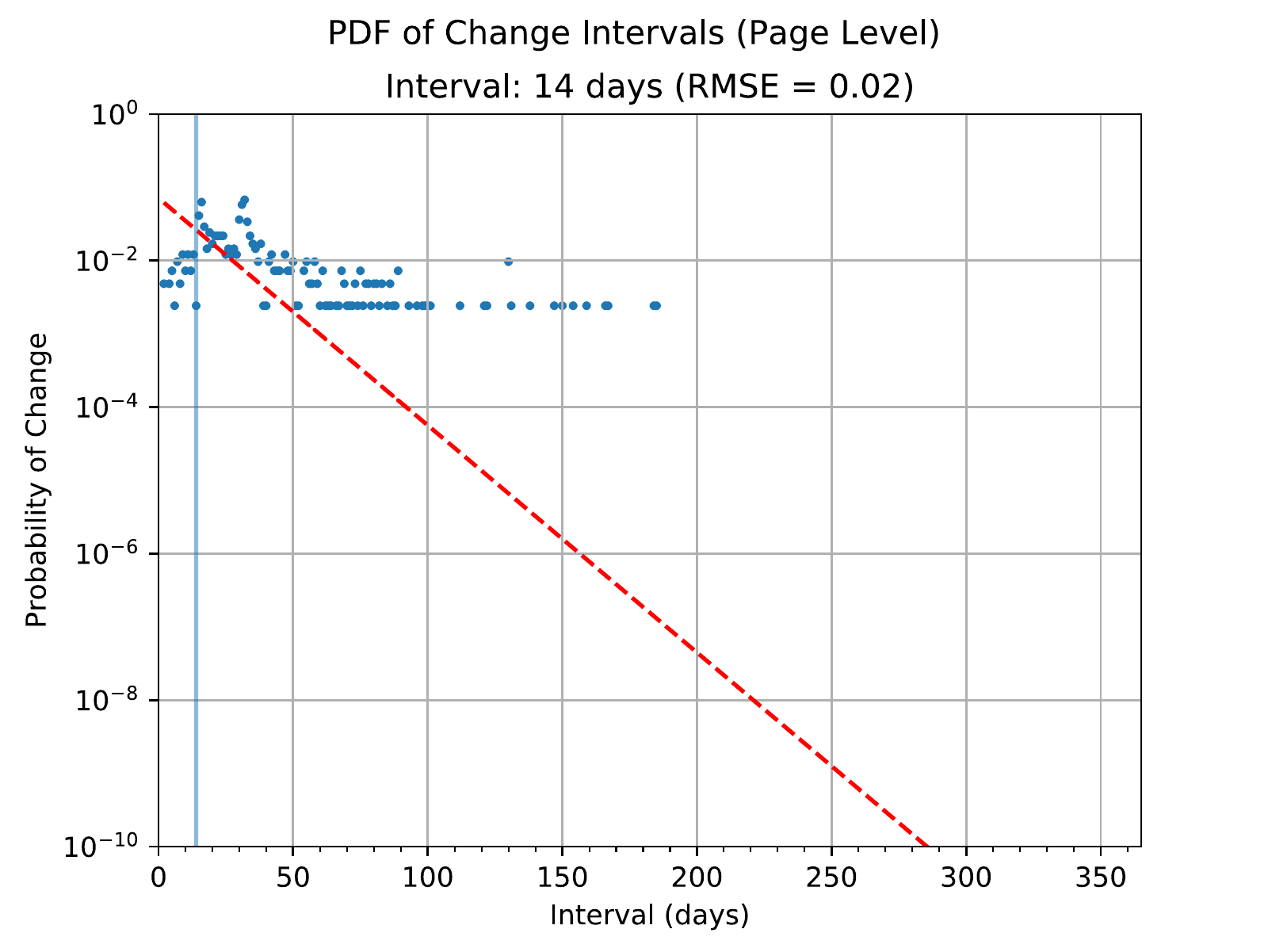}
}
\enskip
\subfigure[Homepage-level, $1/\widetilde{\lambda}=21$ days]{
\includegraphics[width=.44\linewidth,trim={0 0 0 40},clip]{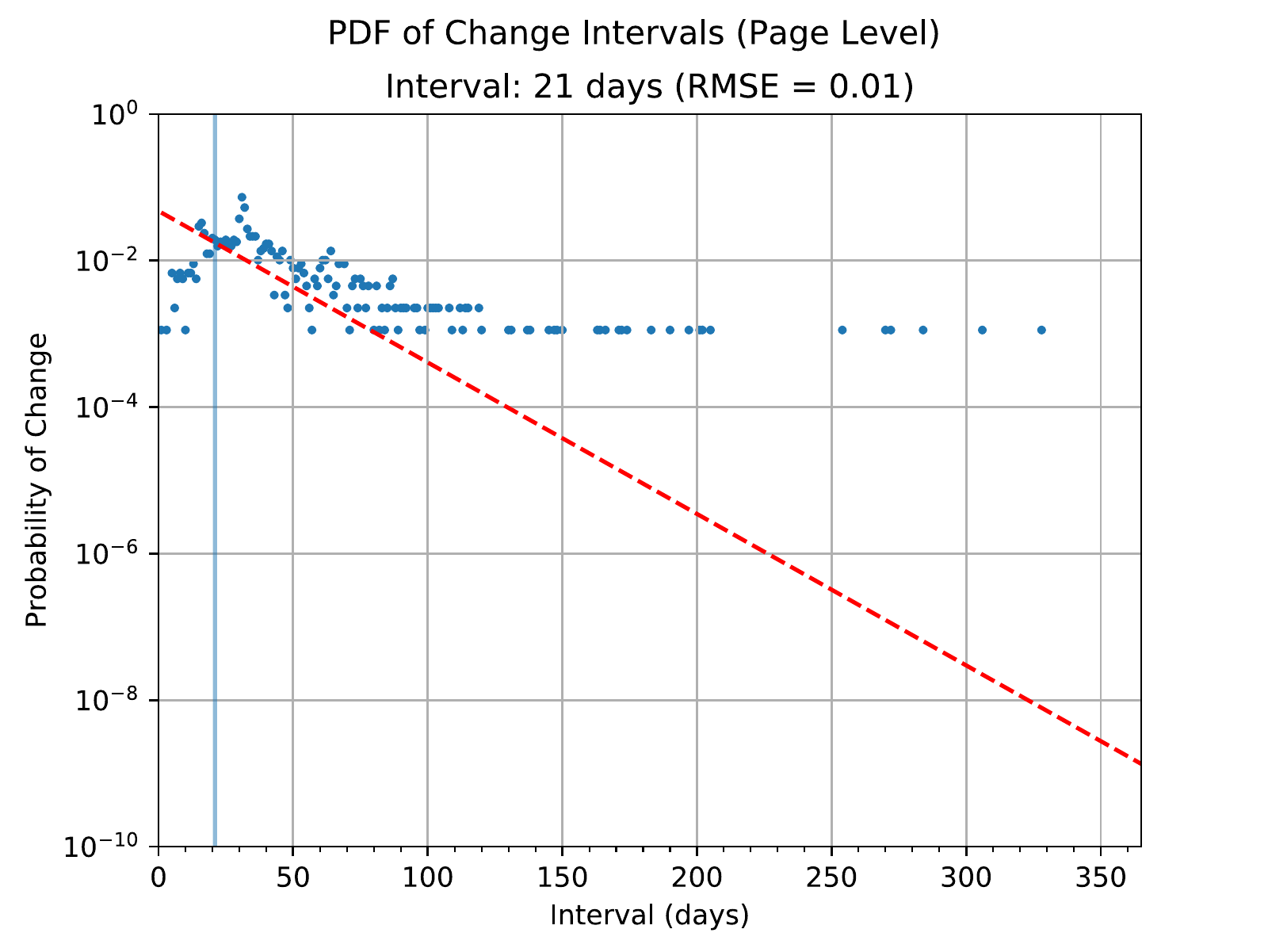}
}
\enskip
\subfigure[Homepage-level, $1/\widetilde{\lambda}=28$ days]{
\includegraphics[width=.44\linewidth,trim={0 0 0 40},clip]{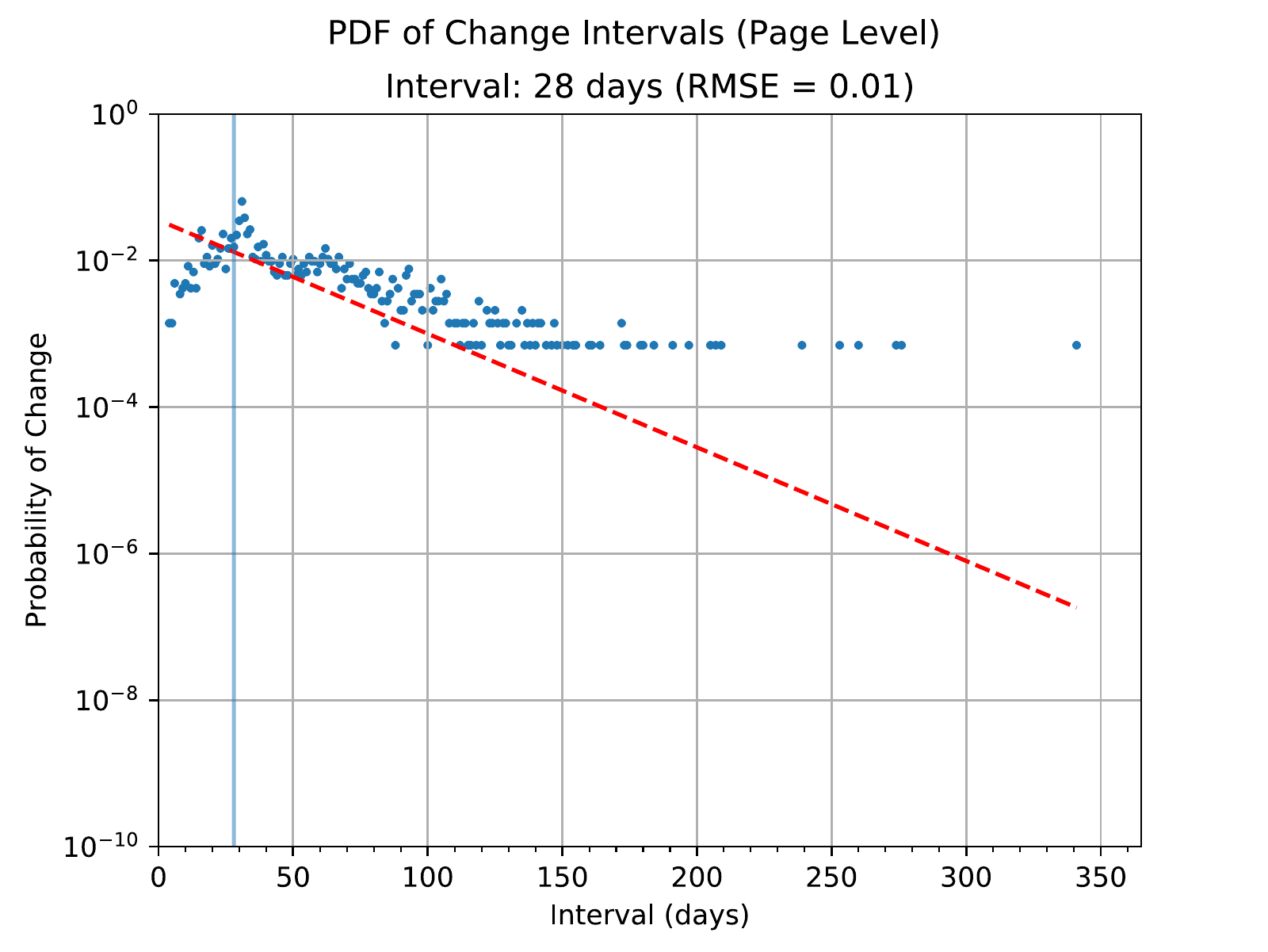}
}
\caption{
Probability (y-axis) of finding author websites with an interpolated update interval~($\Delta{t}$) of $d$ days (x-axis) at homepage-level, among author websites having $1/\widetilde{\lambda}$ across different bin sizes.
The vertical blue line shows where $d=1/\widetilde{\lambda}$.}
\end{figure}
\begin{figure}[ht]
\subfigure[Homepage-level, $1/\widetilde{\lambda}=42$ days]{
\includegraphics[width=.44\linewidth,trim={0 0 0 40},clip]{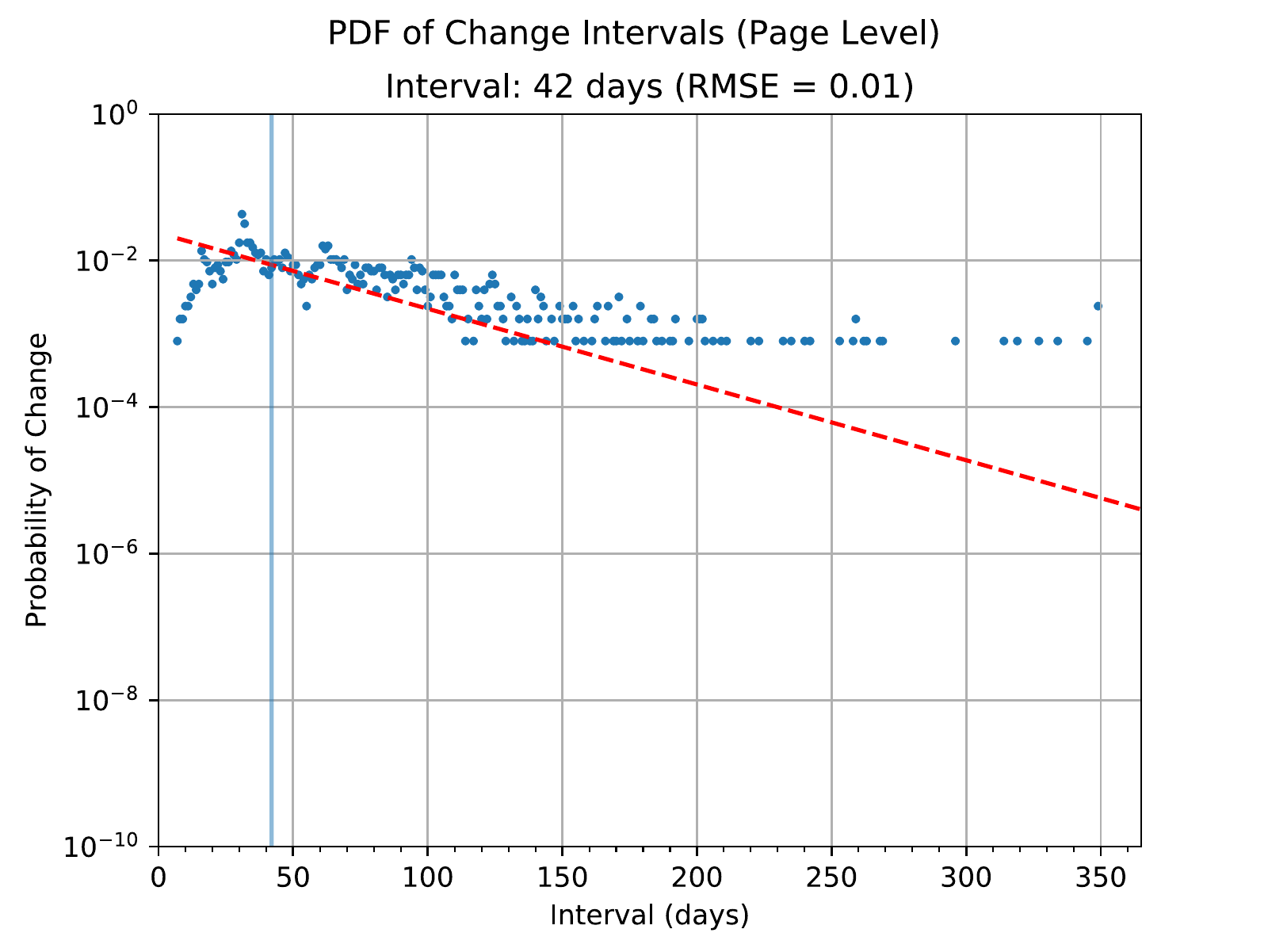}
}
\enskip
\subfigure[Homepage-level, $1/\widetilde{\lambda}=49$ days]{
\includegraphics[width=.44\linewidth,trim={0 0 0 40},clip]{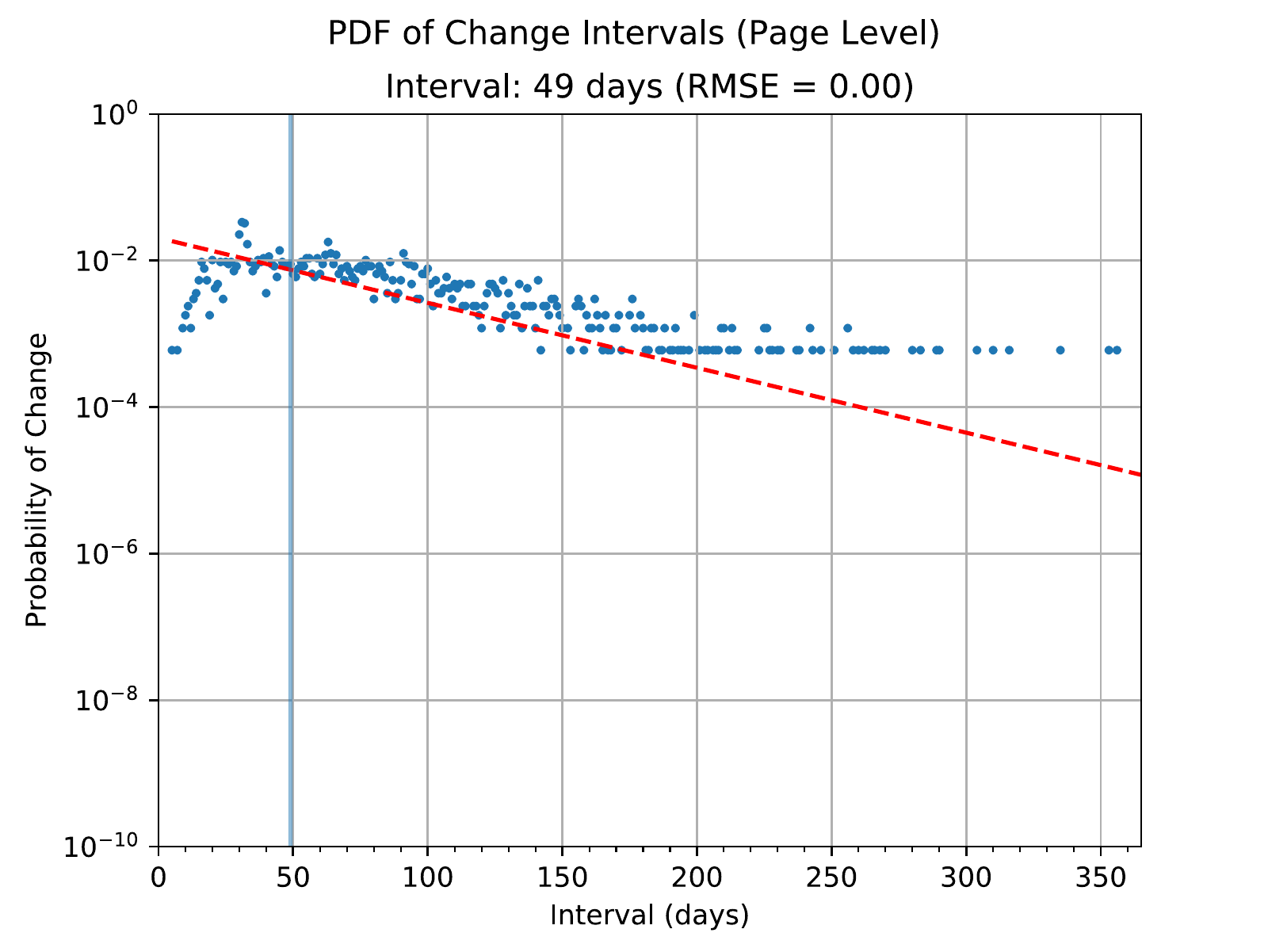}
}
\enskip
\subfigure[Homepage-level, $1/\widetilde{\lambda}=56$ days]{
\includegraphics[width=.44\linewidth,trim={0 0 0 40},clip]{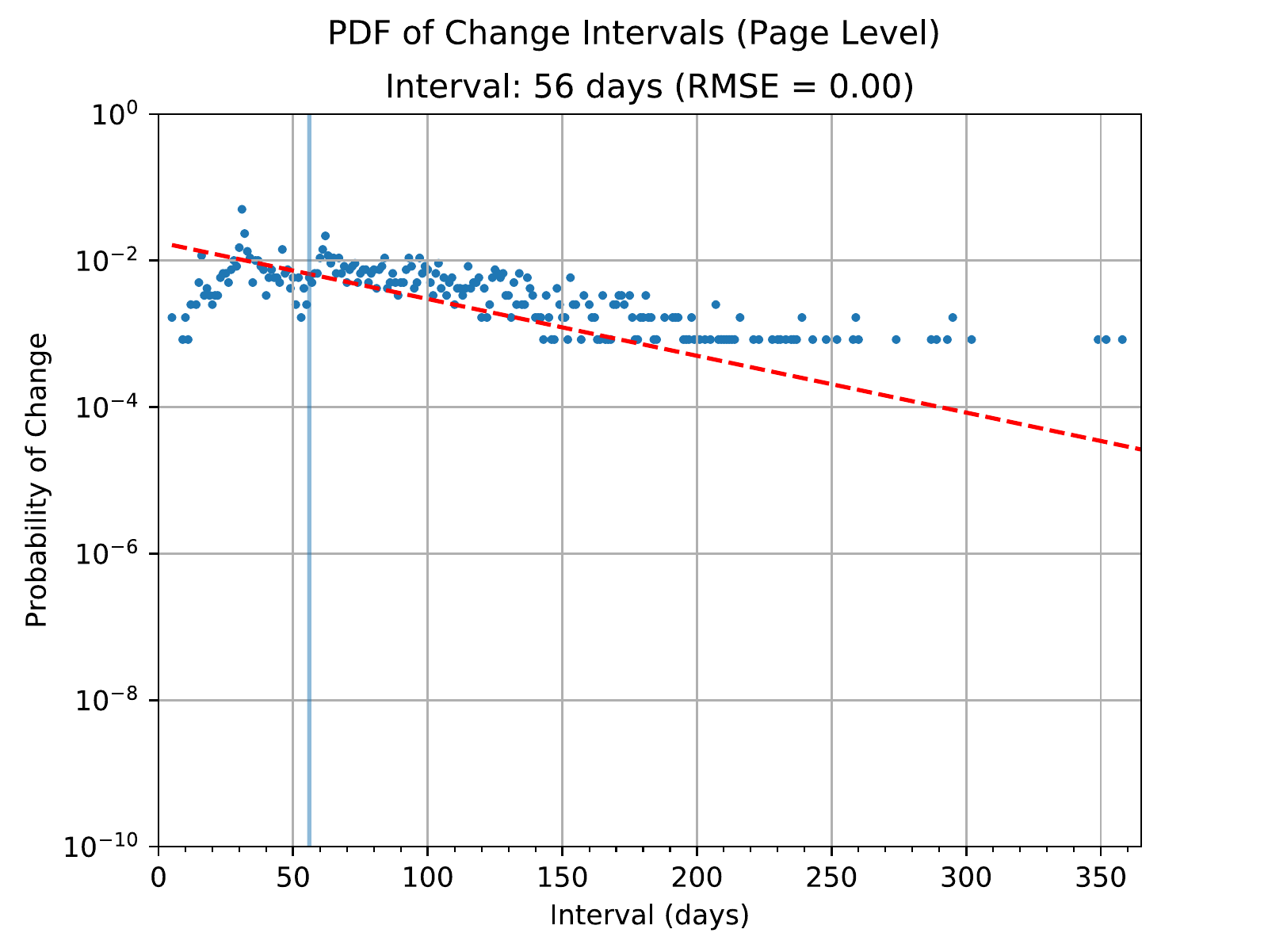}
}
\enskip
\subfigure[Homepage-level, $1/\widetilde{\lambda}=63$ days]{
\includegraphics[width=.44\linewidth,trim={0 0 0 40},clip]{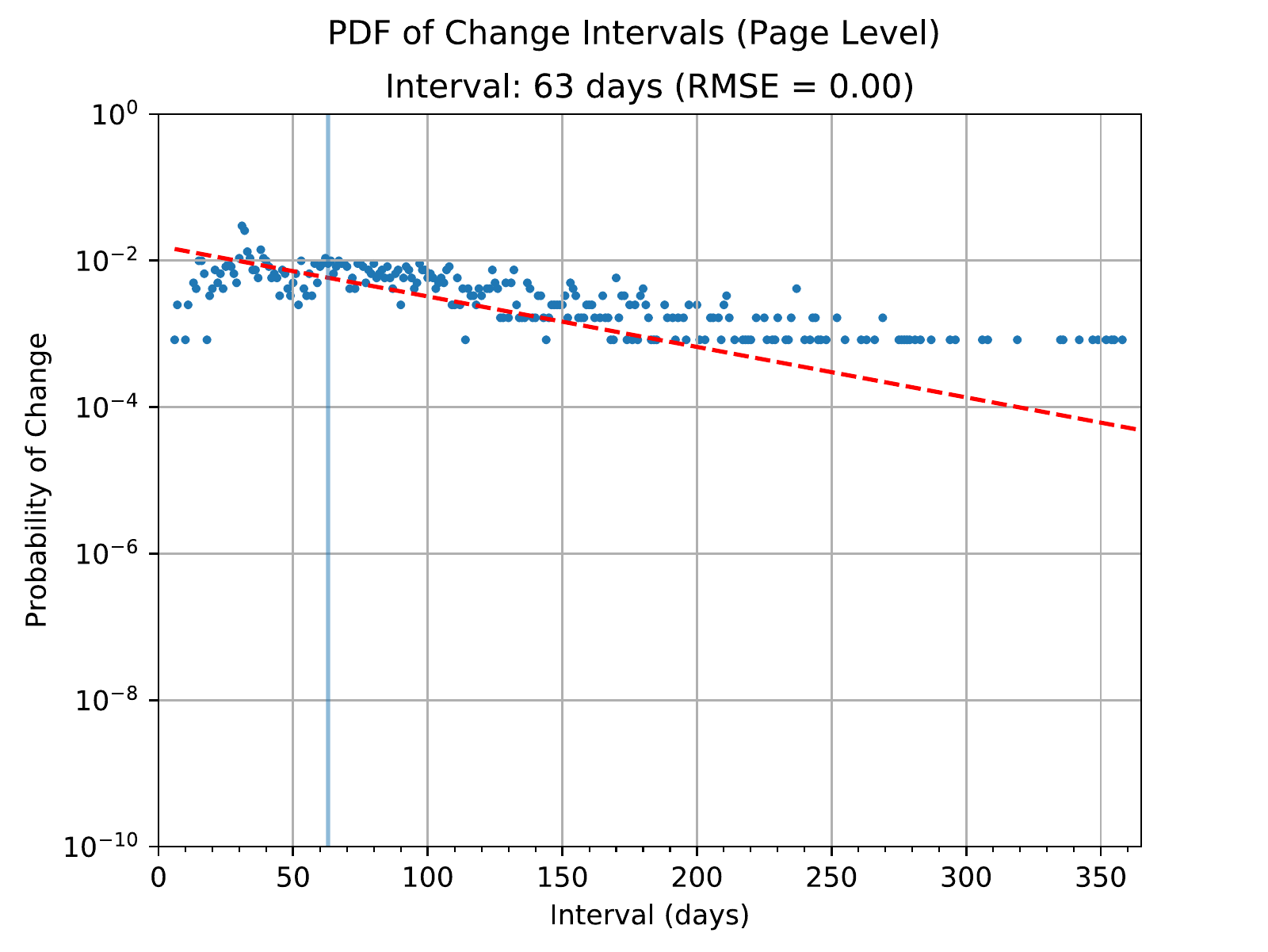}
}
\caption{
Probability (y-axis) of finding author websites with an interpolated update interval~($\Delta{t}$) of $d$ days (x-axis) at homepage-level, among author websites having $1/\widetilde{\lambda}$ across different bin sizes.
The vertical blue line shows where $d=1/\widetilde{\lambda}$.}
\end{figure}

\begin{figure}[ht]
\centering
\subfigure[Website-level, $1/\widetilde{\lambda}=7$ days]{
\includegraphics[width=.44\linewidth,trim={0 0 0 40},clip]{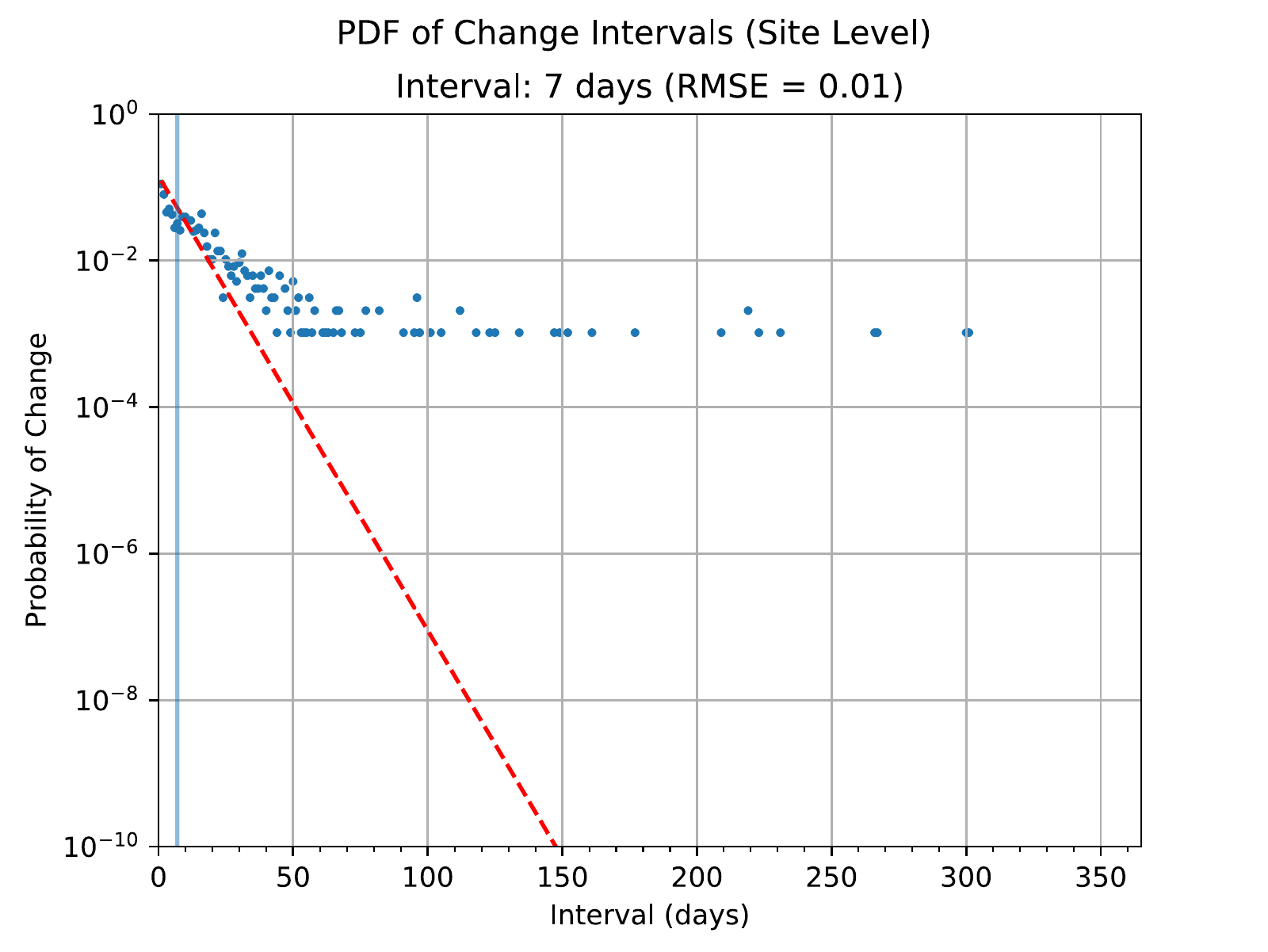}
}
\enskip
\subfigure[Website-level, $1/\widetilde{\lambda}=14$ days]{
\includegraphics[width=.44\linewidth,trim={0 0 0 40},clip]{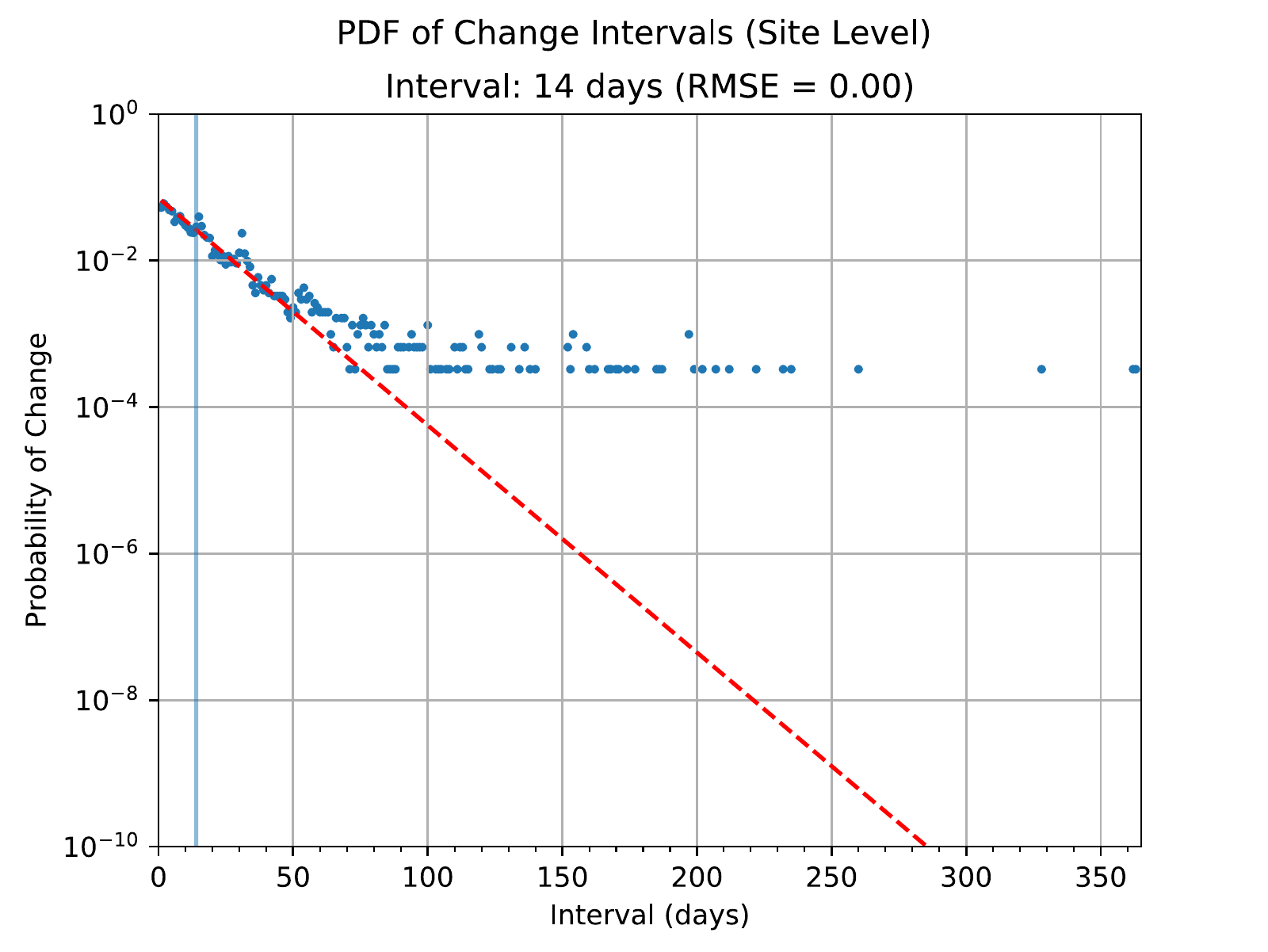}
}
\enskip
\subfigure[Website-level, $1/\widetilde{\lambda}=21$ days]{
\includegraphics[width=.44\linewidth,trim={0 0 0 40},clip]{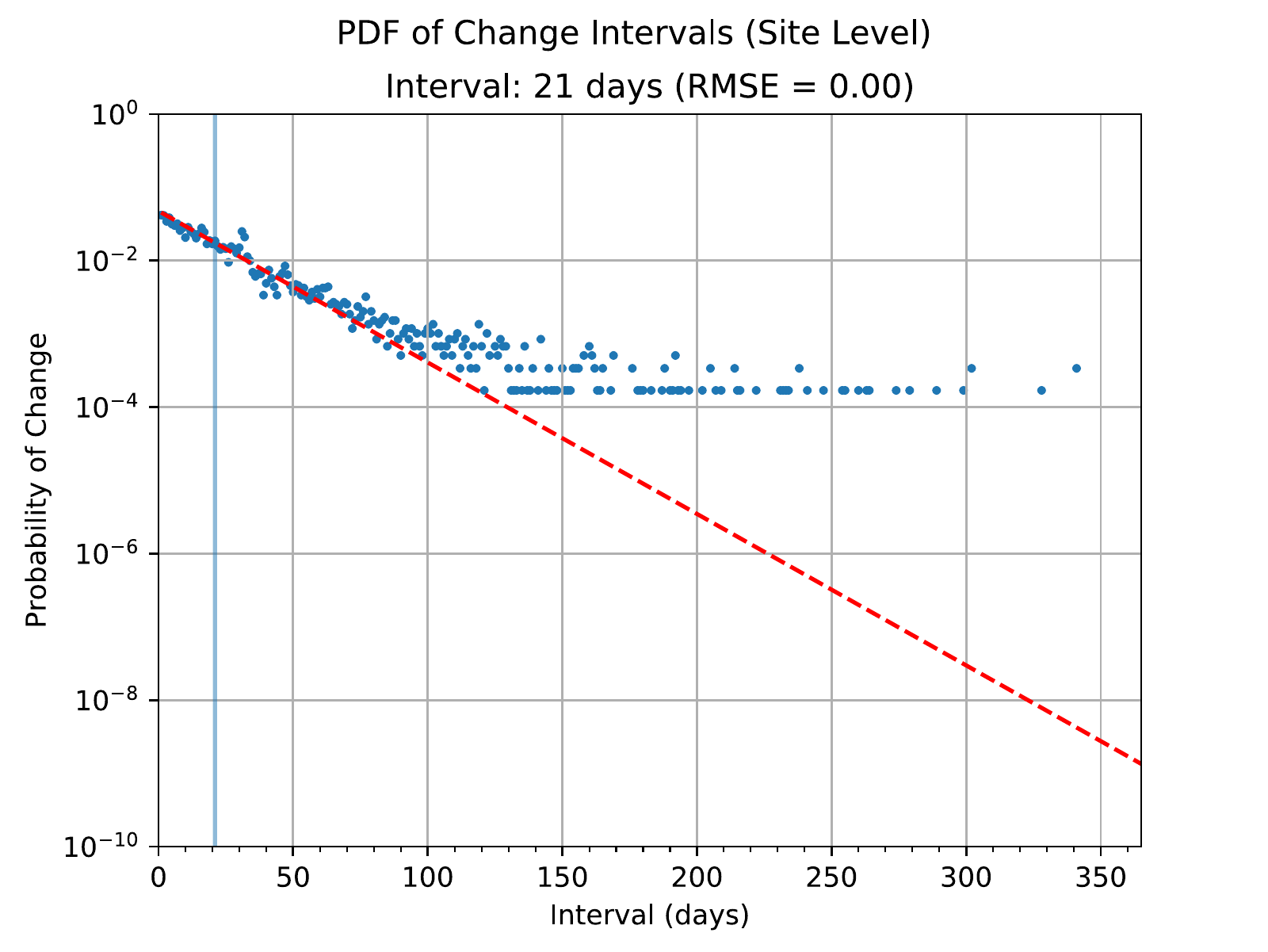}
}
\enskip
\subfigure[Website-level, $1/\widetilde{\lambda}=28$ days]{
\includegraphics[width=.44\linewidth,trim={0 0 0 40},clip]{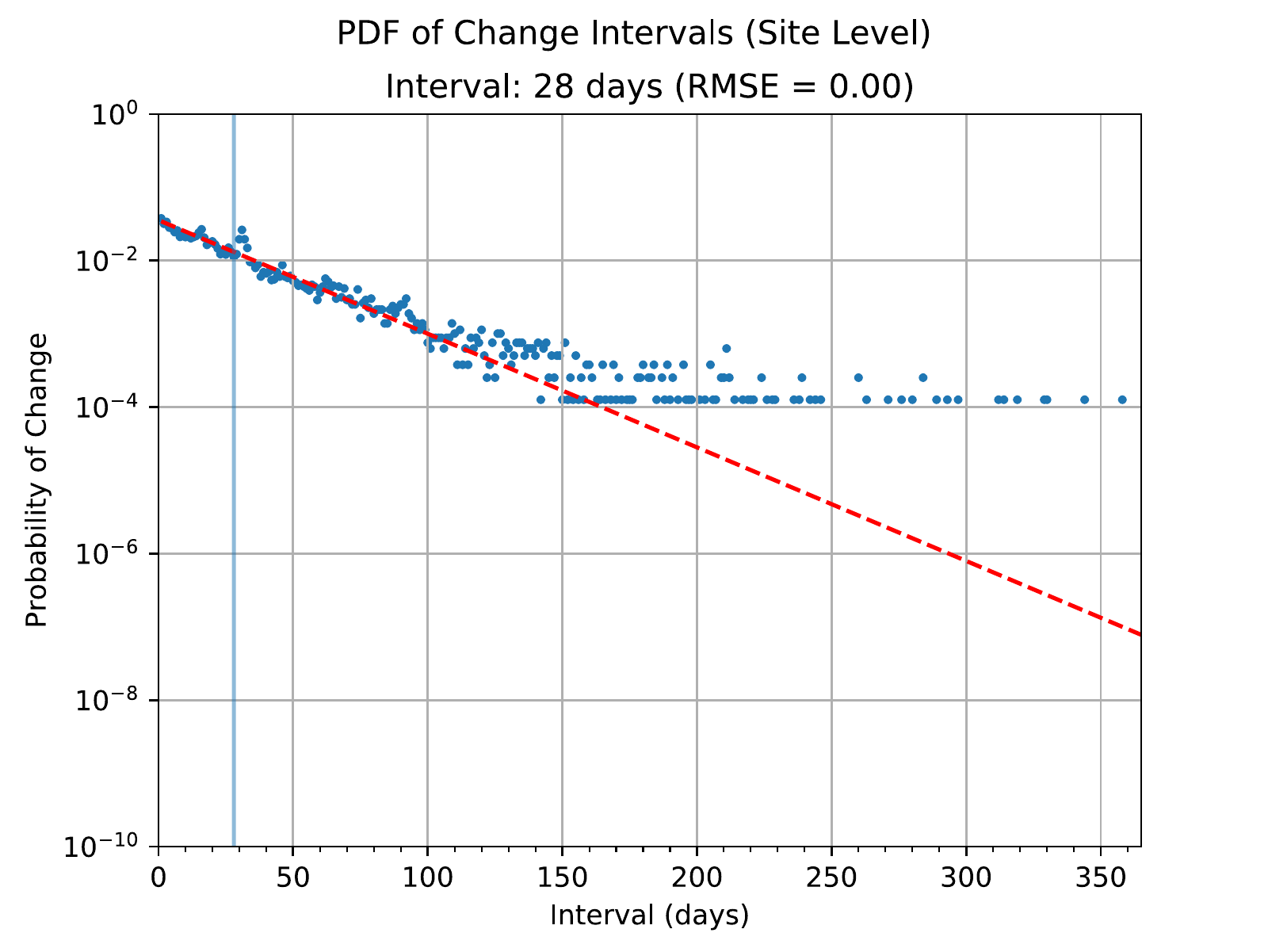}
}
\caption{
Probability (y-axis) of finding author websites with an interpolated update interval~($\Delta{t}$) of $d$ days (x-axis) at website-level, among author websites having $1/\widetilde{\lambda}$ across different bin sizes.
The vertical blue line shows where $d=1/\widetilde{\lambda}$.}
\end{figure}
\begin{figure}[ht]
\subfigure[Website-level, $1/\widetilde{\lambda}=42$ days]{
\includegraphics[width=.44\linewidth,trim={0 0 0 40},clip]{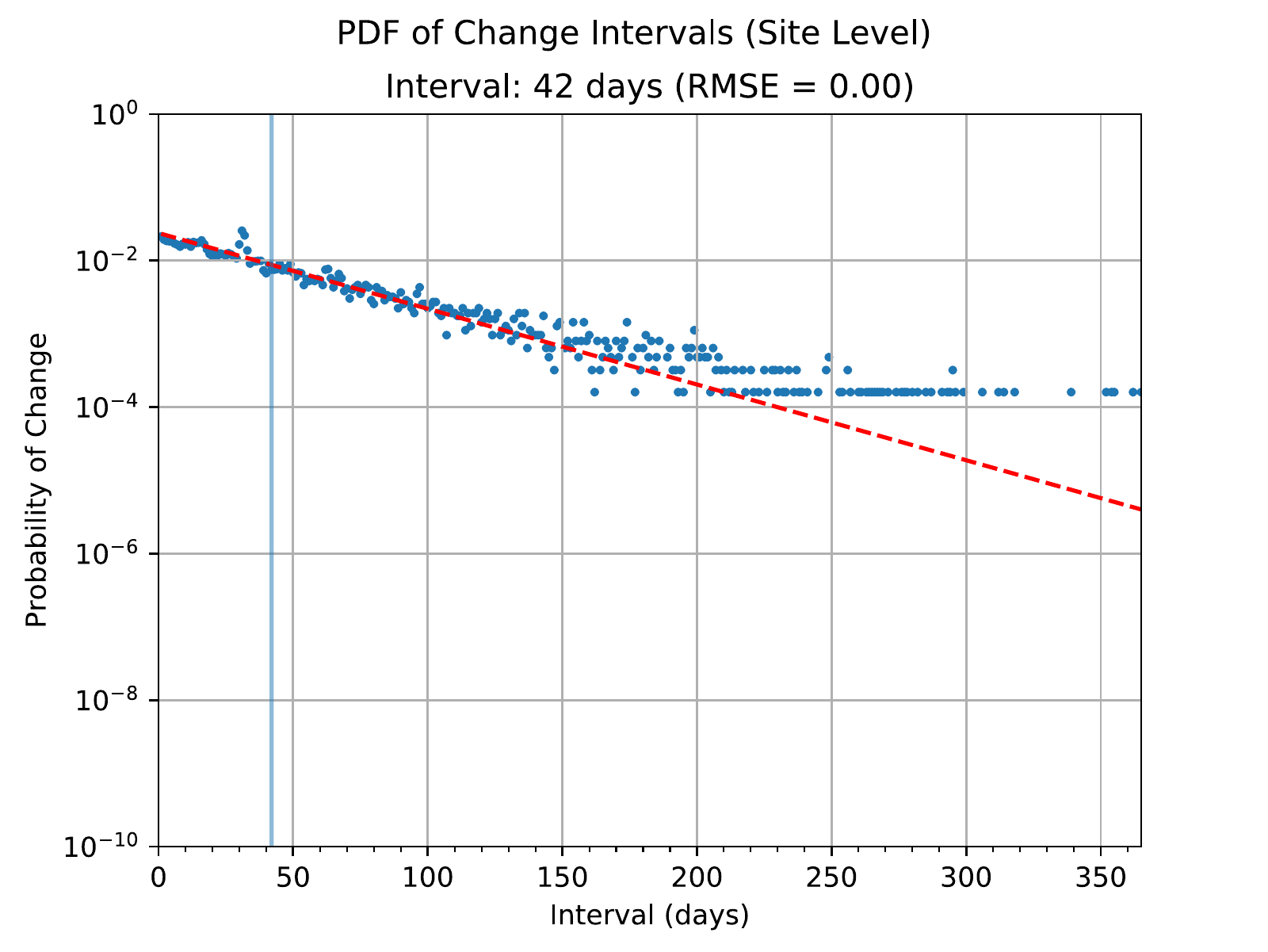}
}
\enskip
\subfigure[Website-level, $1/\widetilde{\lambda}=49$ days]{
\includegraphics[width=.44\linewidth,trim={0 0 0 40},clip]{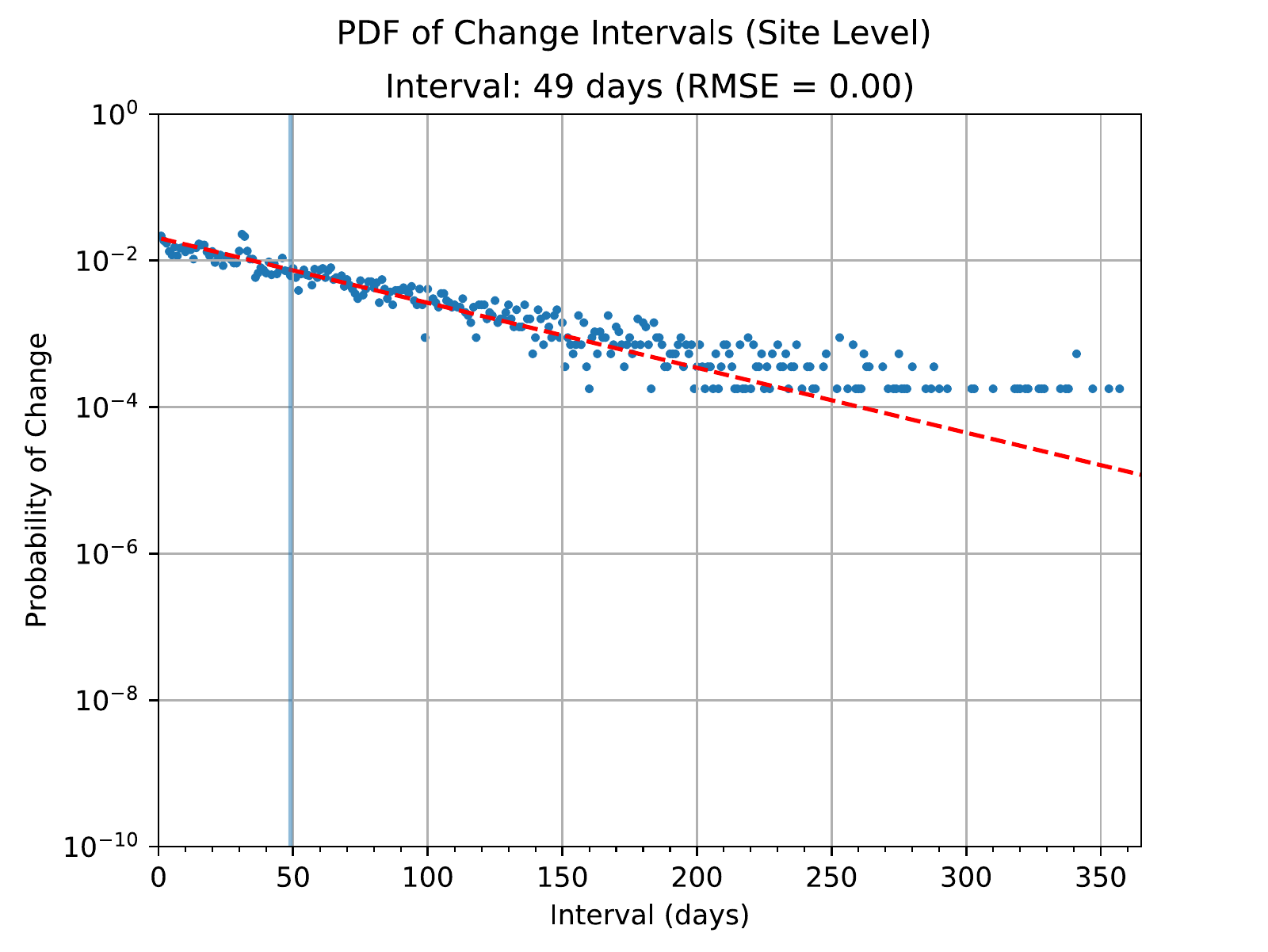}
}
\enskip
\subfigure[Website-level, $1/\widetilde{\lambda}=56$ days]{
\includegraphics[width=.44\linewidth,trim={0 0 0 40},clip]{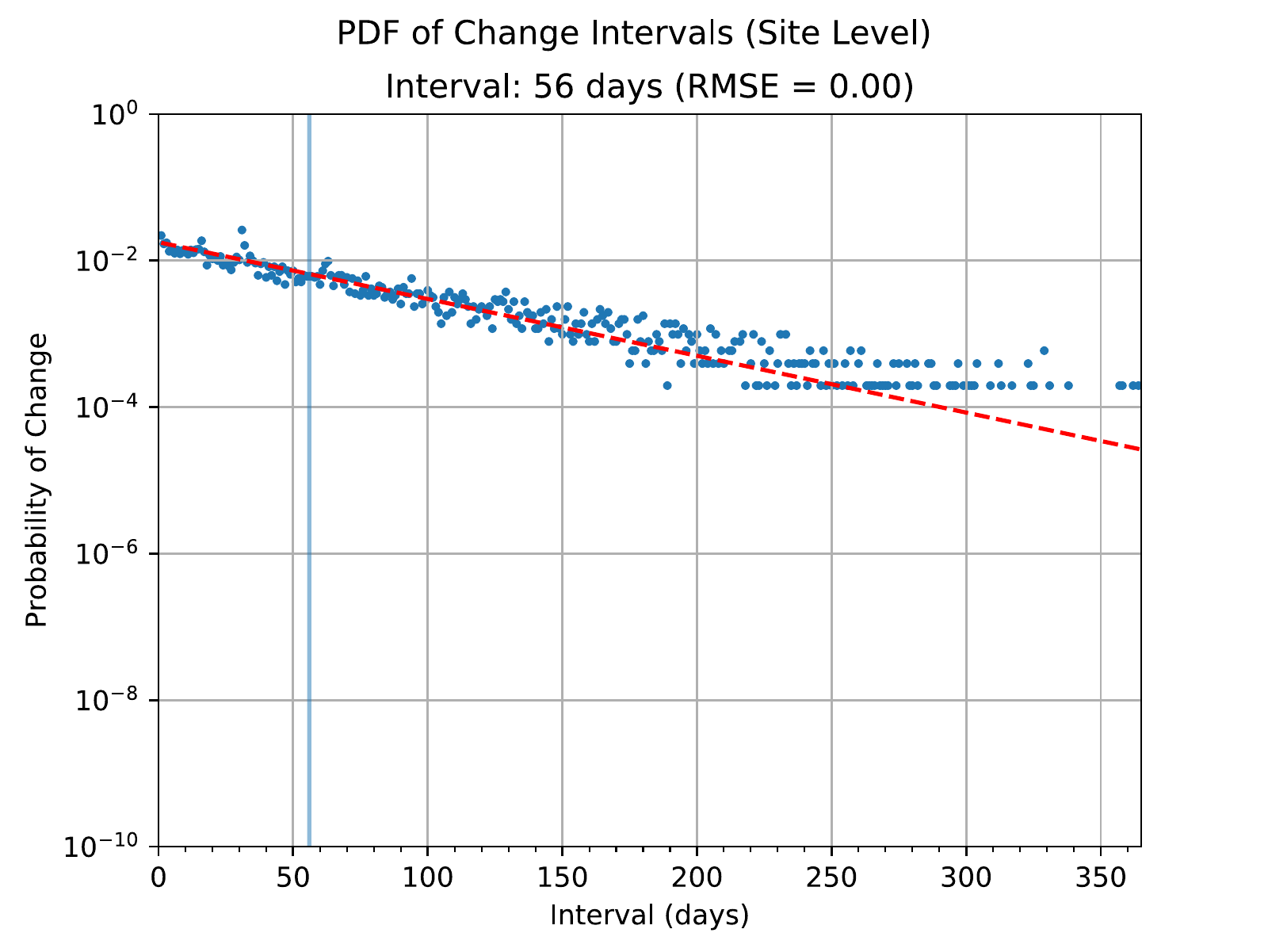}
}
\enskip
\subfigure[Website-level, $1/\widetilde{\lambda}=63$ days]{
\includegraphics[width=.44\linewidth,trim={0 0 0 40},clip]{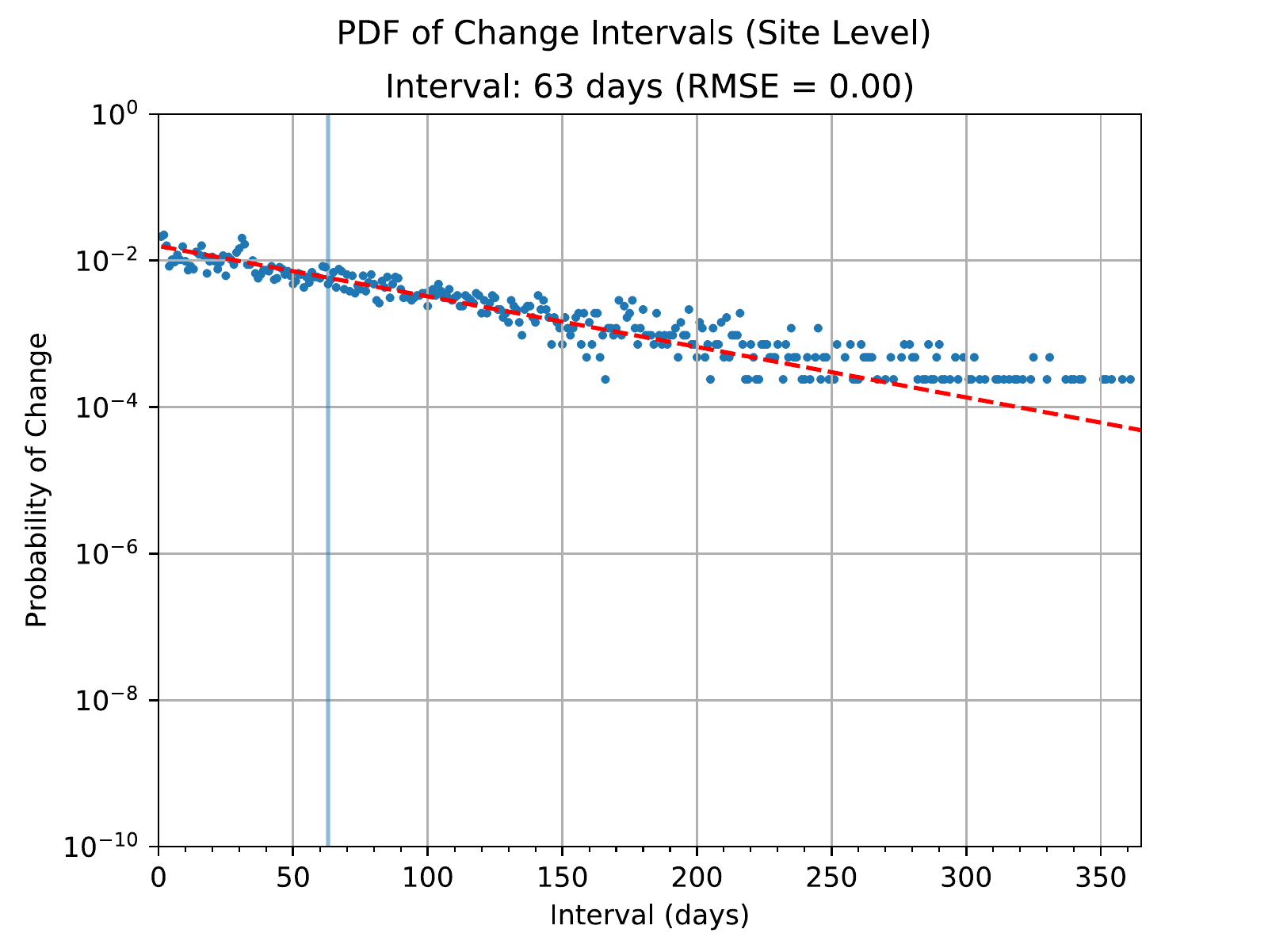}
}
\caption{
Probability (y-axis) of finding author websites with an interpolated update interval~($\Delta{t}$) of $d$ days (x-axis) at website-level, among author websites having $1/\widetilde{\lambda}$ across different bin sizes.
The vertical blue line shows where $d=1/\widetilde{\lambda}$.}
\end{figure}

Figures 15, 16, 17, and 18 illustrates the probability of finding author websites with an interpolated update interval of $d$ days for additional values of $d$, ranging from 7 days to 70 days, at both homepage-level (see Figure 15) and webpage-level (see Figure 16).
The results suggest that as $d$ increases, the probability distribution gets closer to the expected poisson distribution in both cases.

\end{document}